\documentclass[pra,10pt,twocolumn,floatfix,groupedaddress]{revtex4-1} 

\usepackage[dvips]{graphicx}
\usepackage{subfigure}
\usepackage{amsmath}
\usepackage{amssymb}
\usepackage{bm}

\graphicspath{{./}}
\def\Eq{Eq.~}
\def\Eqs{Eqs.~}
\def\Fig{Fig.~}
\def\Figs{Figs.~}
\def\Ref{Ref.~}

\def\Sec{Sec.~}

\def\be{\begin{equation}}
\def\ee{\end{equation}}
\def\bea{\begin{eqnarray}}
\def\eea{\end{eqnarray}}

\newcommand{\ket}[1]{\left| #1 \right\rangle}

\newcommand{\refeqn}[1]{(\ref{#1})}

\begin{document}

\title{Atom interferometric techniques for measuring uniform magnetic field gradients\\and gravitational acceleration}

\author{B. Barrett}
\author{I. Chan}
\author{A. Kumarakrishnan}
\affiliation{Department of Physics \& Astronomy, York University, Toronto, Ontario M3J 1P3, Canada}

\date{\today}


\begin{abstract}
We discuss techniques for probing the effects of a constant force acting on cold atoms using two configurations of a grating echo-type atom interferometer. Laser-cooled samples of $^{85}$Rb with temperatures as low as 2.4 $\mu$K have been achieved in a new experimental apparatus with a well-controlled magnetic environment. We demonstrate interferometer signal lifetimes approaching the transit time limit in this system ($\sim 270$ ms), which is comparable to the timescale achieved by Raman interferometers. Using these long timescales, we experimentally investigate the influence of a homogeneous magnetic field gradient using two- and three-pulse interferometers, which enable us to sense changes in externally applied magnetic field gradients as small as $\sim 4 \times 10^{-5}$ G/cm. We also provide an improved theoretical description of signals generated by both interferometer configurations that accurately models experimental results. With this theory, absolute measurements of $B$-gradients at the level of $3 \times 10^{-4}$ G/cm are achieved. Finally, we contrast the suitability of the two- and three-pulse interferometers for precision measurements of the gravitational acceleration, $g$.
\end{abstract}

\maketitle

\section{Introduction}

Atom interferometers (AIs) have been employed to investigate a host of inertial effects over the past few decades. Such effects include the acceleration due to gravity \cite{Kasevich-PRL-1991, Peters-Nature-1999, Peters-Metrologia-2001, Hughes-PRL-2009, Poli-PRL-2011}, gravity gradients \cite{Snadden-PRL-1998, McGuirk-PRA-2002, Yu-ApplPhysB-2006}, and rotations \cite{Gustavson-PRL-1997, Wu-PRL-2007, Burke-PRA-2009}. Raman interferometric measurements of gravity \cite{Kasevich-PRL-1991, Peters-Nature-1999, Peters-Metrologia-2001} use cold atoms and transit time limited experiments in an atomic fountain to reach a precision of $\sim 3$ parts per $10^9$ (ppb) with 1 minute of interrogation time. This technique requires two phase-locked lasers to drive Raman transitions between two hyperfine ground states. It also requires state selection into the $m_F = 0$ magnetic sub-level to avoid sensitivity to $B$-fields and $B$-gradients, as well as velocity selection to guarantee that all interfering atoms have the same initial sub-recoil velocity.

In contrast to the Raman interferometer, the grating echo-type AI \cite{Cahn-PRL-1997, Strekalov-PRA-2002} uses a single off-resonant excitation frequency that drives a cycling transition with the same initial and final state. This AI requires no state or velocity selection, and is insensitive to both the AC Stark effect and the Zeeman effect \footnote{This condition is true for far off-resonant excitation fields only. For fields closer to resonance, both the AC Stark effect and the Zeeman effect can induce a relative shift between the ground and excited states, thus affecting the response of the interferometer in a systematic way.}. Additionally, as we will show on the basis of a theoretical model, the intensity of the AI signal is insensitive to uniform $B$-gradients provided the atoms are pumped into a single magnetic sub-level---which need not be $m_F = 0$.

\begin{figure}[!b]
  \centering
  \hspace{-0.10cm}
  \subfigure{\label{fig:1a-RecoilDiagram-TwoPulse}
  \includegraphics[width=0.46\textwidth]{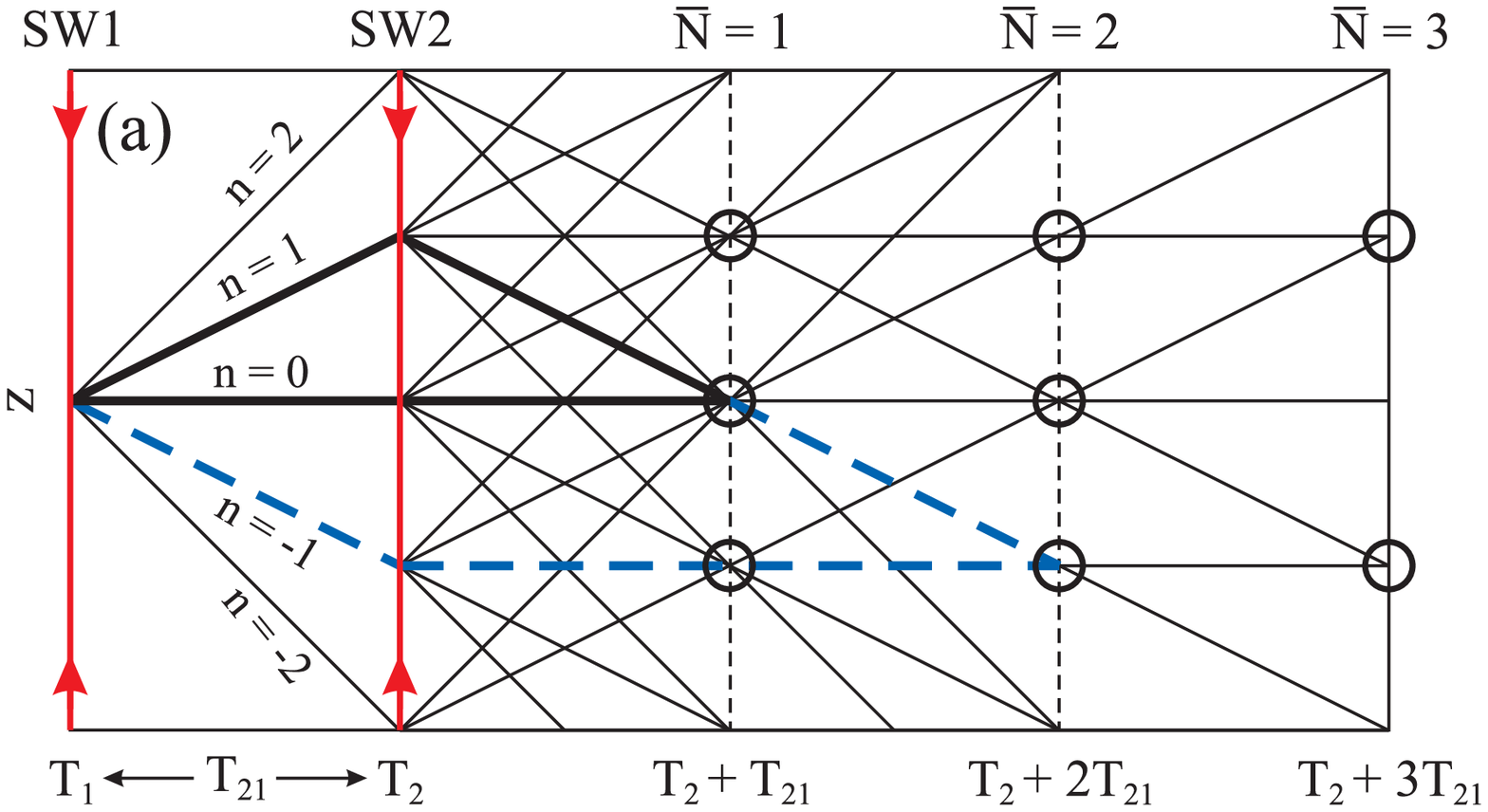}}
  \subfigure{\label{fig:1b-RecoilDiagram-ThreePulse}
  \includegraphics[width=0.45\textwidth]{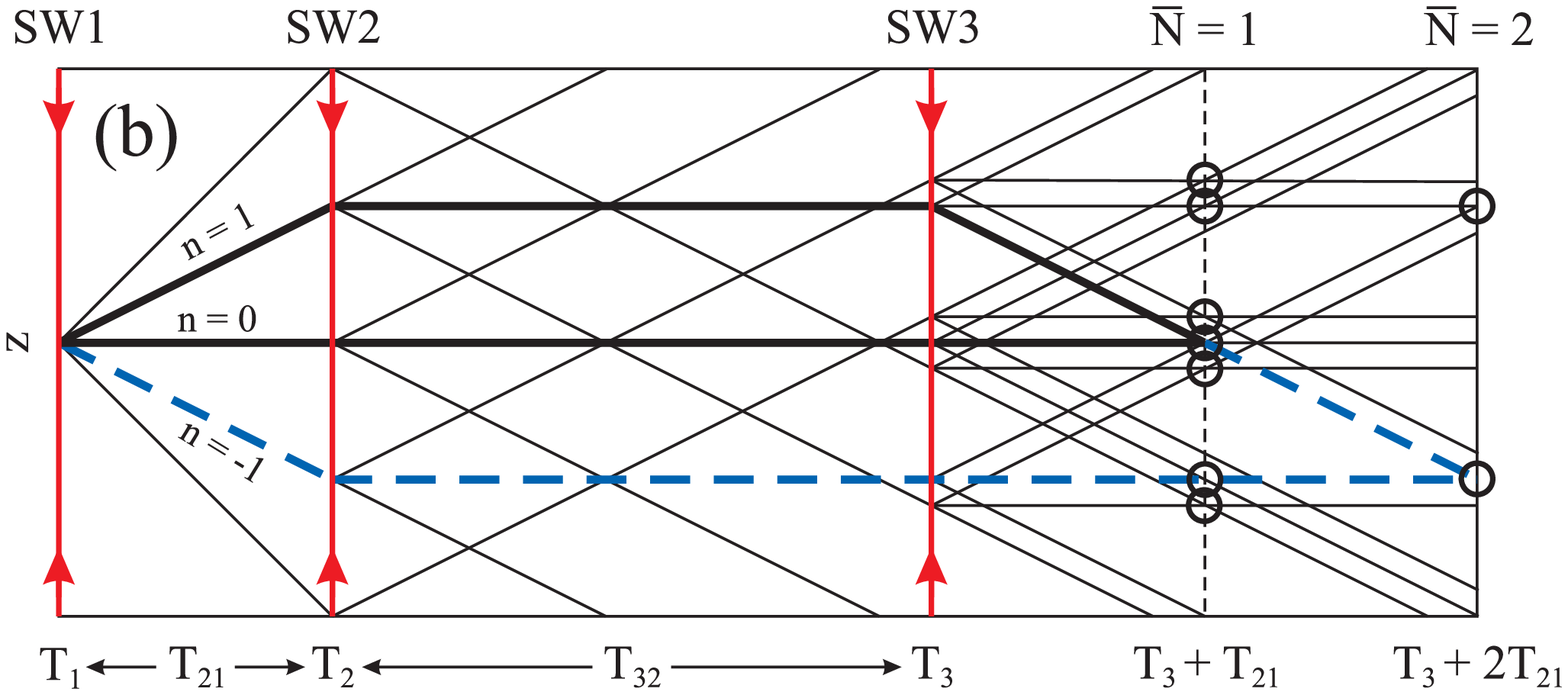}}
  \caption{(Color online) Recoil diagrams for the two-pulse (a) and three-pulse (b) AIs in the absence of any external forces. Standing wave excitations are labeled by SW1 -- SW3. At $t = T_1$ the atom is diffracted into a superposition of momentum states differing by integer ($n$) multiples of $\hbar q$. A second sw pulse is applied at $t = T_2$ which further diffracts the atoms. In the two-pulse case, interference between states differing by $\hbar q$ ($\Delta n = \pm 1$) occurs at times $t^{(2)}_{\rm{echo}} = T_1 + (\bar{N} + 1) T_{21}$. For the three-pulse scheme, a third sw pulse is applied at $t = T_3$ which produces interference at times $t^{(3)}_{\rm{echo}} = T_1 + (\bar{N} + 1) T_{21} + T_{32}$. Examples of interfering trajectories for $\bar{N} = 1$ and 2 are labeled by solid black and dashed blue lines, respectively. Circles indicate locations where interference fringes occur with spatial frequency $q$.}
  \label{fig:1-RecoilDiagrams}
\end{figure}

In this work, we use two configurations of the grating echo AI to demonstrate experiments with timescales comparable to those of Raman AIs. An improved theoretical description of the echo AI has enabled accurate modeling of experimental data from which sensitive measurements of an externally applied $B$-gradient can be extracted. This model is sufficiently general to describe all time-domain configurations of grating echo AIs, while accounting for a constant force on the atoms, as well as the sub-level structure of the atomic ground state. Recent work \cite{Su-PRA-2010} with a particular configuration of this AI, has shown that measurements of the phase of the electric field are less sensitive to mirror vibrations. Investigations of the influence of $B$-gradients on this AI configuration validate our predictions of gravitational effects, and indicate that this AI is particularly well-suited for precise measurements of the gravitational acceleration, $g$.

We begin with a review of the two AI configurations used in this work, which are illustrated in \Fig \ref{fig:1-RecoilDiagrams}. The two-pulse echo AI \cite{Cahn-PRL-1997, Strekalov-PRA-2002, Beattie-PRA-2008, Barrett-PRA-2010} utilizes short (Raman-Nath) standing wave (sw) pulses to diffract a sample of laser cooled atoms at $t = T_1$ into a superposition of momentum states: $\ket{n\hbar\bm{q}}$. Here, $n$ is an integer and $\bm{q} = \bm{k}_1 - \bm{k}_2 \approx 2\bm{k}$ is the difference between the traveling wave vectors comprising the sw. At $t = T_2$, a second sw pulse further diffracts the atomic wave packets---creating sets of center-of-mass trajectories that overlap and produce interference in the form of a density modulation in the vicinity of $t_{\rm{echo}}^{(2)} = T_1 + (\bar{N}+1) T_{21}$, where $T_{21} \equiv T_2 - T_1$ and $\bar{N} = 1, 2, \ldots$ is the order of the echo, as shown in \Fig \ref{fig:1a-RecoilDiagram-TwoPulse}. The induced density modulation is coherent for $\tau_{\rm{coh}} = 2/q\sigma_v \sim 3$ $\mu$s about these ``echo'' times, beyond which the modulation dephases due to the distribution of velocities in the sample. Here, $\sigma_v = (2 k_B \mathcal{T}/M)^{1/2}$ characterizes the width of the velocity distribution along $\hat{\bm{z}}$. A traveling wave pulse is applied along the $\hat{\bm{z}}$-direction in the vicinity of $t_{\rm{echo}}^{(2)}$ to ``read out'' the amplitude of the grating by coherently Bragg scattering light along the $-\hat{\bm{z}}$-direction. The duration of this signal is limited by the coherence time, $\tau_{\rm{coh}}$. Due to the nature of Bragg diffraction, this back-scattered light is proportional to the Fourier component of the density distribution with spatial frequency $q$. This harmonic is only produced by interference of momentum states that differ by $\hbar q$ ($\Delta n = \pm 1$). As a result, the two-pulse AI exhibits a temporal modulation at the atomic recoil frequency, $\omega_q = \hbar q^2/2M$, and is therefore sensitive to recoil effects.

The three-pulse ``stimulated'' grating echo AI (henceforth referred to as the three-pulse AI) was first demonstrated in \Ref \citenum{Strekalov-PRA-2002} using a single hyperfine ground state, and was termed a ``stimulated'' echo due to similarities in pulse geometry with the stimulated photon echo scheme \cite{Mossberg-PRA-1979, Borde-PRA-1984, Allen-BOOK-1987, Dubetsky-PRA(R)-1992}. Recent work involving this interferometer \cite{Su-PRA-2010} has shown certain advantages over the two-pulse scheme for phase measurements of the atomic grating. The three-pulse AI involves applying two sw pulses at $t = T_1$ and $t = T_2$, followed by a third pulse applied at $t = T_3 = T_2 + T_{32}$, where $T_{32} \equiv T_3 - T_2$. This pulse geometry produces an echo in the vicinity of $t_{\rm{echo}}^{(3)} = T_1 + (\bar{N}+1) T_{21} + T_{32}$, as shown in \Fig \ref{fig:1b-RecoilDiagram-ThreePulse}. However, unlike the two-pulse AI where all pairs of trajectories produced by the second pulse interfere at the echo times, for the three-pulse AI only momentum states of the same order $(\Delta n = 0)$ after the second pulse produce interference at the echo times for arbitrary $T_{21}$ and $T_{32}$. For this reason, the signal produced by this interferometer as a function of $T_{32}$ (with $T_{21}$ fixed) is insensitive to atomic recoil effects (i.e. no temporal modulation) and is therefore ideal for probing other effects---such as those due to a constant force on the atoms.

Reference \citenum{Barrett-Advances-2011} extensively reviews the grating echo AI and discusses applications relating to atomic recoil \cite{Weel-PRA(R)-2003, Beattie-PRA-2008, Beattie-PRA(R)-2009, Beattie-PRA-2009, Barrett-PRA-2010, Barrett-SPIE-2011}, gravity and magnetic gradients \cite{Weel-PRA-2006}.

Previous experiments based on this AI \cite{Cahn-PRL-1997, Strekalov-PRA-2002, Weel-PRA(R)-2003, Beattie-PRA-2008, Beattie-PRA(R)-2009, Beattie-PRA-2009, Andersen-PRL-2009, Barrett-PRA-2010} were typically limited to $T_{21} < 10$ ms by decoherence effects due to spatially and temporally varying $B$-fields. Additionally, the sample temperature (typically $\sim 50$ $\mu$K) and excitation beam configuration (fixed frequency sw with $\sim 0.5$ cm diameter) limited the transit time in these experiments. In this work, we have improved the level of $B$-field and $B$-gradient suppression by using a non-magnetic vacuum chamber, which has enabled the extension of AI signal lifetimes. The magnetically controlled environment allows a sample of $^{85}$Rb atoms to be cooled to temperatures as low as 2.4 $\mu$K. By expanding the excitation beam diameter to $\sim 2$ cm, and chirping the sw pulses to cancel Doppler shifts, echo AI signal lifetimes of $\sim 220$ ms and transit times of $\sim 270$ ms have been achieved. These timescales are comparable to those of fountain experiments involving Raman AIs \cite{Kasevich-PRL-1991, Peters-Nature-1999, Wicht-PhysScr-2002}. In contrast, long-lived echo AI signals have been observed only by using magnetic guides to limit transverse cloud expansion \cite{Su-PRA-2010}.

The experimental apparatus presented here has made it possible to exploit the aforementioned advantages of the echo AI for a variety of precision measurements, such as the atomic recoil frequency \cite{Barrett-SPIE-2011} and the gravitational acceleration \cite{Barrett-Advances-2011}, that are currently underway. Additionally, we recently utilized this apparatus to perform a coherent transient experiment with cold Rb atoms to achieve a precise determination of the atomic $g$-factor ratio \cite{Chan-PRA-2011}.

In this Article, we apply long timescales to understanding and detecting the effects of $B$-gradients using the two-pulse grating echo AI, as well as a three-pulse ``stimulated'' grating echo AI \cite{Strekalov-PRA-2002, Su-PRA-2010}. The passive detection of magnetic anomalies is of interest for various applications, such as submarine and mine detection where the ambient magnetic noise of the environment is large compared to the sensitivity of the instrument \cite{Davis-JModOpt-2008}. The influence of gravity and $B$-gradients on AI experiments has been considered in the past. Reference \citenum{Roach-JPhysB-2004} calculates how such forces affect the visibility of interference patterns in atomic diffraction experiments. In previous work \cite{Weel-PRA-2006}, we demonstrated the effect of both gravity and $B$-gradients on the two-pulse AI. A theoretical description of these effects based on a spin-1/2 system was able to explain the basic signal dependence on the pulse separation, $T_{21}$, but was insufficient to model experimental data.

This work relies on an improved theoretical description of a generalized echo AI that includes an arbitrary number of sw excitation pulses, the effects of a constant force on the atoms, spontaneous emission and the sub-level structure of the atomic ground state (the 5S$_{1/2}$ $F = 3$ state of $^{85}$Rb is used in the experiment). Coupled with these theoretical predictions, we achieve sensitivity to \emph{changes} in $B$-gradients at the level of $\sim 0.04$ mG/cm. In addition, absolute measurements of $B$-gradients as small as $\sim 0.3$ mG/cm, and sensitivity to the curvature of $B$-fields are demonstrated. These results are consistent with independent measurements of the spatial variation in the $B$-field using a flux-gate magnetometer. These studies help place limits on the sensitivity of a broad class of time-domain AIs to $B$-gradients.

We also consider implications for achieving precise measurements of $g$ using the two- and three-pulse echo AIs. In particular, analysis of the three-pulse AI suggests there are significant advantages for measuring $g$ over the two-pulse AI. Although the experimental apparatus used in this work is not designed to detect gravitational effects, predictions of the grating phase modulation due to gravity for both AIs have been validated by measuring the effects of externally applied $B$-gradients. Measurements of $g$ using these AIs will be presented elsewhere.

This Article is organized as follows. In \Sec \ref{sec:Theory} we present theoretical predictions for the two- and three-pulse AI signals in the presence of a homogeneous $B$-gradient. In \Sec \ref{sec:Setup}, we describe details related to the experimental apparatus. We present the results of experiments related to long timescales in \Sec \ref{sec:Results}, and discuss measurements of $B$-gradients using both the two-pulse and three-pulse techniques. Section \ref{sec:Gravity} discusses the feasibility of a precise measurement of $g$ using the formalism developed to describe $B$-gradients. We conclude in \Sec \ref{sec:Conclusions}. The Appendix presents a calculation of the signal generated by a generalized echo AI---encompassing the two- and three-pulse AIs---in the presence of a constant force.

\section{Theory}
\label{sec:Theory}

In this section, we present the key results of calculations for both the two- and three-pulse AI signals in the presence of a homogeneous $B$-gradient. Details of the calculations---which are sufficiently general to account for any constant force on the atoms, and an arbitrary number of excitation pulses---are presented in the Appendix.

In general, the sensitivity of these interferometers can be characterized by the space-time area they enclose. Since only those states differing by $\hbar q$ at the echo time contribute to the signal, the area of both AIs is primarily controlled by $T_{21}$. In the absence of any external forces, the areas of the two- and three-pulse AIs can be calculated by inspecting their recoil diagrams [\Figs \ref{fig:1a-RecoilDiagram-TwoPulse} and \ref{fig:1b-RecoilDiagram-ThreePulse}, respectively]
\begin{subequations}
\bea
  \label{eqn:A(2)}
  A^{(2)} & = & \frac{\hbar q}{2M} \bar{N} (\bar{N} + 1) \left( T^{(2)}_{21} \right)^2, \\
  \label{eqn:A(3)}
  A^{(3)} & = & \frac{\hbar q}{2M} \left[ \bar{N} (\bar{N} + 1) \left( T^{(3)}_{21} \right)^2 + 2 \bar{N} T_{32} T^{(3)}_{21} \right],
\eea
\end{subequations}
where $M$ is the mass of the atom. Henceforth, quantities containing superscripts $(2)$ or $(3)$ indicate the interferometer for which that quantity applies. At first glance, it might appear that the three-pulse AI encloses a larger area than the two-pulse AI due to the extra term in \Eq \refeqn{eqn:A(3)}. However, one must compare the enclosed areas at the same echo times, which are given by $t^{(2)}_{\rm{echo}} = T_1 + (\bar{N} + 1)T^{(2)}_{21}$ and $t^{(3)}_{\rm{echo}} = T_1 + (\bar{N} + 1) T^{(3)}_{21} + T_{32}$ for the two- and three-pulse schemes, respectively. By setting $t^{(2)}_{\rm{echo}} = t^{(3)}_{\rm{echo}}$, it can be shown that $A^{(2)} - A^{(3)} = \hbar q \bar{N} T_{32}^2 / 2M(\bar{N} + 1)$. This suggests that the two-pulse AI is always more sensitive to external forces than the three-pulse AI. Nevertheless, the three-pulse AI offers a unique feature: the spatial separation between interfering wave packets remains constant between the application of the second and third sw pulses. This is advantageous because larger spatial separations leads to increased decoherence, and therefore reduced timescale in the experiment \cite{Su-PRA-2010}. Since the separation can be precisely controlled by the pulse separation $T_{21}$, one can increase the signal lifetime by using smaller $T_{21}$.

Additionally, since the signal generated by the two-pulse AI is modulated at the recoil frequency, $\omega_q$, there are periodic regions where the signal-to-noise ratio is less than one and not well-suited for accurate phase measurements. However, the three-pulse technique is insensitive to atomic recoil if $T_{21}$ is fixed. Therefore, the scattered field amplitude has no additional modulation at $\omega_q$ as $T_{32}$ is varied---allowing regions of low signal-to-noise ratio to be avoided.

Both the gravitational force and a constant $B$-gradient produce a constant force on the atoms, $\bm{\mathcal{F}} = \mathcal{F} \hat{\bm{z}}$, which generates a phase shift in the atomic interference pattern. The basic physical mechanism that produces this phase shift is a difference in potential energy between the two arms of the AI. One can compute the relative phase between the two arms $\Delta \phi = (\mathcal{S}_{\rm{B}} - \mathcal{S}_{\rm{A}})/\hbar$ using the classical action \cite{Peters-Metrologia-2001}
\be
  \mathcal{S}(t) = \int_0^{t} \mathcal{L}[z(t'), \dot{z}(t')] dt',
\ee
where $\mathcal{L} = M\dot{z}^2/2 + \mathcal{F} z$ is the Lagrangian in this case. If $\mathcal{S}_{\rm{B}}$ and $\mathcal{S}_{\rm{A}}$ represent, respectively, the action along the upper and lower arms of the two-pulse AI, it can be shown that the phase shift between these arms is
\be
  \label{eqn:DeltaPhi(2)}
  \Delta \phi^{(2)}
  = \bar{N}(\bar{N} + 1) \left[ \omega_q T_{21} + \frac{q \mathcal{F}}{M} T_{21}^2 \right] + \bar{N}^2 q v_0 T_{21},
\ee
where $v_0$ is the initial velocity of the atom along the $\hat{\bm{z}}$-direction. The term proportional to $v_0$ is due to the relative Doppler shift between the two arms of the AI. Since the atomic sample has a finite velocity distribution (characterized by a $1/e$ radius, $\sigma_v$, and temperature, $\mathcal{T}$), this term is responsible for the coherence time of the echo: $\tau_{\rm{coh}} = 2/q \sigma_v$. As expected, the contribution to the phase shift from the potential energy (the term proportional to $\mathcal{F}$) is independent of the initial velocity of the cloud.

A similar calculation for the relative phase shift between the arms of the three-pulse AI yields
\begin{align}
\begin{split}
  \label{eqn:DeltaPhi(3)}
  & \Delta \phi^{(3)} = \bar{N}(\bar{N} + 1) \left[ \omega_q T_{21} + \frac{q \mathcal{F}}{M} T_{21}^2 \right] \\
  & \;\;\;\;\; + \bar{N} \frac{q \mathcal{F}}{M} T_{32} T_{21} + \bar{N} q v_0 (T_{32} + \bar{N} T_{21}).
\end{split}
\end{align}
This expression is similar to \Eq \refeqn{eqn:DeltaPhi(2)}, with additional terms proportional to the pulse spacing $T_{32}$. One can vary either $T_{21}$ or $T_{32}$ to detect phase modulation produced by an external force, $\mathcal{F}$. However, since there are no terms containing the phase $\omega_q T_{32}$, one can effectively turn off the sensitivity to atomic recoil by fixing $T_{21}$. This makes the three-pulse AI ideal for investigating the effects due to $\mathcal{F}$, especially when $q\mathcal{F} T_{21}/M \gg \omega_q$ since no additional modulation at $\omega_q$ is present. This is particularly advantageous for measurements of gravity, as discussed in \Sec \ref{sec:Gravity}.

Since the two-pulse AI is intrinsically more sensitive than the three-pulse AI, it is better suited to measurements of $\mathcal{F}$ when $q\mathcal{F} T_{21}/M < \omega_q$. In this work, we demonstrate this feature by measuring externally applied $B$-gradients. The sensitivity of the three-pulse AI to $\mathcal{F}$ can be enhanced by utilizing the additional phase proportional $T_{32} T_{21}$ in \Eq \refeqn{eqn:DeltaPhi(3)}. Experimentally, this can be accomplished by varying \emph{both} pulse separations, $T_{21}$ and $T_{32}$, with $T_{21}$ varied in integer multiples of the recoil period: $\tau_q = \pi/\omega_q$.

To determine the response of the grating echo AI in the presence of a constant force, we assume a potential energy with the form
\be
  \label{eqn:U(r)}
  \hat{U}(z) = -\hat{\mathcal{M}} \, z,
\ee
where $\hat{\mathcal{M}} = -\partial \hat{U}/\partial z$ is a matrix operator with units of force that commutes with both the position ($z$) and momentum ($p$) operators. In the case of a constant $B$-gradient, the potential is
\be
  \hat{U}(z) = -\bm{\mu} \cdot \bm{B}(z) = - \frac{g_F \mu_B \beta}{\hbar} \hat{F}_z z,
\ee
where $g_F$ is the Land\'{e} $g$-factor, $\mu_B$ is the Bohr magneton, $\bm{B}(z) = \beta \bm{z}$ is the magnetic field vector with gradient $\beta$ along the $z$-direction (also assumed to be the quantization axis), and $\hat{F}_z$ is the projection operator for total angular momentum, $\bm{F}$. In this case, $\hat{\mathcal{M}} = \mathcal{F} \hat{F}_z/\hbar$ and the force is $\mathcal{F} = g_F \mu_B \beta$, where $\hat{F}_z$ operates on the basis states $\ket{F\,m_F}$ and has eigenvalues $\hbar m_F$.

In both interferometer schemes, the phase of the grating is imprinted on the electric field back-scattered from a traveling-wave read-out pulse applied in the vicinity of an echo time. For the two-pulse AI, measuring the phase of the scattered field is equivalent to measuring the relative position of the grating, since both the position and the phase scale as $T_{21}^2$. Similarly, in the three-pulse case, the phase measured as a function of $T_{32}$ is proportional to the velocity of the grating---which scales as $T_{21}$.

We first examine the effects of $B$-gradients on the two-pulse AI, followed by a comparison with the three-pulse AI.

\subsection{Two-Pulse Interferometer}
\label{sec:Theory-2Pulse}

In general, the electric field scattered by the atoms at the time of an echo is proportional to the amplitude of the Fourier harmonic of the atomic density grating with spatial frequency $q$. For the case of an external $B$-gradient, $\beta$, the scattered field has distinct contributions from each magnetic sub-level:
\be
  \label{eqn:E(2)_beta}
  E_{\beta}^{(2)}(t;\bm{T}) = \sum_{m_F} E^{(2)}_{m_F}(t;\bm{T}) e^{i m_F \phi^{(2)}_{\beta}(t;\bm{T})},
\ee
where $E^{(2)}_{m_F}$ is the field scattered by the state $\ket{F\,m_F}$ [given by \Eq \refeqn{eqn:E(2)_mF}] and $m_F \phi^{(2)}_{\beta}$ is the phase shift of the density grating produced by the same state in the presence of the $B$-gradient. For the $\bar{N}^{\rm{th}}$ order echo at $t_{\rm{echo}}^{(2)} = T_1 + (\bar{N} + 1)T_{21}$, with the set of onset times $\bm{T} = \{ T_1, T_1 + T_{21} \}$ and $\Delta t = t - t_{\rm{echo}}^{(2)}$, the phase shift of the grating $\phi^{(2)}_{\beta}(t;\bm{T})$ is given by
\begin{align}
\begin{split}
  \label{eqn:phi(2)_beta}
  & \phi^{(2)}_{\beta}(\Delta t;T_{21}) = \frac{q g_F \mu_B \beta}{2M} \left\{ \bar{N}(\bar{N}+1)T_{21}^2 \right. \\
  & \;\;\;\;\; \left. + \, 2 \big[T_1 + (\bar{N}+1) T_{21}\big] \Delta t + \Delta t^2 \right\}.
\end{split}
\end{align}
The general form of this equation for a constant force, $\mathcal{F}$, is given by \Eq \refeqn{eqn:phi(2)_F} in the Appendix. In the discussions that follow, we take $\Delta t = 0$ which corresponds to the echo time. Since the echo lasts for $\tau_{\rm{coh}} \sim 3$ $\mu$s about $\Delta t = 0$, the signal is obtained by integrating the back-scattered field over this time.

Equations \refeqn{eqn:E(2)_beta} and \refeqn{eqn:phi(2)_beta} indicate that the field amplitude scattered from state $\ket{F\,m_F}$ exhibits phase modulation as a function of $T_{21}$ at a frequency $m_F \omega^{(2)}_{\beta}(T_{21})$ due to the presence of the gradient, where
\begin{align}
\begin{split}
  \label{eqn:omega(2)_beta}
  & \omega^{(2)}_{\beta}(T_{21}) = \left| \frac{\partial \phi^{(2)}_{\beta}}{\partial T_{21}} \right| \\
  & \;\;\;\;\; = \frac{q g_F \mu_B \beta}{M} \left[ \bar{N}(\bar{N} + 1) T_{21} + (\bar{N} + 1) \Delta t \right].
\end{split}
\end{align}
This modulation frequency depends linearly on $\beta$ and the pulse separation, $T_{21}$ (i.e. the frequency is chirped with $T_{21}$). The phase modulation of the grating produced by state $\ket{F\;m_F}$ also scales linearly with the magnetic quantum number, $m_F$, as shown in \Eq \refeqn{eqn:E(2)_beta}. For an arbitrary set of magnetic sub-level populations, the total scattered field [\Eq \refeqn{eqn:E(2)_beta}] contains all harmonics $m_F \omega^{(2)}_{\beta}(T_{21})$, where $m_F = -F,\ldots,F$. If more than one sub-level is populated, interference between the fields scattered off of each state produces modulation in the total scattered field. This effect can then be detected in the field amplitude, $E_{\beta}^{(2)}$, or the field intensity, $|E_{\beta}^{(2)}|^2$, by varying $\beta$ or the pulse separation, $T_{21}$. The amplitude of each harmonic comprising this modulation is determined by the sub-level populations, as well as the transition probabilities between ground and excited state sub-levels.

\begin{figure}[!t]
  \centering
  \subfigure{\label{fig:2a-GradientSignal}\includegraphics[width=0.23\textwidth]{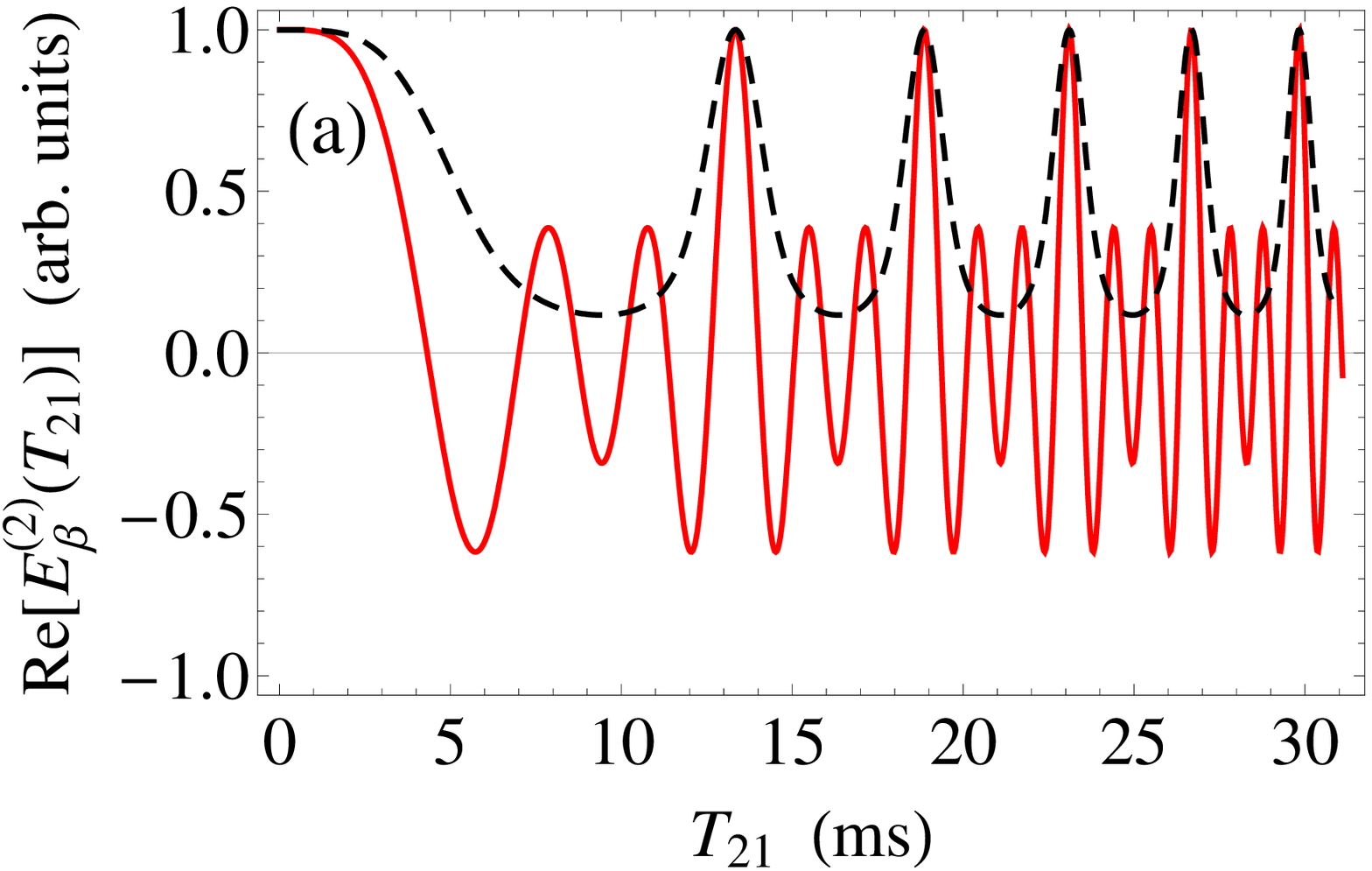}}
  \subfigure{\label{fig:2b-GradientSignal}\includegraphics[width=0.23\textwidth]{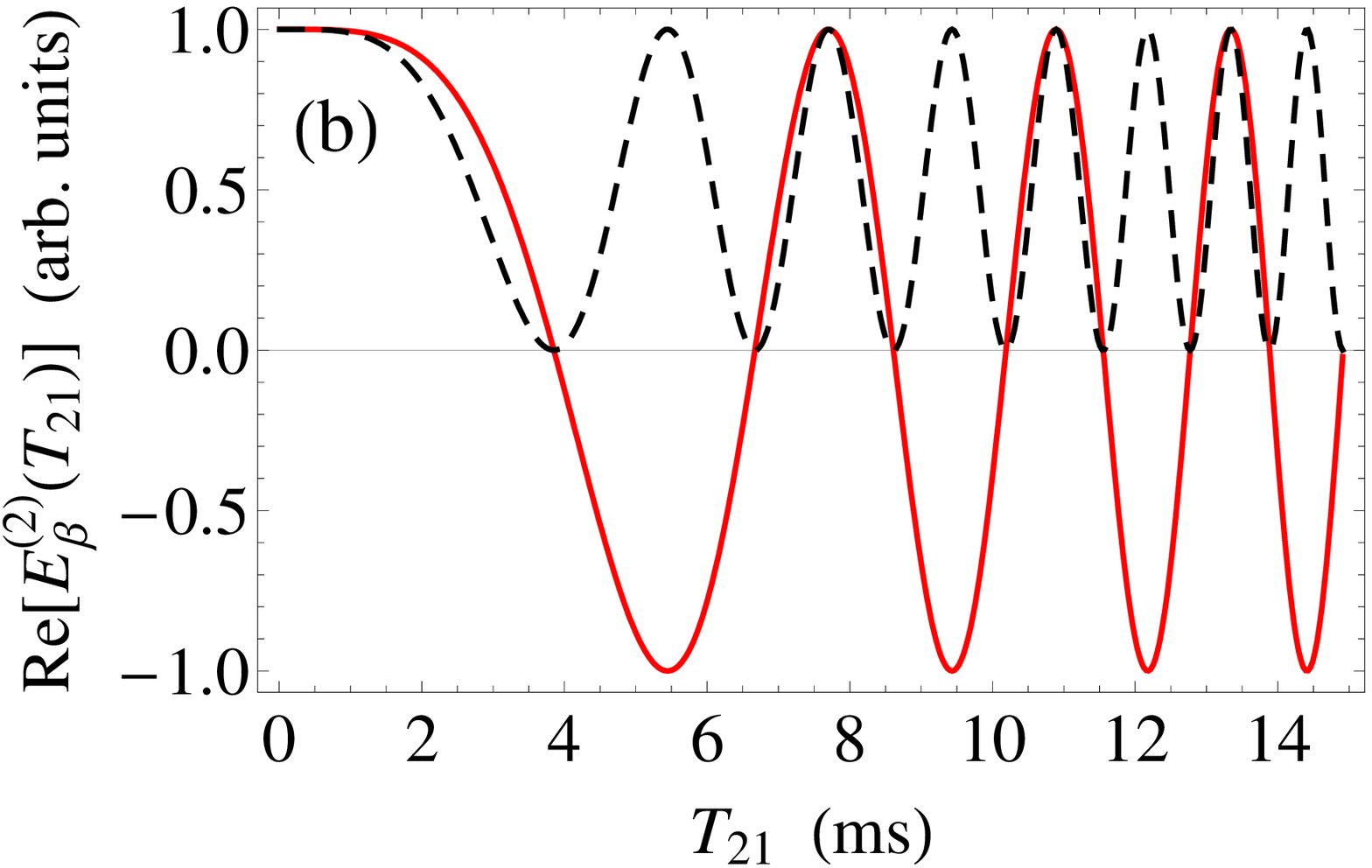}}
  \subfigure{\label{fig:2c-GradientSignal}\includegraphics[width=0.23\textwidth]{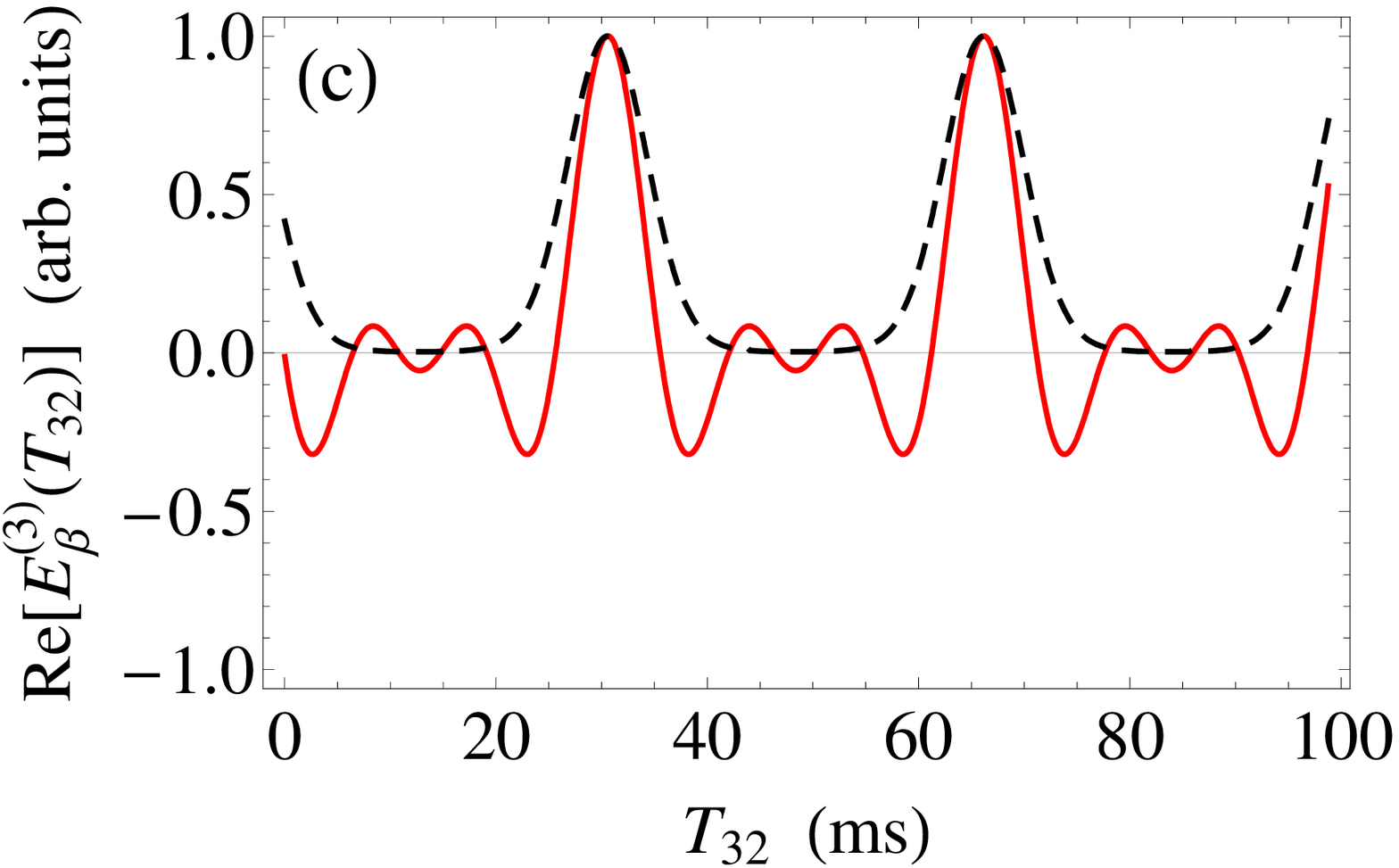}}
  \subfigure{\label{fig:2d-GradientSignal}\includegraphics[width=0.23\textwidth]{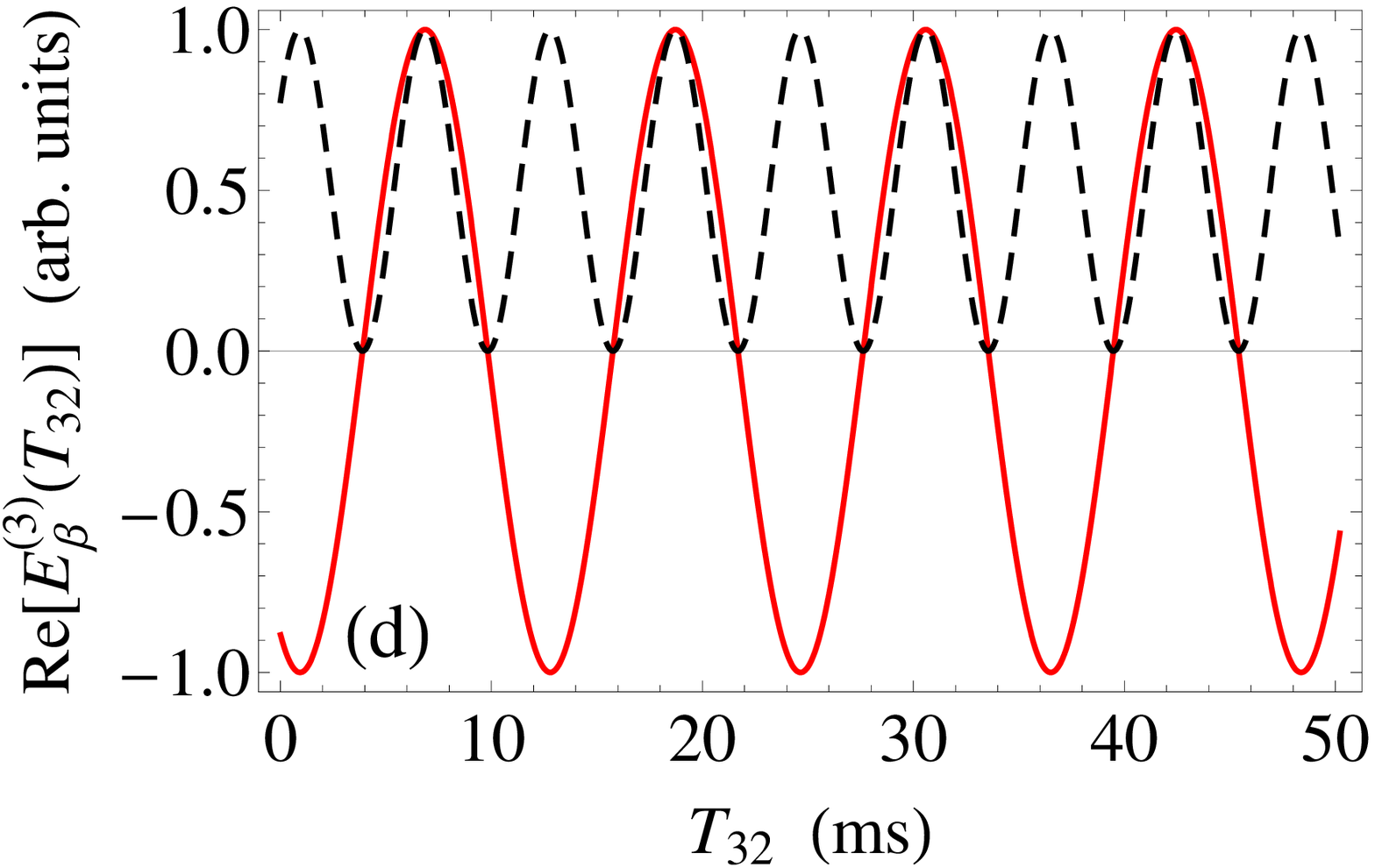}}
  \caption{(Color online) The predicted $B$-gradient signal for the two-pulse AI (a,b) [based on \Eq \refeqn{eqn:E(2)_beta}], where $T_{21}$ is varied in steps of $\tau_q \sim 32$ $\mu$s, and the three-pulse AI (c,d), where $T_{32}$ is varied. In all plots the solid red line is the real part of the scattered field, the black dashed line shows the field intensity (which is measured in the experiment), the $B$-gradient is fixed at $\beta = 10$ mG/cm, and the first order echo ($\bar{N} = 1$) is used. In parts (a,c) the excitation beams are circularly polarized ($|q_L| = 1$) and the magnetic sub-level populations are equally distributed among the $F = 3$ ground state of $^{85}$Rb. In this case, the field undergoes amplitude modulation with multiple frequency components, and the field intensity exhibits modulation with a contrast $< 100\%$. For the two-pulse AI (a), the field intensity exhibits modulation at a chirped frequency, whereas for the three-pulse AI (c) the modulation is not chirped. Parts (b,d) show the field generated by a sample that is optically pumped equally into the $|3\;$$-3\rangle$ and $\ket{3\;3}$ states with linearly polarized excitation beams ($q_L = 0$). Here, both the field and the field intensity exhibit modulation with only one frequency component, and the contrast of the oscillations is 100\%. For both (c,d), $T_{21}$ was fixed at a typical experimental value of 5 ms.}
  \label{fig:2-GradientSignal}
\end{figure}

If the system is optically pumped into a single sub-level, such as the extreme state: $\ket{F\;F}$, then the phase modulation of the grating only affects the phase of the electric field---which cannot be observed using intensity detection. Instead, one can use heterodyne detection to measure the electric field amplitude and obtain the relative phase of the scattered light \cite{Cahn-PRL-1997, Weel-PRA(R)-2003, Weel-PRA-2006}. Furthermore, if the system is optically pumped into the $\ket{F\;0}$ state, there is no phase modulation due to $B$-gradients since this state is insensitive to magnetic fields.

Figures \ref{fig:2a-GradientSignal} and \ref{fig:2b-GradientSignal} show the expected two-pulse AI signal as a function of $T_{21}$ in steps of the recoil period, $\tau_q = \pi/\omega_q$ ($\sim 32$ $\mu$s for $^{85}$Rb). Since $\omega^{(2)}_{\beta} < \omega_q$, incrementing $T_{21}$ in this fashion eliminates additional modulation due to atomic recoil. Figure \ref{fig:2a-GradientSignal} shows the signal for equally distributed sub-level populations, while \Fig \ref{fig:2b-GradientSignal} is for an optically pumped system in the two extreme states: $|3\;$$-3\rangle$ and $\ket{3\;3}$. Both of these figures show amplitude modulation, but in the optically pumped case there is only one frequency component present and the modulation occurs with maximum contrast---increasing the sensitivity to gradients.

Eliminating the amplitude modulation in the signal due to $B$-gradients [shown by the dashed lines in \Fig \ref{fig:2a-GradientSignal}] is a key requirement for precision measurements of $\omega_q$. We will show in \Sec \ref{sec:Results} that these conditions can be realized with sufficient suppression of ambient $B$-gradients in a glass cell. It is also possible to eliminate sensitivity to $B$-gradients using intensity detection if the atoms are pumped into a single magnetic sub-level.

\subsection{Three-Pulse Interferometer}
\label{sec:Theory-3Pulse}

The effects due to $B$-gradients manifest themselves differently in the three-pulse interferometer. We derive the expression for the signal in the Appendix [see \Eqs \refeqn{eqn:E(N)} and \refeqn{eqn:E(3)_mF}] and find that the amplitude of the scattered field does not depend on the time between the second and third sw pulses, $T_{32}$, but only on $T_{21}$---similar to the two-pulse AI. However, the phase of the grating in the three-pulse case depends on both $T_{21}$ and $T_{32}$:
\bea
  \label{eqn:phi(3)_beta}
  & & \phi_{\beta}^{(3)}(\Delta t; \bm{T})
  = \frac{q g_F \mu_B \beta}{2M} \left\{ \bar{N} (\bar{N} + 1) T_{21}^2 + 2\bar{N} T_{32} T_{21} \right. \notag \\
  & & \;\;\;\;\; \left. + \, 2 \big[ T_1 + T_{32} + (\bar{N}+1) T_{21} \big] \Delta t + \Delta t^2 \right\}.
\eea
In this case, the set of pulse onset times is given by $\bm{T} = \{ T_1, T_1 + T_{21}, T_1 + T_{21} + T_{32} \}$ and $\Delta t = t - t^{(3)}_{\rm{echo}}$. This phase is identical to \Eq \refeqn{eqn:phi(2)_beta} for the two-pulse interferometer with the addition of the two terms proportional to $T_{32}$. Equation \refeqn{eqn:phi(3)_beta} suggests that the force can be determined by measuring the phase modulation of the grating as a function of either $T_{21}$ or $T_{32}$, or by varying both pulse separations simultaneously. Varying $T_{32}$ produces a phase modulation of the atomic grating at a frequency that is proportional to $T_{21}$:
\be
  \label{eqn:omega(3)_beta}
  \omega^{(3)}_{\beta}(T_{21})
  = \left| \frac{\partial \phi^{(3)}_{\beta}}{\partial T_{32}} \right|
  = \frac{q g_F \mu_B \beta}{M} \left( \bar{N} T_{21} + \Delta t \right).
\ee

Figures \ref{fig:2c-GradientSignal} and \ref{fig:2d-GradientSignal} show the expected three-pulse signal as a function of $T_{32}$, with $T_{21}$ fixed at a typical experimental value of 5 ms, in the presence of a $B$-gradient $\beta = 10$ mG/cm. When the sub-level populations are equally distributed [\Fig \ref{fig:2c-GradientSignal}] the phase of the total scattered field contains multiple frequency components---one for each sub-level: $m_F \omega_{\beta}^{(3)}$. The interference between these components produces a modulation in the total scattered field amplitude. This is similar to the two-pulse case shown in \Fig \ref{fig:2a-GradientSignal}, except that the modulation occurs at a single frequency that is fixed by $\beta$, $T_{21}$ and $\bar{N}$. For a sample that is optically pumped equally into the two extreme states: $|3\;$$-3\rangle$ and $\ket{3\;3}$, as shown in \Fig \ref{fig:2d-GradientSignal}, there is only one frequency component present in the scattered field. In this case, the amplitude modulation occurs with greater contrast than for any other configuration of sub-level populations.

\section{Experimental Setup}
\label{sec:Setup}

We now review the experimental setup that has made possible long-lived grating echo AI signals. This setup is substantially different from previous echo experiments \cite{Weel-PRA-2006, Beattie-PRA-2008, Beattie-PRA(R)-2009, Beattie-PRA-2009} after implementing many improvements. These include suppression of stray magnetic gradients using a non-magnetic chamber, increasing the trapped atom number with large diameter beams, extending the transit time by cooling the sample to $\sim 10$ $\mu$K and implementing large diameter excitation beams, and by chirping the excitation frequencies to eliminate Doppler shifts associated with the falling cloud.

\begin{figure}[!t]
  \centering
  \includegraphics[width=0.35\textwidth]{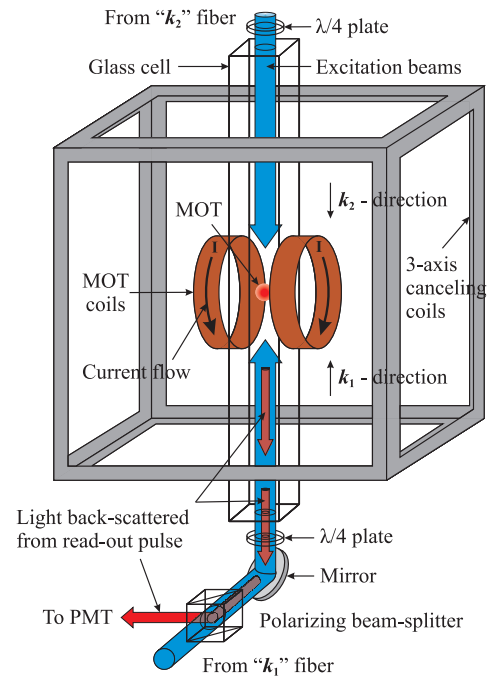}
  \caption{(Color online) Diagram of the experiment. The excitation beams are both $\sigma^+$-polarized by the $\lambda/4$-plates. The glass cell has approximate dimensions $7.6 \times 7.6 \times 84$ cm. Each pair of square quadrupole coils have a side length of $\sim 66$ cm and contain overlapped coils wired in both Helmholtz and anti-Helmholtz configurations for canceling $B$-fields and $B$-gradients, respectively.}
  \label{fig:3-ExpSetup}
\end{figure}

The experiment utilizes a sample of laser-cooled $^{85}$Rb atoms in a magneto-optical trap (MOT) containing approximately $10^9$ atoms in a Gaussian spatial distribution with a horizontal $e^{-1}$ radius of $\sim 1.7$ mm. The MOT is contained in a borosilicate glass cell maintained at a pressure of $\sim 1 \times 10^{-9}$ Torr. In addition to the anti-Helmholtz coils used for trapping, three pairs of square quadrupole coils are centered on the MOT, as shown in \Fig \ref{fig:3-ExpSetup}. Each square frame contains two overlapping coils, one connected in the Helmholtz configuration with the coil in the opposite frame, and the other in the anti-Helmholtz configuration. These sets of coils are used to cancel ambient magnetic fields and field gradients over the volume of the MOT at the level of $\sim 1$ mG and $\sim 0.1$ mG/cm, respectively. The initial set points for the currents in the canceling coils that produced $\sim 1$ mG of $B$-field suppression were determined using an atomic magnetometer experiment \cite{Chan-PRA-2011} that allowed the field at the location of the MOT to be measured.

Light derived from a Ti:sapph laser with frequency $\nu_L$ and linewidth $\sim 1$ MHz is locked to the 5S$_{1/2}$ $F=3 \to F'=4$ transition ($\nu_L = \nu_0$) using saturated absorption spectroscopy. The light is then shifted 130 MHz above resonance by an acousto-optic modulator (AOM) operating in dual-pass mode such that $\nu_L = \nu_0 + 130$ MHz. A separate ``trapping'' AOM shifts this light by $-148$ MHz such that the detuning is $\Delta = -14$ MHz ($\nu_L = \nu_0 - 14$ MHz). Approximately 370 mW of this light is transmitted through an anti-reflection-coated, single-mode optical fiber (operating at 60\% efficiency) and expanded to a diameter of $\sim 5.4$ cm for trapping atoms from background vapor.

An external cavity diode laser is used to derive repump light for the trapping setup. It is locked to the 5S$_{1/2}$ $F = 2 \to F'=(2,3)$ crossover transition and up-shifted by $\sim 32$ MHz using an AOM. Approximately 25 mW of repump light is obtained after coupling through the same optical fiber as the trapping light. At $t = 0$, the MOT coils are pulsed off in $\sim 100$ $\mu$s, while the trapping and repump beams are left on for 6 ms of molasses cooling. For $\sim 3$ ms of this time, the detuning of the trapping light is linearly chirped from $\Delta = -14$ MHz to $-50$ MHz to further cool the atoms, and the power is simultaneously ramped down in order to reduce heating due to photon scattering. With this procedure we achieve temperatures as low as $\mathcal{T} = 2.4$ $\mu$K.

\begin{figure}[!t]
  \centering
  \includegraphics[width=0.45\textwidth]{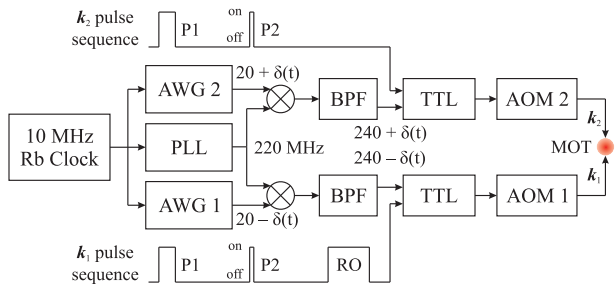}
  \caption{(Color online) Schematic of the rf chain used for chirped AI pulses. A phase-locked loop (PLL) generates a 220 MHz rf signal which is split and mixed with the output of two separate arbitrary waveform generators (AWGs). The AWGs are triggered at the start of the experiment to output a frequency sweep from 20 MHz to $20 \mbox{ MHz} \pm \delta(t)$ after a time $t$, where $\delta(t) = g t/\lambda$, $g \sim 9.8$ m/s$^2$ and $\lambda \sim 780$ nm. The sum frequency from the mixers is isolated using a band-pass filter (BPF) with a center frequency of 240 MHz and a 5 dB pass-band of 4 MHz. The outputs of the BPFs are pulsed using a set of transistor-transistor logic (TTL) switches, which are then sent to the $\bm{k}_1$ and $\bm{k}_2$ AOMs. The two-pulse AI sequence for both $\bm{k}_1$ and $\bm{k}_2$ are shown. Here, P1 and P2 refer to traveling wave components comprising the first and second sw pulses, and RO denotes  the traveling wave read-out pulse sent along $\bm{k}_1$. Both AWGs and the PLL are externally referenced to a 10 MHz Rb clock.}
  \label{fig:4-RFSchematic}
\end{figure}

Light from the Ti:sapph laser is also used to derive the AI pulses. A ``gate'' AOM operating in dual-pass configuration shifts the undiffracted light from the ``trapping'' AOM from $\nu_L = \nu_0 + 130$ MHz to $\nu_L = \nu_0 + 290$ MHz. The gate AOM is also pulsed so as to serve as a high-speed shutter during the experiment. The light from the gate AOM is split and sent into two separate AOMs (referred to as the ``$\bm{k}_1$'' and ``$\bm{k}_2$'' AOMs) operating at $240 \mbox{ MHz} \pm \delta(t)$ that produce the sw pulses. Here, $\delta(t) = g t/\lambda$ is a time-dependent frequency shift that is added to (subtracted from) the radio frequency (rf) driving the $\bm{k}_1$ ($\bm{k}_2$) AOM using an arbitrary waveform generator, as shown in \Fig \ref{fig:4-RFSchematic}. Chirping the excitation pulses in this manner cancels the Doppler shift of the atoms falling under gravity. The rf driving these AOMs is also phase locked to a 10 MHz rubidium clock to eliminate any electronically induced phase shifts. Light entering the $\bm{k}_1$ AOM is downshifted by $240 \mbox{ MHz} - \delta(t)$ and sent into an optical fiber that carries the light toward the MOT. Similarly, the $\bm{k}_2$ AOM downshifts the light by $240 \mbox{ MHz} + \delta(t)$. In this configuration, the detuning of the $\bm{k}_1$ ($\bm{k}_2$) pulse is $\Delta_1 = 50 \mbox{ MHz} - \delta(t)$ [$\Delta_2 = 50 \mbox{ MHz} + \delta(t)$]. This light is coupled into a separate fiber and aligned through the MOT along the vertical direction, as illustrated in \Fig \ref{fig:3-ExpSetup}. The output of both fibers is expanded to a $e^{-2}$ diameter of $\sim 2$ cm. The rf pulses driving the $\bm{k}_1$ and $\bm{k}_2$ AOMs are controlled using TTL switches with an isolation ratio of 100 dB, which produces optical pulses with rise times of $\sim 20$ ns. The ``gate'' AOM is turned off between excitation pulses to further reduce background light from reaching the atoms.

In the vicinity of any given echo (see \Fig \ref{fig:1-RecoilDiagrams}), the read-out pulse is applied to the sample along the $\bm{k}_1$-direction and a coherent back-scattered field from the atoms occurs along the direction of $\bm{k}_2$. The power of the scattered field is recorded as a function of time using a photo-multiplier tube (PMT) that is gated on for 9 $\mu$s. The echo signal lasts $\tau_{\rm{coh}} \sim 3$ $\mu$s before coherence is lost due to Doppler dephasing. For $T_{21} \lesssim 10$ ms, the scattered field can reach powers greater than 100 $\mu$W. However, for $T_{21} > 10$ ms, the signal size decreases exponentially. The noise floor for the PMT is approximately 0.1 $\mu$W. Typically, one computes the time-integrated area of the echo signal as a measure of the signal size for a given set of parameters. Since this quantity has units of energy, it is henceforth referred to as the echo energy.

\section{Results and Discussion}
\label{sec:Results}

We now review the main experimental results of this work relating to long-lived AI signals and sensing externally applied $B$-gradients.

\subsection{Investigations of AI Timescale}

\begin{figure*}[!t]
  \centering
  \includegraphics[width=0.80\textwidth]{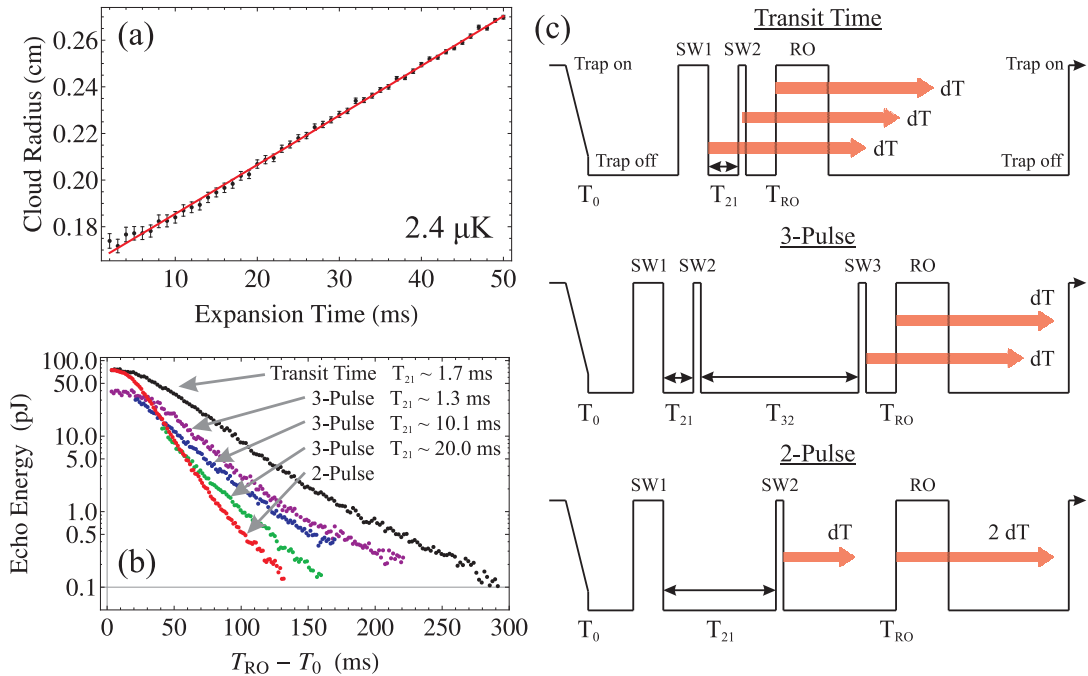}
  \caption{(Color online) (a) Temperature measurement of the laser-cooled sample. The horizontal $e^{-1}$ cloud radius is measured as a function of expansion time using a CCD camera. A hyperbolic fit to these data (shown as the solid red line) yields a measurement of $\mathcal{T} = 2.4 \pm 0.2$ $\mu$K. (b) Data showing the signal lifetime for various pulse configurations. The horizontal axis is the time of the read-out pulse, $T_{\rm{RO}}$, relative to the time of trap turn-off, $T_0$, which signifies the start of the experiment. An echo energy of $\sim 0.1$ pJ is equivalent to the level of noise in the detector. These data were obtained with pulse durations for the two-pulse (three-pulse) AI: $\tau_1 = 800$ (800) ns, $\tau_2 = 100$ (70) ns, $\tau_3 = 0$ (70) ns; pulse intensity $I \sim 64$ mW/cm$^2$; and a sample temperature of $\mathcal{T} \sim 20$ $\mu$K. Pulse separations for each configuration are shown in the figure. (c) Pulse configurations for the data shown in (b). The red arrows distinguish which pulses were varied and the time step, $dT$, indicates the amount each pulse was incremented relative to the others. For the two-pulse AI, $dT = n \tau_q$, where $n = 10$. Pulses without arrows are fixed in time.}
  \label{fig:5-SignalLifetime}
\end{figure*}

Figure \ref{fig:5-SignalLifetime}(a) shows a measurement of the temperature of the laser cooled sample. At $t = T_0$, all optical and magnetic fields associated with the MOT are switched off and the atoms are allowed to thermally expand in the dark. At $t = T_0 + T_{\rm{exp}}$, the trapping and repump beams are turned back on and a calibrated charged-coupled device (CCD) is triggered to photograph the cloud with an exposure time of 100 $\mu$s. This process is repeated for various expansion times, $T_{\rm{exp}}$, and the $e^{-1}$ radius of the cloud, $R$, is measured by fitting to the Gaussian intensity profiles obtained from each image. The temperature is obtained by fitting to a hyperbola \cite{Weiss-JOSAB-1989, Vorozcovs-JOSAB-2005} with the form $R = [R_0^2 + \sigma_v^2 (T_{\rm{exp}} - t_0)^2]^{1/2}$, where $R_0$ is the initial cloud radius, $\sigma_v = (2 k_B \mathcal{T}/M)^{1/2}$ is the $e^{-1}$ radius of the velocity distribution and $t_0$ is a phenomenological offset from $T_{\rm{exp}} = 0$. The data shown in \Fig \ref{fig:5-SignalLifetime}(a) give a temperature of $\mathcal{T} \sim 2.4$ $\mu$K in $^{85}$Rb. This relatively low MOT temperature is attributed to the well-controlled magnetic environment within the glass cell, as well as the molasses cooling procedure described above.

Measurements of the AI signal lifetime under different pulse configurations are shown in \Fig \ref{fig:5-SignalLifetime}(b), with each configuration explained schematically in \Fig \ref{fig:5-SignalLifetime}(c). For the transit time measurement, the two-pulse AI configuration was used with $T_{21}$ fixed. The excitation and read-out pulses were incremented synchronously. The signal lifetime for the three-pulse AI was determined by fixing $T_{21}$ and varying the third sw pulse and read-out synchronously. For the two-pulse AI, the lifetime was measured by fixing the first sw pulse and incrementing the second sw pulse and read-out in steps $n \tau_q$ and $2 n \tau_q$, respectively, where $n = 10$.

Here, all ambient $B$-fields and $B$-gradients are canceled along all three axes at the level of $\sim 1$ mG and $\sim 0.1$ mG/cm, respectively. The transit time data was obtained by using the two-pulse AI with $T_{21}$ fixed at $\sim 1.690$ ms and varying the time of all sw pulses relative to the time of trap turn-off, $T_0$. In this measurement, the AI signal is proportional to the number of atoms that remain in the volume defined by the $\sim 2$ cm diameter excitation beams during the thermal expansion of the cloud. Although the echo energy spans almost three orders of magnitude as it decays exponentially, signals are clearly distinguishable from the noise floor ($\sim 0.1$ pJ) at times as large as $\sim 270$ ms, as shown in \Fig \ref{fig:5-SignalLifetime}(b). This time represents the transit time limit for the conditions of our experiment---corresponding to a drop height of $\sim 36$ cm. This distance nearly coincides with the bottom viewport of the vacuum system. We emphasize here that such lifetimes are not possible with this interferometer unless the frequencies of the $\bm{k}_1$ and $\bm{k}_2$ beams are oppositely chirped such that the Doppler shift due to gravity [$\delta(t) = g t / \lambda$] is canceled \emph{or} the bandwidth of the sw pulses is large enough to account for such a shift. The frequency chirp puts the sw pulses on resonance for the two-photon transition back to the same ground state for all times during the sample's free-fall.

The signal lifetime for the two-pulse AI configuration is shown as the red curve in \Fig \ref{fig:5-SignalLifetime}(b). Here, the signal lasts approximately 130 ms, corresponding to $T_{21} \sim 65$ ms. To the best of our knowledge, this is the largest timescale observed with this interferometer, corresponding to more than a factor of 6 improvement over our previous work \cite{Weel-PRA-2006, Beattie-PRA-2008, Beattie-PRA(R)-2009, Beattie-PRA-2009}. However, the lifetime of the two-pulse echo is still limited by decoherence from a small, inhomogeneous $B$-gradient that the atoms sample over the $\sim 8$ cm they have fallen in 130 ms. A non-linear $B(z)$ produces a spatially-dependent force between interfering trajectories---resulting in a differential phase shift between paths of the interferometer that causes dephasing and, therefore, a loss of signal. Such a non-linearity in $B(z)$ has been measured to be $\partial^2 B/\partial z^2 \sim -0.4$ mG/cm$^2$ with a flux-gate magnetometer placed at different spatial locations around the glass chamber. This curvature is produced by a combination of non-ideal coil configurations and the presence of nearby ferromagnetic materials.

\begin{figure*}[!t]
  \centering
  \subfigure{\label{fig:6a-EchoSignalVsGradient}\includegraphics[width=0.48\textwidth]{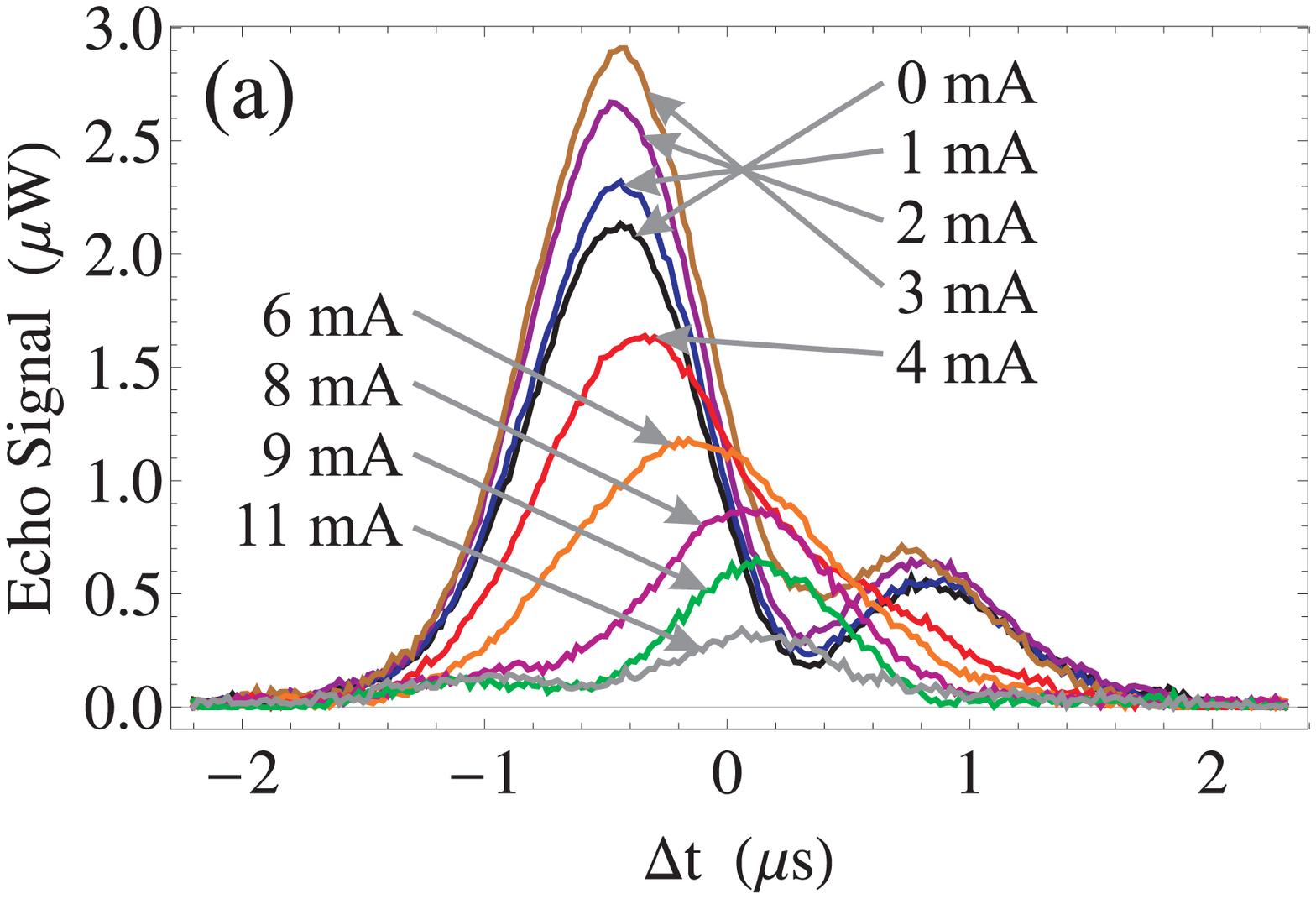}}
  \hspace{0.2cm}
  \subfigure{\label{fig:6b-EchoEnergyVsGradient}\includegraphics[width=0.48\textwidth]{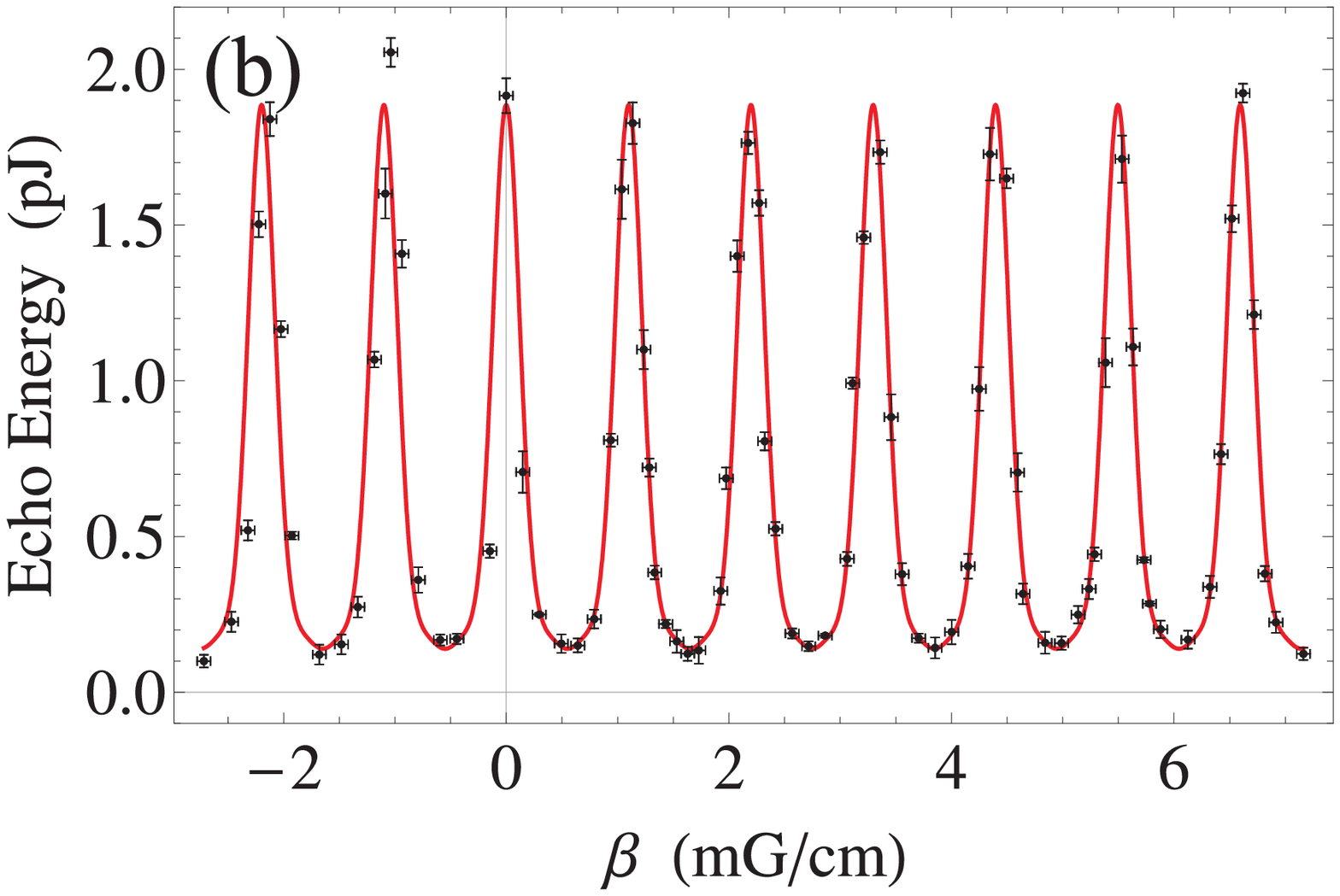}}
  \caption{(Color online) (a) First order ($\bar{N} = 1$) two-pulse echo signal for various applied $B$-field gradients, $\beta$. The theoretically expected echo time is at $\Delta t = t - 2 T_{21} = 0$. Each mA of current corresponds to a change of $\sim 0.04$ mG/cm in the applied gradient. (b) First order ($\bar{N} = 1$) two-pulse echo energy as a function of $\beta$. The solid red line is a fit based on \Eq \refeqn{eqn:ModelFit-2Pulse}. The value of $\beta$ is obtained from a calibration using a flux-gate magnetometer. Pulse parameters for both (a) and (b): $\tau_1 = 800$ ns, $\tau_2 = 130$ ns, $I \sim 64$ mW/cm$^2$, $T_{21} \sim 40.6$ ms.}
  \label{fig:6-EchoVsGradient}
\end{figure*}

There are two important features that should be recognized from the data for the three-pulse AI shown in \Fig \ref{fig:5-SignalLifetime}(b). First, at $T_{\rm{RO}} - T_0 \approx 0$, the echo energy for the three-pulse AI is a factor of $\sim 2$ smaller than that of the two-pulse AI. This comes about because the additional Kapitza-Dirac pulse involved in the three-pulse AI produces fewer pathways that result in interference at the echo time compared to the two-pulse AI. Second, the lifetime of the three-pulse echo depends strongly on the value of $T_{21}$. As $T_{21}$ increases, the signal lifetime approaches that of the two-pulse AI. This feature comes about because, between the second and third sw pulses, the wave packets that interfere at the echo times have a constant spatial separation [see \Fig \ref{fig:1b-RecoilDiagram-ThreePulse}], which is given by $\Delta z = \bar{N} \hbar q T_{21}/M$. From this expression, it is clear that $\Delta z$ can be controlled by $T_{21}$ and the choice of echo order, $\bar{N}$. By decreasing this separation, the interferometer becomes less sensitive to decoherence from non-linear $B$-fields since phase shifts produced by this effect become approximately common mode between interfering momentum states. Reference \citenum{Su-PRA-2010} also used this interferometer and a magnetic guide to show that smaller spatial separations lead to increased timescales.

In general, the lifetime for the three-pulse echo can be tailored to last much longer than that of the two-pulse echo, which is advantageous for precisely measuring the effects of external forces. For example, we achieve timescales as large as $\sim 220$ ms for $T_{21}$ fixed at $\sim 1.3$ ms---which is much closer to the transit time limit than the lifetime of the two-pulse echo. To the best of our knowledge, the only experiment that has achieved longer timescales for this AI have employed magnetic guides \cite{Su-PRA-2010} to limit transverse expansion of the sample and thereby extending the transit time.

\subsection{Investigations of External $B$-Gradients}

When $T_{21}$ is large, the two-pulse AI can be used to explore the sensitivity to small external $B$-gradients. We demonstrate the detection of changes in the $B$-gradient as small as $\sim 4 \times 10^{-5}$ G/cm in \Fig \ref{fig:6a-EchoSignalVsGradient}. Here, the $\bar{N} = 1$ echo signal was recorded with $T_{21}$ fixed at $\sim 40.6$ ms for various applied gradients. Changes in the gradient were facilitated by varying the current through the set of vertical quadrupole coils centered on the MOT (see \Fig \ref{fig:3-ExpSetup}). The smallest controllable increment in current we could achieve was 1 mA, which corresponds to a change of $\sim 0.04$ mG/cm as estimated from an independent calibration based on a flux-gate magnetometer.

In a similar experiment, the $\bar{N} = 1$ echo energy was measured for $T_{21}$ fixed at $\sim 40.6$ ms as a function of $\beta$, as shown in \Fig \ref{fig:6b-EchoEnergyVsGradient}. Here, it is clear that the echo energy has a strong periodic dependence on the applied $B$-gradient. These data provide confirmation of the theoretical prediction given by \Eqs \refeqn{eqn:E(2)_beta} and \refeqn{eqn:phi(2)_beta}. This dependence is produced by the interference between electric fields scattered off of gratings produced by different magnetic sub-levels. For example, for a given $\beta$, gratings produced by states $\ket{F\,m_F}$ and $\ket{F\,m_F'}$ undergo phase shifts $m_F \phi^{(2)}_{\beta}$ and $m_F' \phi^{(2)}_{\beta}$, respectively, where $\phi^{(2)}_{\beta}$ is given by \Eq \refeqn{eqn:phi(2)_beta}. For constructive interference between fields scattered by these states, the $B$-gradient must satisfy $(m_F - m_F') \phi^{(2)}_{\beta} / 2 = 2 n \pi$, where $n$ is an integer. Thus, as $\beta$ is varied, the phase shift induced in the $\ket{F\,m_F}$ and $\ket{F\,m_F'}$ gratings produces periodic constructive (destructive) interference in the total scattered field, and therefore, maxima (minima) in the echo energy. This process occurs simultaneously in all $2F+1$ sub-levels. As a result, the observed signal is a weighted sum of the scattered fields from all states. Here, there are $2F (2F+1) = 42$ pairs of states that produce interference---although not all pairs have unique contributions. Since the excitation beams were circularly polarized in the experiment, the fields scattered from the extreme states ($\ket{3\,3}$ or $|3\,$$-3\rangle$) dominate the signal.

\begin{figure*}[!t]
  \centering
  \subfigure{\label{fig:7a-2Pulse-Gradients-N1-N2}\includegraphics[width=0.48\textwidth]{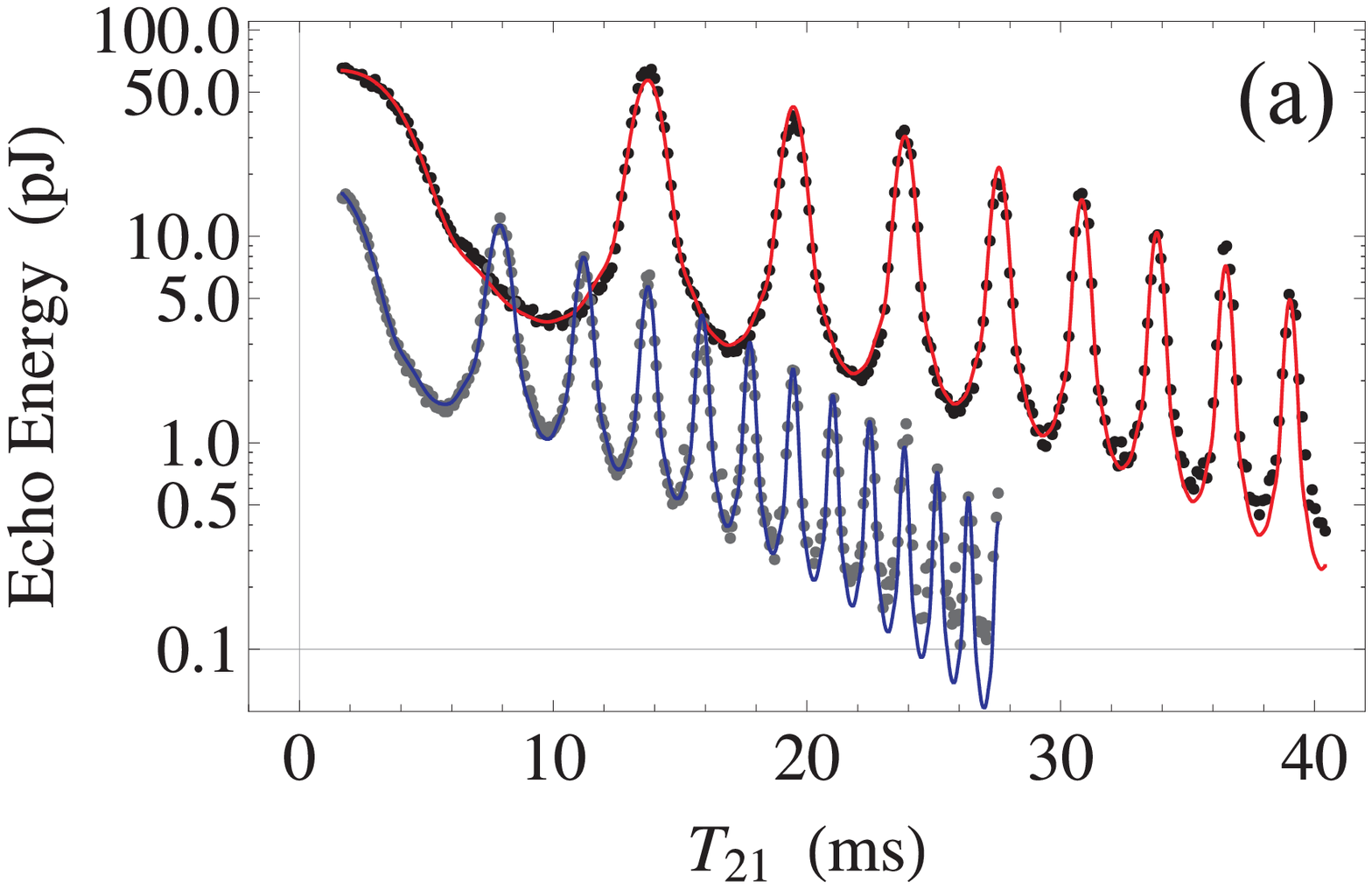}}
  \hspace{0.2cm}
  \subfigure{\label{fig:7b-3Pulse-Gradients-N1-N2}\includegraphics[width=0.48\textwidth]{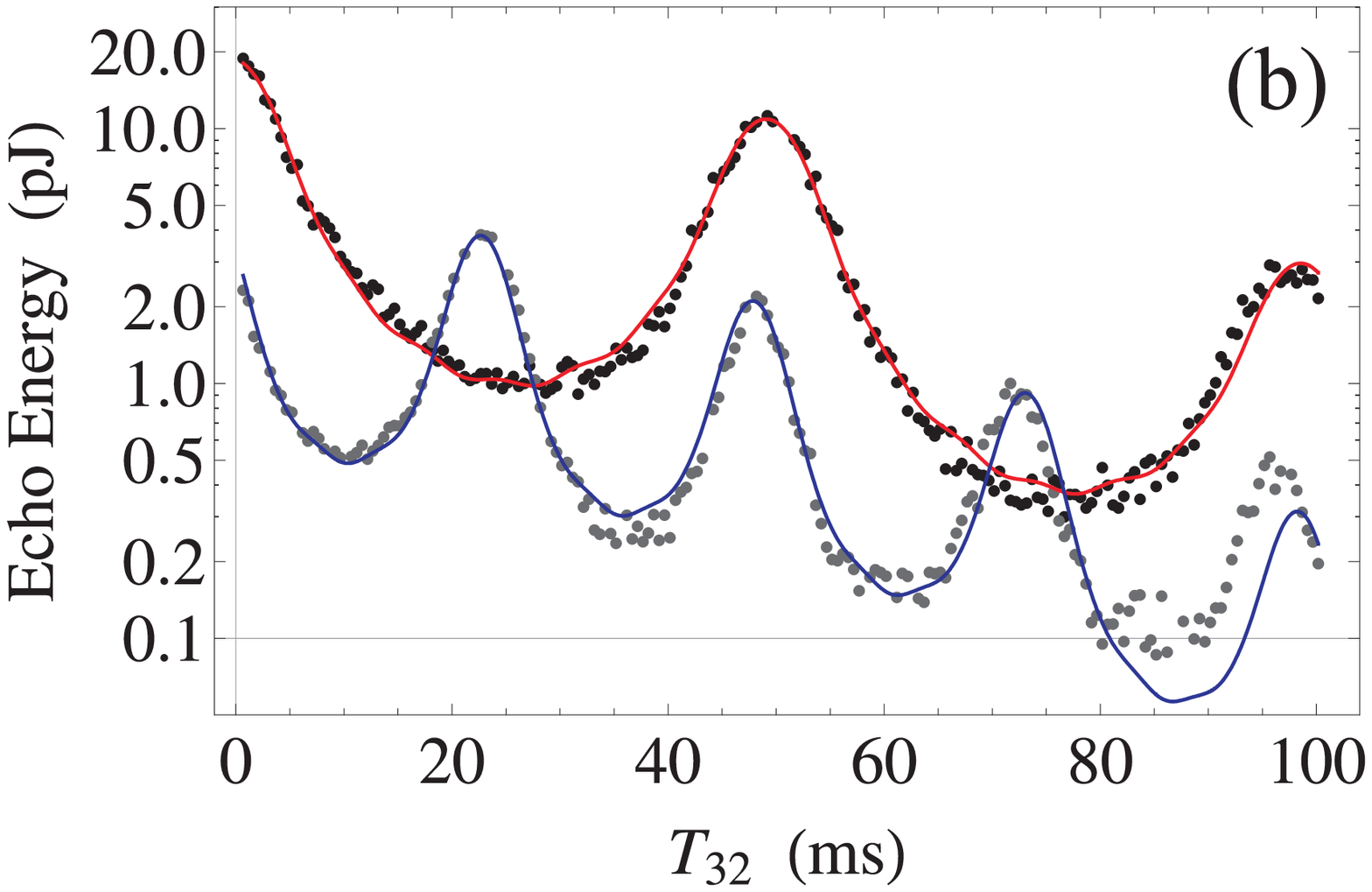}}
  \caption{(Color online) (a) Two-pulse echo energy as a function of $T_{21}$ in the presence of a fixed $\beta$. Gradient-induced oscillations are shown for the first two echo orders (black points: $\bar{N} = 1$; gray points: $\bar{N} = 2$). Here, $T_{21}$ is varied in integer multiples of $\tau_q$ to avoid sensitivity to atomic recoil effects. Fits based on \Eq \refeqn{eqn:ModelFit-2Pulse}, shown as the solid lines, give $|\beta| = 9.30(1)$ mG/cm and $|\beta| = 9.34(1)$ mG/cm, respectively. (b) Data analogous to (a) obtained using a slightly larger gradient with the three-pulse AI. Here, $T_{21}$ is fixed at 2.0 ms and $T_{32}$ is varied to map out the modulation for the first two echo orders (black points: $\bar{N} = 1$; gray points: $\bar{N} = 2$). Fits based on \Eq \refeqn{eqn:ModelFit-3Pulse} give $|\beta| = 17.50(4)$ mG/cm and $|\beta| = 17.78(5)$ mG/cm, respectively. Two-pulse (three-pulse) AI parameters: $\tau_1 = 800$ (800) ns; $\tau_2 = 100$ (70) ns; $\tau_3 = 0$ (70) ns; $I \sim 64$ mW/cm$^2$.}
  \label{fig:7-Gradients-N1-N2}
\end{figure*}

We use the following model, based on the squared modulus of \Eq \refeqn{eqn:E(2)_beta}, to fit the data shown in \Fig \ref{fig:6b-EchoEnergyVsGradient}:
\begin{align}
\begin{split}
  \label{eqn:ModelFit-2Pulse}
  S^{(2)}(\beta, T_{21})
  & = S_0 e^{-(T_{21} - t_0)^2/\tau^2} \sum_{m_F, m_F'} a_{m_F} a_{m_F'} \\
  & \times e^{i A (m_F - m_F') \beta \bar{N} (\bar{N} + 1) (T_{21} - t_1)^2},
\end{split}
\end{align}
where $S_0$, $t_1$ and the set of $\{a_{m_F}\}$ are free parameters, $A = q g_F \mu_B/2M$ is a constant and $t_0$ was set to $T_{21}$ for this data. In this model, the Gaussian factor outside the sum is added phenomenologically to account for signal loss due to both the transit time and any decoherence in the system. Also, each $a_{m_F}$ is proportional to the magnetic sub-level population, $|\alpha_{m_F}|^2$, through \Eq \refeqn{eqn:E(2)_mF}. As a result, these parameters are constrained to be positive. All other fit parameters are unconstrained. In principle, it should be possible to obtain the sub-level populations from the set of best fit parameters $\{a_{m_F}\}$. However, determining the constant of proportionality between the $a_{m_F}$, the populations and the scattered field intensity is complicated \cite{Slama-PRA-2006, Schilke-PRL-2011} and not addressed by the theory presented here. We emphasize, however, that fits to data presented in this work give similar results for the set of $\{a_{m_F}\}$, which are consistent with our expectations for circularly polarized excitation beams.

It is interesting that a measurement of the parameter $A$ from data similar to that shown in \Fig \ref{fig:6b-EchoEnergyVsGradient} can be used to test the theory of magnetic interactions \cite{Chan-PRA-2011, Anthony-PRA-1994}.

Surveys of gradient-induced modulation on the echo signal shown in \Fig \ref{fig:7-Gradients-N1-N2} provide additional confirmation of the theory outlined in \Sec \ref{sec:Theory} and the Appendix. Figure \ref{fig:7a-2Pulse-Gradients-N1-N2} indicates that, in the presence of a $B$-gradient, the two-pulse echo energy becomes modulated at a frequency that increases linearly with $T_{21}$ (i.e. the modulation is chirped), as predicted by \Eq \refeqn{eqn:omega(2)_beta}. This figure shows gradient oscillations for both the $\bar{N} = 1$ and the $\bar{N} = 2$ orders of the two-pulse echo. Since the chirp rate increases as $\bar{N} (\bar{N} + 1)$, the second order echo is modulated at a rate three times that of the first order echo. Confirmation of this is provided by a least-squares fit to the data based on \Eq \refeqn{eqn:ModelFit-2Pulse}, as shown by the solid lines in \Fig \ref{fig:7a-2Pulse-Gradients-N1-N2}. Since the gradient was held fixed in the experiment, the fits to the two data sets should provide similar measurements of $|\beta|$. The two measurements yield $|\beta| = 9.30(1)$ mG/cm for $\bar{N} = 1$ and $|\beta| = 9.34(1)$ mG/cm for $\bar{N} = 2$ \footnote{Measurements of the $B$-gradient from the scattered field intensity are not sensitive to the sign of $\beta$. However, the sign can be determined using a heterodyne technique to measure the scattered electric field amplitude.}, where the quoted error is the $1\sigma$ statistical uncertainty generated by the fit. These measurements are in good agreement with each other and an independent measurement from a flux-gate magnetometer. We emphasize that accurate fits to these data and the extraction of $\beta$ were possible only through the development of the multi-level formalism presented in the Appendix. In particular, since the oscillations shown in \Fig \ref{fig:7a-2Pulse-Gradients-N1-N2} do not occur with 100\% contrast (i.e. each oscillation minima does not reach the level of the noise), a model including only two magnetic sub-levels with equal excitation probabilities, such as that described in \Ref \citenum{Weel-PRA-2006}, is insufficient to model the data.

Figure \ref{fig:7b-3Pulse-Gradients-N1-N2} shows data similar to that shown in \Fig \ref{fig:7a-2Pulse-Gradients-N1-N2}, but for the first two orders of the three-pulse echo and a slightly larger $B$-gradient. This data illustrates that the three-pulse AI is less sensitive to gradients than the two-pulse AI. Since $T_{21}$ is fixed at 2.0 ms, the modulation frequency is constant and proportional to $\bar{N}$ and $T_{21}$---confirming the predictions of \Eq \refeqn{eqn:omega(3)_beta}. The data is fit to the following model:
\bea
  \label{eqn:ModelFit-3Pulse}
  & & S^{(3)}(\beta, T_{32}, T_{21}) = S_0 e^{-(T_{32} - t_0)^2/\tau^2} \sum_{m_F, m_F'} a_{m_F} a_{m_F'} \notag \\
  & & \;\;\;\; \times e^{i A (m_F - m_F') \beta [\bar{N} (\bar{N} + 1) T_{21}^2 + 2 \bar{N} T_{21} (T_{32} - t_1)]},
\eea
which is based on \Eqs \refeqn{eqn:E(N)}, \refeqn{eqn:E(N)_mF} and \refeqn{eqn:phi(N)}, with a Gaussian decay factor added phenomenologically. All other parameters in this model are similar to those discussed in reference to \Eq \refeqn{eqn:ModelFit-2Pulse}. Measurements of the magnitude of the gradient from fits to these data yield $|\beta| = 17.50(4)$ mG/cm and $|\beta| = 17.78(5)$ mG/cm for the $\bar{N} = 1$ and $\bar{N} = 2$ echoes, respectively. These two measurements differ by more than $5\sigma$, which deserves some explanation. By inspecting the fit to the $\bar{N} = 2$ echo, it is clear that the data is not well-modeled by a single frequency sinusoid as $T_{32}$ becomes large. This provides evidence that the atoms are sampling different gradients as they drop under gravity---an effect that is not accounted for in the theory. By analyzing different sections of this data, we estimate that the gradient varies by as much as $\sim 1.6$ mG/cm between $T_{32} \sim 40$ ms and 100 ms---during which time atoms fall $\sim 4$ cm. Independent measurements of the curvature of the $B$-field, where $|\beta|$ was found to change by $\sim 0.4$ mG/cm every centimeter, are consistent with the variation in $\beta$ detected by atoms.

\begin{figure}[!t]
  \centering
  \subfigure{\label{fig:8a-2Pulse-SmallGradient}\includegraphics[width=0.46\textwidth]{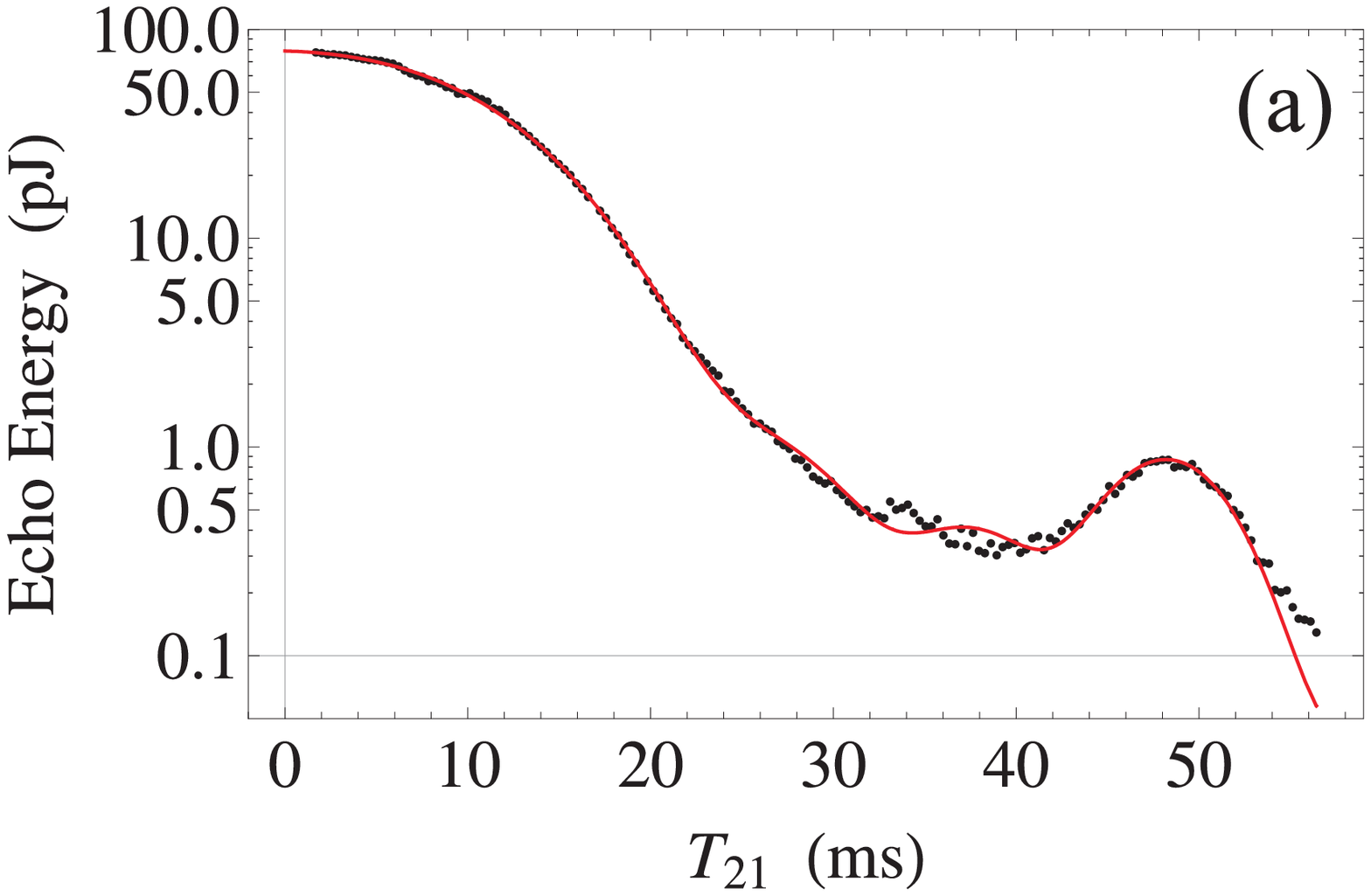}}
  \subfigure{\label{fig:8b-3Pulse-SmallGradient}\includegraphics[width=0.46\textwidth]{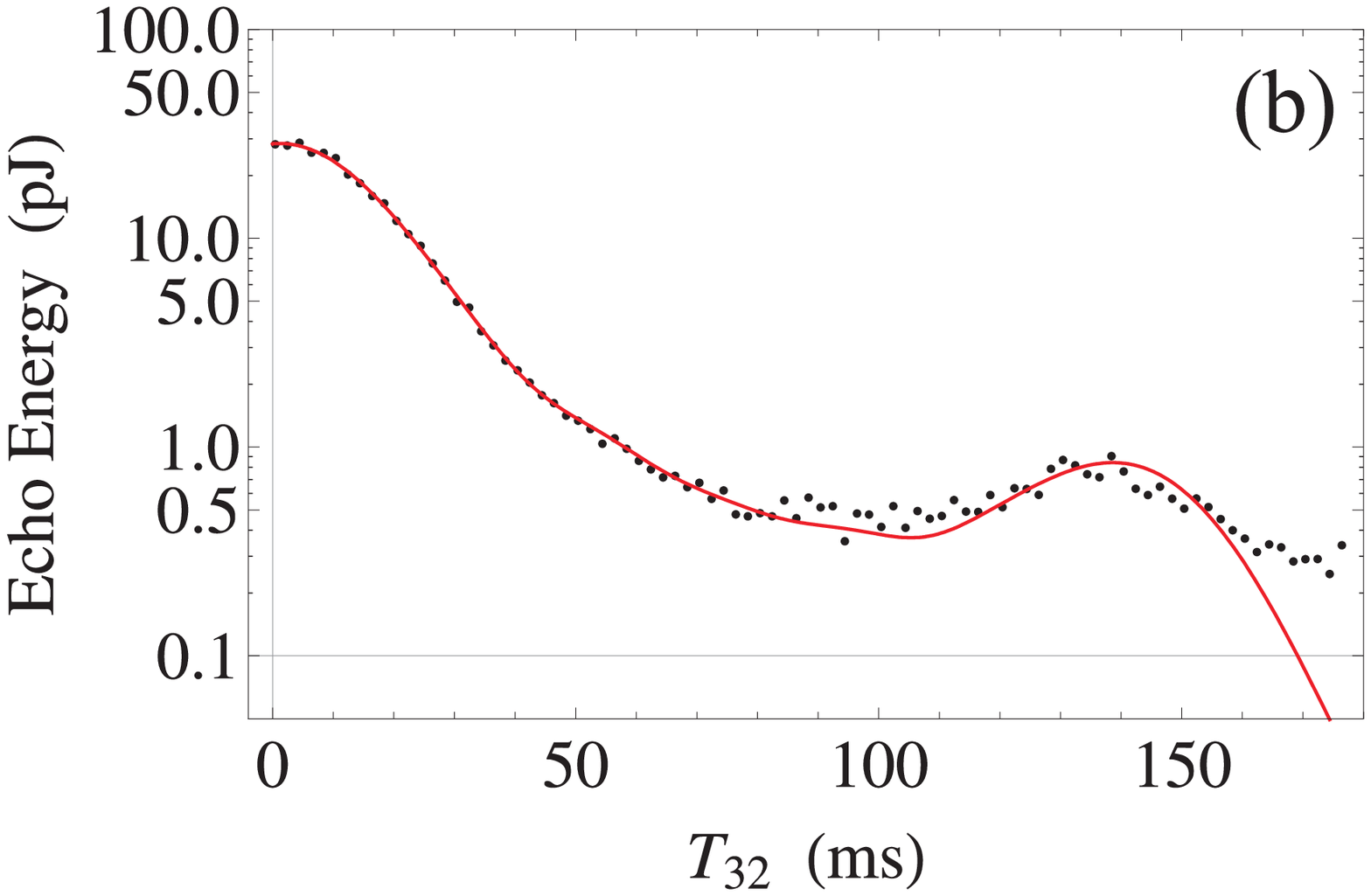}}
  \caption{(Color online) Smallest direct measurement of a $B$-gradient using the two-pulse (a) and three-pulse (b) AIs. Here, the applied fields were tuned separately for each AI such that the first revival in the $\bar{N} = 1$ echo energy occurred at the largest time. A fit based on \Eq \refeqn{eqn:ModelFit-2Pulse} for the two-pulse AI gave $|\beta| = 0.26(3)$ mG/cm. Similarly, a fit based on \Eq \refeqn{eqn:ModelFit-3Pulse} yielded $|\beta| = 9.5(1)$ mG/cm for the three-pulse AI, with $T_{21} \sim 1.27$ ms. Two-pulse (three-pulse) AI parameters: $\tau_1 = 800$ (800) ns, $\tau_2 = 100$ (70) ns, $\tau_3 = 0$ (70) ns, $I \sim 64$ mW/cm$^2$.}
  \label{fig:8-SmallGradient}
\end{figure}

Although we have demonstrated sensitivity to \emph{changes} in the $B$-gradient as small as $\sim 4 \times 10^{-5}$ G/cm using $T_{21} \sim 40$ ms with the two-pulse AI, our ability to measure the absolute magnitude of the applied gradient is less sensitive. This is primarily because the measurement is based on fitting data to an oscillatory model and extracting the modulation rate---which cannot be done accurately without the presence of an oscillatory component in the data. To estimate the smallest measurable $B$-gradient with the two interferometers, we tuned the applied fields for each AI separately such that the first revival in the $\bar{N} = 1$ echo energy occurred at the largest time. The resulting data are shown in \Fig \ref{fig:8-SmallGradient}, which yielded measurements of $|\beta| = 0.26(3)$ mG/cm for the two-pulse AI [\Fig \ref{fig:8a-2Pulse-SmallGradient}] and $|\beta| = 9.5(1)$ mG/cm for the three-pulse AI [\Fig \ref{fig:8b-3Pulse-SmallGradient}].

\section{Applications to Gravity}
\label{sec:Gravity}

The apparatus shown in \Fig \ref{fig:3-ExpSetup} is designed for measurements of the atomic recoil frequency \cite{Barrett-SPIE-2011}. As a result, it is not isolated from external vibrations and is unsuitable for measurements of the optical phase of the scattered read-out light using heterodyne detection. For this reason, a measurement of $g$ from the phase of the atomic grating \cite{Weel-PRA-2006, Barrett-Advances-2011} is beyond the scope of this Article and will be presented elsewhere. However, the aforementioned results relating to $B$-gradients validate theoretical predictions that can be applied to precise measurements of gravity. In this section, we discuss the feasibility of such a measurement by applying the formalism presented in the Appendix.

The best portable gravimeter \cite{Niebauer-Metrologia-1995} uses an optical Mach-Zehnder interferometer where one arm contains a free-falling corner-cube for position-sensitive measurements of $g$ at the level of $\sim 1$ ppb over a few minutes. The position sensitivity in these devices comes from detecting interference fringes as a function of the drop time of the cube relative to an inertial frame defined by a stationary mirror. The frequency at which the fringes accumulate scales linearly with the drop time (i.e. the frequency is chirped). The matter-wave analog of this gravimeter is the two-pulse echo AI \cite{Weel-PRA-2006, Barrett-Advances-2011}, where changes in the phase of the grating due to gravity are detected relative to the nodes of a pulsed sw---which serves as the inertial reference frame. In this case, the accumulation of fringes due to matter-wave interference is also described by a chirped-frequency sinusoid.

We now review the main results of the grating echo theory that pertain to gravity. The gravitational potential can be written as
\be
  \hat{U}(z) = M g \hat{I} z,
\ee
where the force is $\mathcal{F} = - M g$ and $\hat{I}$ is the $(2F+1) \times (2F+1)$ identity matrix. The effect on the echo AI is similar to that of the $B$-gradient on a sample that has been optically pumped into a single state. Since gravity acts equally on all states, the phase shift of the grating produced by each state is the same. Therefore, the expression for the field scattered from the grating simplifies significantly compared to \Eq \refeqn{eqn:E(2)_beta}:
\be
  \label{eqn:E(2)_g}
  E_g^{(2)}(t;\bm{T}) = \left[ \sum_{m_F} E^{(2)}_{m_F}(t;\bm{T}) \right] e^{i \phi^{(2)}_g(t;\bm{T})},
\ee
where the grating phase due to gravity is
\begin{align}
\begin{split}
  \label{eqn:phi(2)_g}
  & \phi^{(2)}_g(\Delta t;T_{21}) = -\frac{q g}{2} \left\{ \bar{N}(\bar{N}+1)T_{21}^2 \right. \\
  & \;\;\;\;\; \left. + \, 2 \big[ T_1 + (\bar{N}+1)T_{21} \big] \Delta t + \Delta t^2 \right\},
\end{split}
\end{align}
as determined by \Eq \refeqn{eqn:phi(2)_F}. This phase cannot be detected from the intensity of the scattered light because there is no differential phase shift between magnetic sub-levels---thus, there is no amplitude modulation of the grating \cite{Weel-PRA-2006}. The scaling of the grating phase with $T_{21}^2$ in \Eq \refeqn{eqn:phi(2)_g} shows the similarity between the two-pulse AI and the optical Mach-Zehnder interferometer discussed in \Ref \citenum{Niebauer-Metrologia-1995}.

\begin{figure}[!t]
  \centering
  \subfigure{\label{fig:9a-GravitySignal}\includegraphics[width=0.23\textwidth]{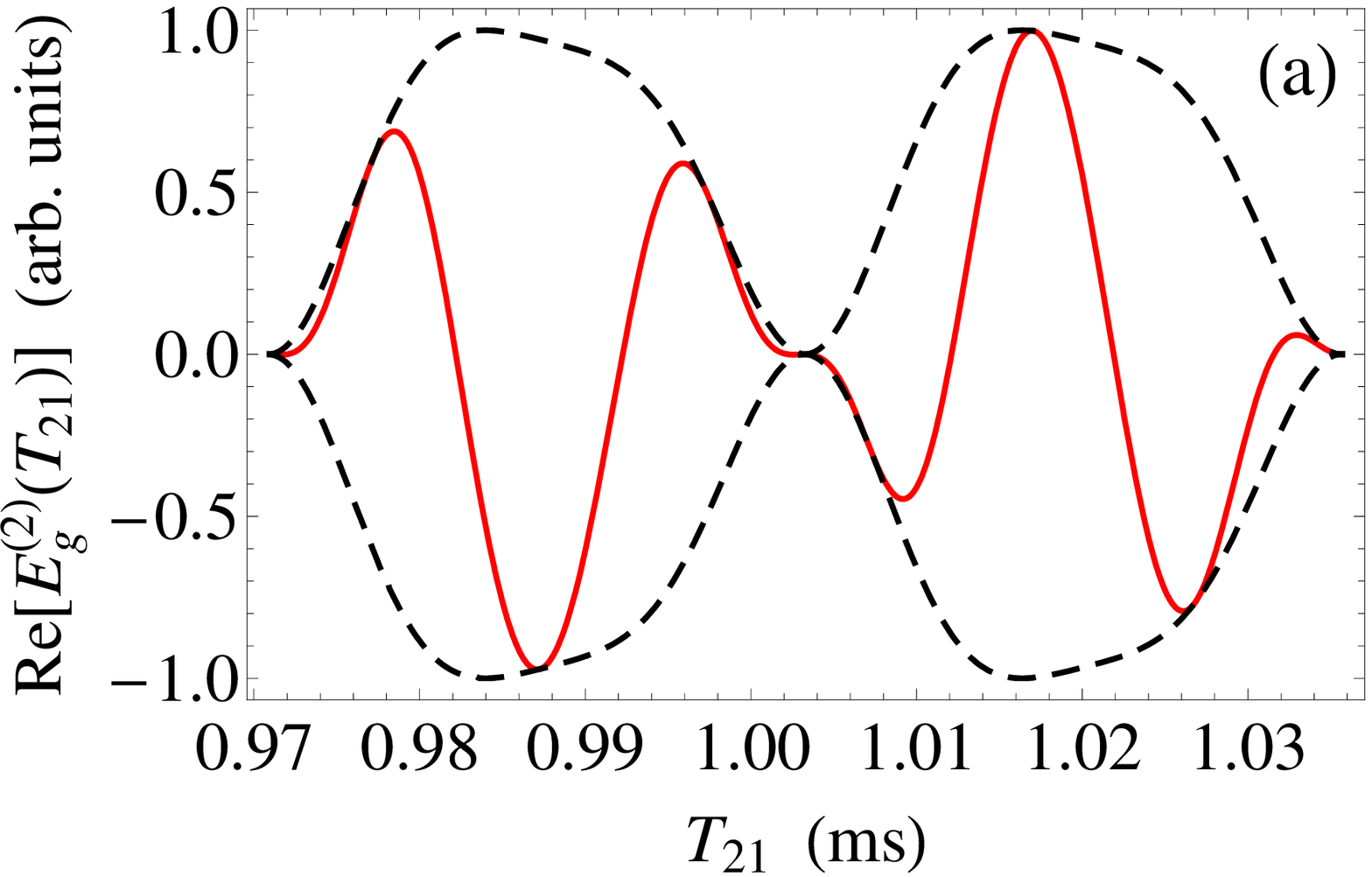}}
  \subfigure{\label{fig:9b-GravitySignal}\includegraphics[width=0.23\textwidth]{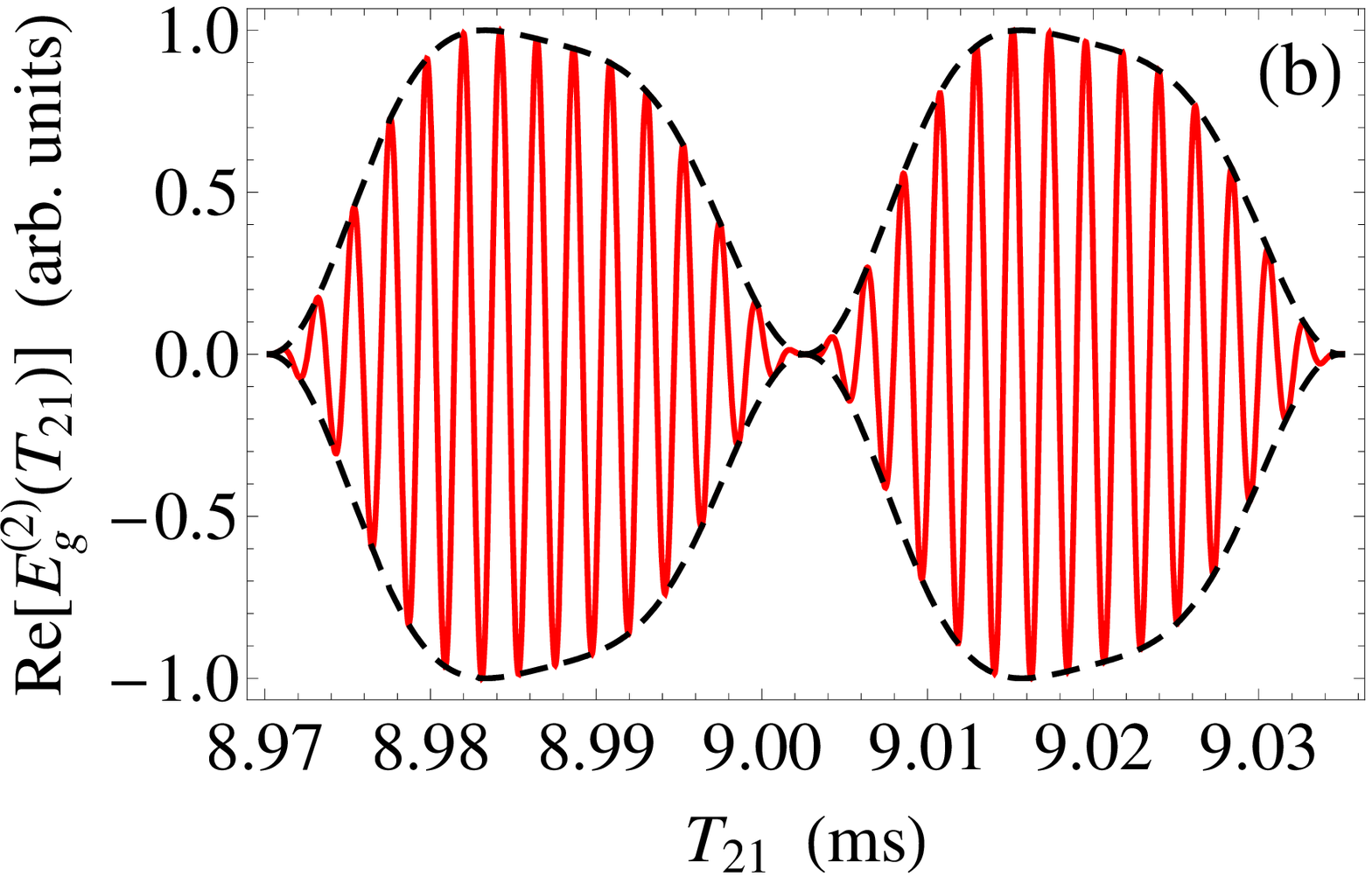}}
  \subfigure{\label{fig:9c-GravitySignal}\includegraphics[width=0.23\textwidth]{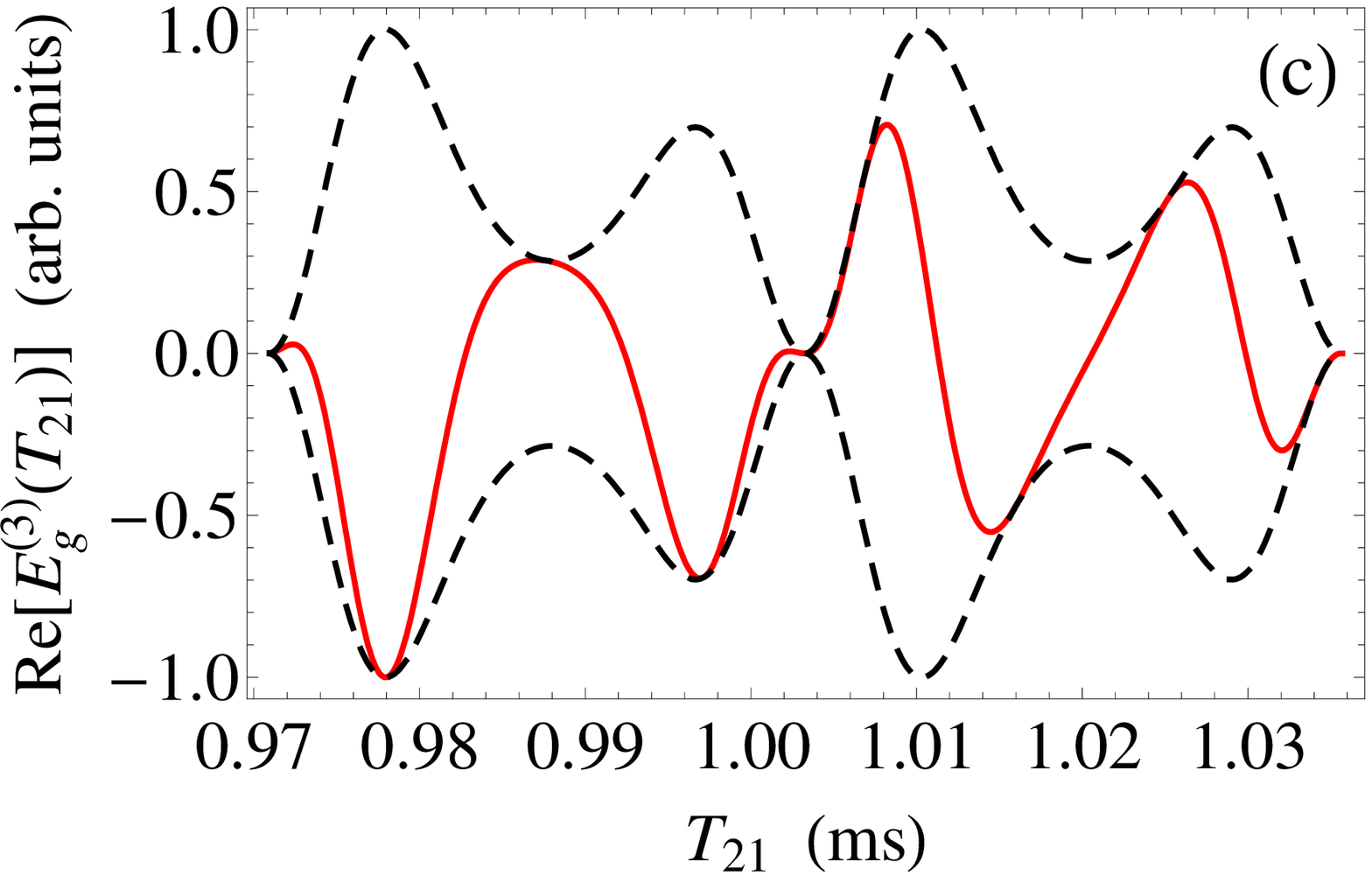}}
  \subfigure{\label{fig:9d-GravitySignal}\includegraphics[width=0.23\textwidth]{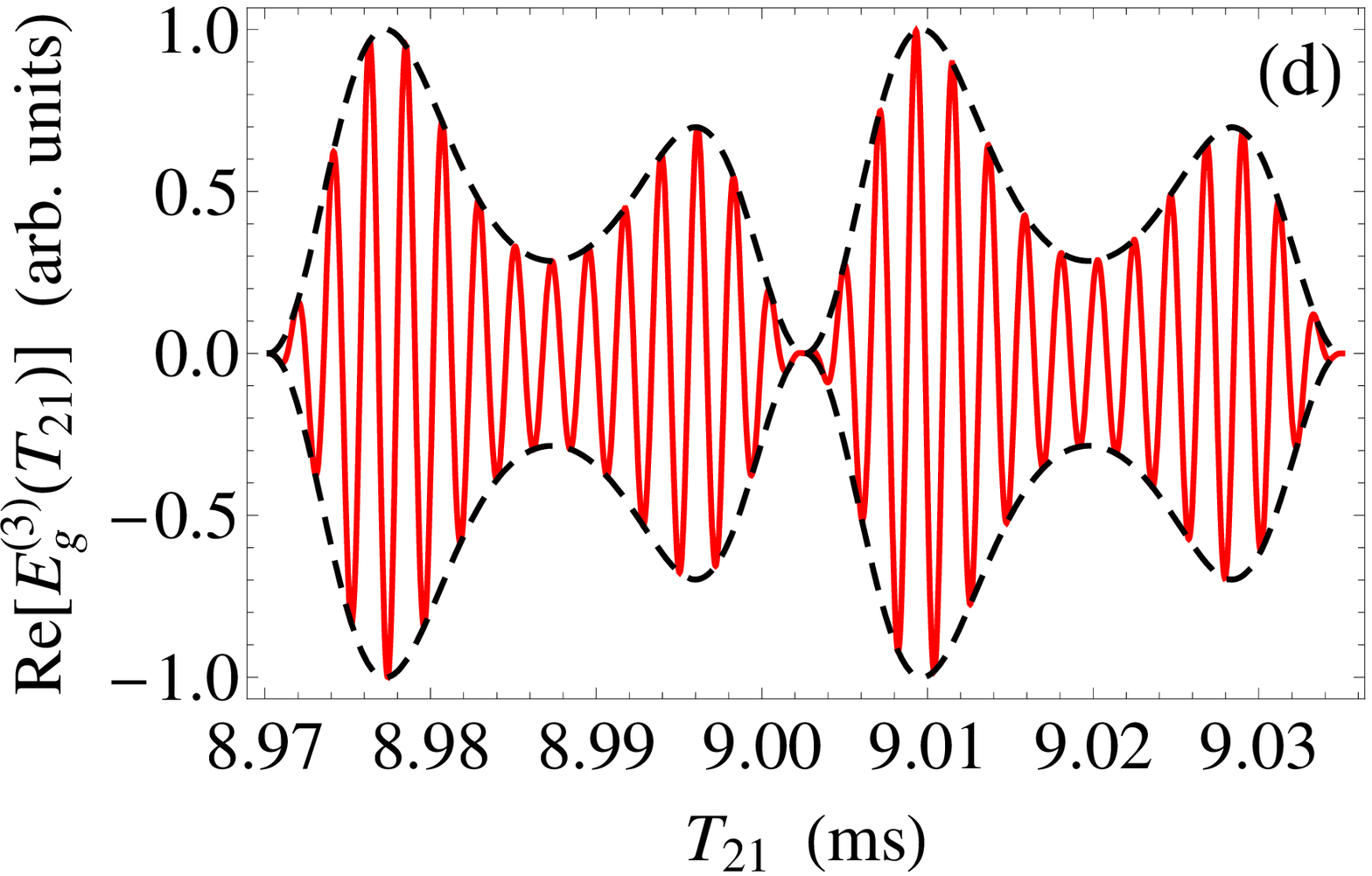}}
  \subfigure{\label{fig:9e-GravitySignal}\includegraphics[width=0.23\textwidth]{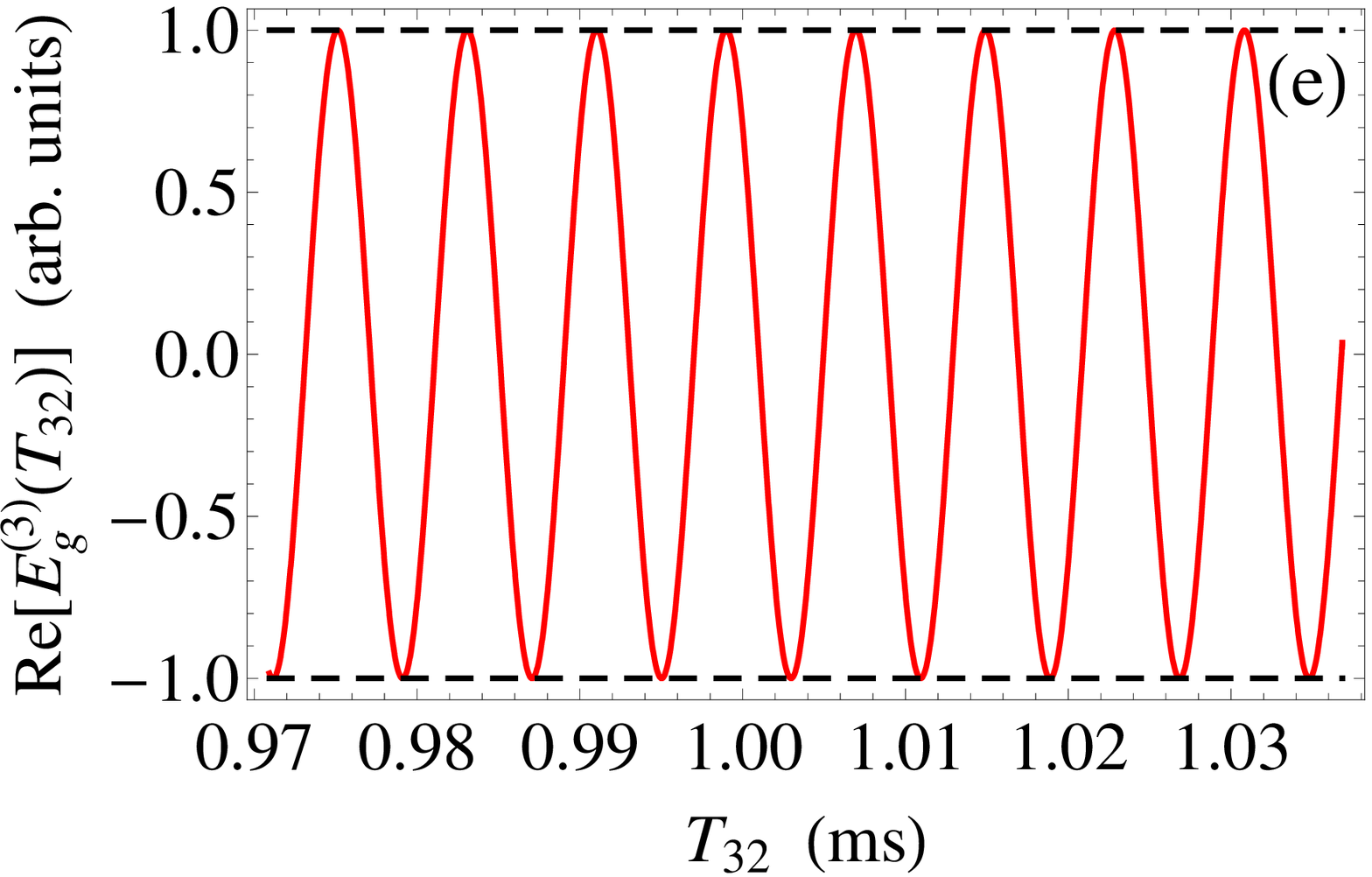}}
  \subfigure{\label{fig:9f-GravitySignal}\includegraphics[width=0.23\textwidth]{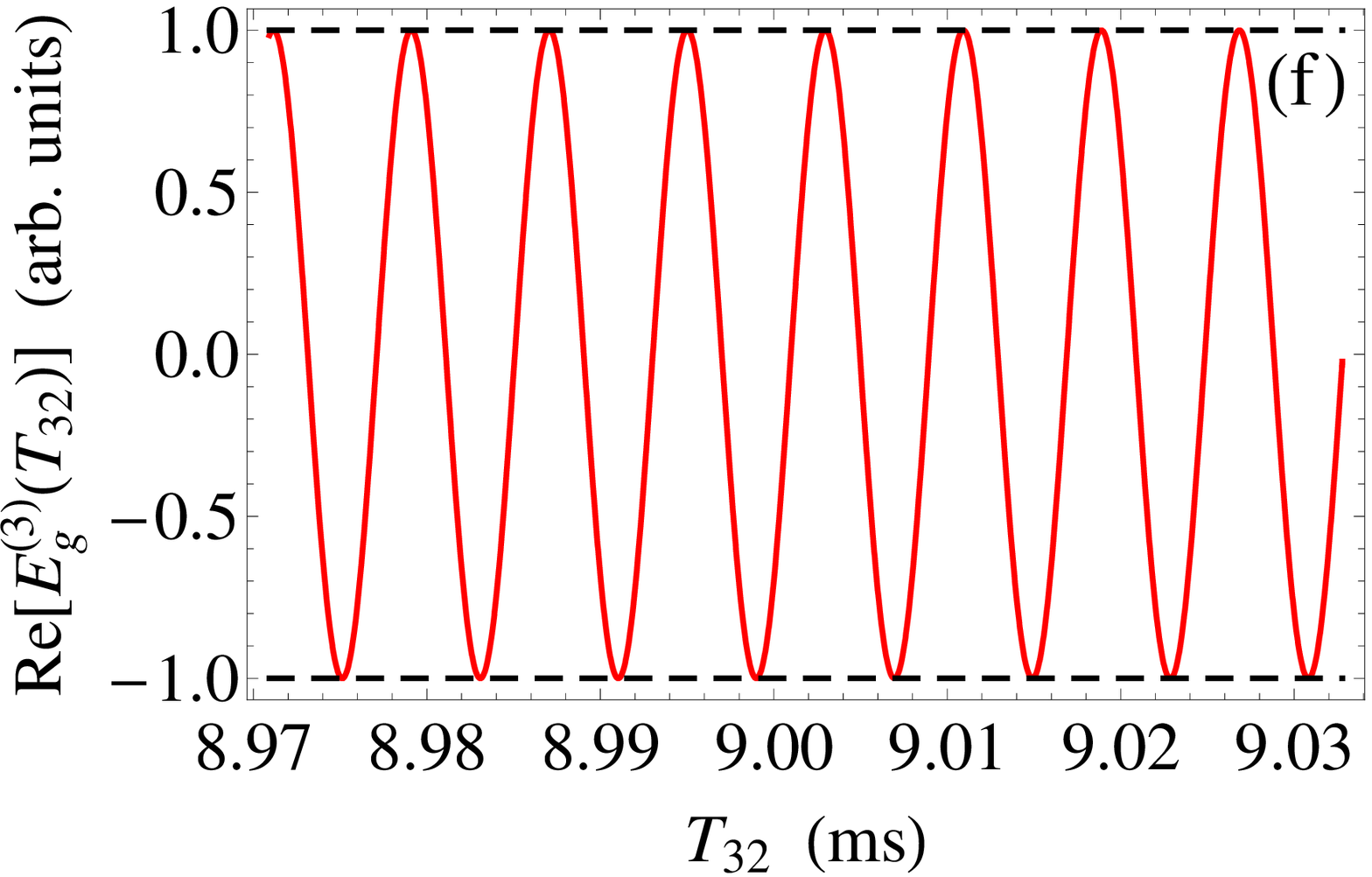}}
  \caption{(Color online) The real part of the scattered electric field (red solid line) in the presence of gravity for the two-pulse (a,b) and the three-pulse (c--f) AIs. In parts (a--d), the black dashed line shows the signal envelope [based on \Eq \refeqn{eqn:E(2)_mF} for (a,b) and \Eq \refeqn{eqn:E(3)_mF} for (c,d)], which is modulated at $2\omega_q \approx 2\pi \times 31$ kHz. (a) The field oscillates at a frequency $\omega^{(2)}_g(T_{21}) \sim 2\pi \times 50$ kHz in the vicinity of $T_{21} \sim 1$ ms. (b) The modulation frequency increases to $\omega^{(2)}_g \sim 2\pi \times 450$ kHz at $T_{21} \sim 9$ ms. Parts (c,d) show results for the three-pulse AI as a function of $T_{21}$, with $T_{32}$ fixed at 100 $\mu$s. The field oscillates at approximately the same frequency in (a,c), and in (b,d), since $T_{32}$ is small. Parts (e,f) show the three-pulse signal as a function of $T_{32}$ with $T_{21} = 5$ ms. Since $T_{21}$ is fixed, there is no sensitivity to atomic recoil and the signal envelope is not modulated. As $T_{32}$ increases, the modulation frequency of the scattered field amplitude remains fixed at $\omega^{(3)}_g(T_{21}) \sim 2\pi \times 125$ kHz.}
  \label{fig:9-GravitySignal}
\end{figure}

Figures \ref{fig:9a-GravitySignal} and \ref{fig:9b-GravitySignal} show the expected two-pulse AI signal in the presence of gravity as a function of $T_{21}$---illustrating that the modulation frequency is chirped linearly with $T_{21}$ ($\omega^{(2)}_g = \partial \phi^{(2)}_g/ \partial T_{21} \propto T_{21}$). As $T_{21}$ increases, $\omega^{(2)}_g$ becomes larger than the recoil frequency (for the first order echo in $^{85}$Rb, this occurs when $T_{21} \gtrsim 300$ $\mu$s), and $T_{21}$ must be incremented in steps less than $\tau_q$ to avoid undersampling the frequency. However, this effect causes reduced sensitivity to the grating phase, since modulation at the recoil frequency produces periodic regions with small signal amplitude. Additionally, as shown in \Ref \citenum{Su-PRA-2010}, this AI is very sensitive to phase changes due to mirror vibrations, which can be detrimental to measurements of $g$ using this technique.

Figures \ref{fig:9c-GravitySignal} and \ref{fig:9d-GravitySignal} show the expected three-pulse signal as a function of $T_{21}$ in the presence of gravity. It is clear that the envelope of the scattered field has a complicated periodic dependence on $T_{21}$, with a zero every $\tau_q \sim 32$ $\mu$s due to the destructive interference of momentum states differing by the two-photon recoil momentum, $\hbar q$. This is similar to the two-pulse case shown in \Figs \ref{fig:9a-GravitySignal} and \ref{fig:9b-GravitySignal}. Here, the grating phase modulation frequency is given by
\be
  \left| \frac{\partial \phi^{(3)}_g}{\partial T_{21}} \right|
  = q g [\bar{N} (\bar{N} + 1) T_{21} + \bar{N} T_{32} + (\bar{N} + 1) \Delta t],
\ee
which is identical to the two-pulse case, $\omega^{(2)}_g$, with the addition of the term proportional to $T_{32}$.

Figures \ref{fig:9e-GravitySignal} and \ref{fig:9f-GravitySignal} show the expected three-pulse signal as a function of $T_{32}$, with $T_{21}$ fixed at 5 ms. In this case, there is no sensitivity to atomic recoil, so the envelope remains at a constant level as $T_{32}$ is varied. The frequency of the phase modulation is also fixed by $\bar{N}$ and $T_{21}$, as given by
\be
  \omega^{(3)}_g (T_{21})
  = \left| \frac{\partial \phi^{(3)}_g}{\partial T_{32}} \right|
  = q g \left( \bar{N} T_{21} + \Delta t \right).
\ee
For the conditions presented in these figures, $\omega_g^{(3)} \sim 2\pi \times 125$ kHz. The work of \Ref \citenum{Su-PRA-2010} shows that the three-pulse AI is significantly less sensitive to mirror vibrations than the two-pulse AI if $T_{32} \gg T_{21}$. Our results have also shown that this configuration is less sensitive to $B$-gradients. For all these reasons, this AI is particularly well-suited for precise measurements of $g$.

Simulations of the two-pulse AI signal with $\bar{N} = 1$, $T_{21} \sim 150$ ms and a phase error of 1\% suggest the precision of a measurement of $g$ should be $\sim 1.4$ ppb. Similarly, we estimate a precision of $\sim 0.4$ ppb for the three-pulse AI using $\bar{N} = 1$, $T_{21} = 75$ ms, $T_{32}$ varied over 150 ms and the same phase error. From these estimates, it is clear that these AIs can have greater sensitivity than the best industrial sensor \cite{Niebauer-Metrologia-1995}. Since the precision scales linearly with the phase error, we anticipate further improvements in sensitivity without extending the timescale. If systematic effects of such a cold atom gravimeter are characterized, it may be possible for the AI experiment to serve as a reference to calibrate other gravimeters.

\section{Conclusions}
\label{sec:Conclusions}

Measurements of applied $B$-gradients using both the two- and three-pulse techniques are in good agreement with independent measurements of $\beta$ using a flux gate magnetometer. We have demonstrated sensitivity to changes in the $B$-gradient at the level of $\sim 4 \times 10^{-5}$ G/cm. Absolute measurements of $\beta$ as small as $\sim 3 \times 10^{-4}$ G/cm were also possible using the two-pulse AI. These measurements indicate that an accurate description of the data presented above requires the inclusion of multiple magnetic sub-levels. We have also shown sensitivity to spatial variation in the $B$-gradient using a long-lived second order ($\bar{N} = 2$) three-pulse echo. It is this non-linearity in the $B$-field that affects the timescale in echo AIs rather than the presence of a small, uniform $B$-gradient.

As tests of the theoretical results presented in \Sec \ref{sec:Theory}, we have separately confirmed the linear dependence of the $\beta$-induced oscillation frequencies, $\omega^{(2)}_{\beta}$ and $\omega^{(3)}_{\beta}$ [given by \Eqs \refeqn{eqn:omega(2)_beta} and \refeqn{eqn:omega(3)_beta}, respectively], on the $B$-gradient. We have also verified that these frequencies both scale linearly with $T_{21}$, and, for the three-pulse AI, $\omega^{(3)}_{\beta}$ is constant as a function of $T_{32}$.

Since we have achieved signal lifetimes approaching the transit time limit, we have shown that fountain-based experiments are possible with grating echo AIs. The advantage of a fountain configuration is that the spatial extent of the AI ($\sim 11$ cm for 300 ms timescale) can be made small, which reduces the requirements for inhomogeneous $B$-field suppression. Such a configuration is ideal for precise measurements of gravity, particularly with the three-pulse AI. Passive suppression of $B$-fields with larger cancelation coils, or optically pumping into the $m_F = 0$ sub-level, represent two ways in which such a measurement can be realized.

Despite the widespread use of Raman-type AIs for inertial sensing \cite{Yu-ApplPhysB-2006, LeGouet-ApplPhysB-2006, Young-OSA-2007}, grating echo-type AIs---which offer reduced experimental complexity---are also excellent candidates for precision measurements of $\omega_q$ and $g$. This work has brought about understanding of systematic effects produced by $B$-gradients on these measurements.

In summary, we have developed a complete understanding of the effects of a constant force that applies to all time-domain AIs. Although the sensitivity for AI-based gradient detection cannot compete with commercial magnetic gradiometers (which offer sensitivities of the order of $\sim 1$ pT/m), the technique is useful for absolute measurements of gradients in cold atom experiments.

\acknowledgements

This work was supported by the Canada Foundation for Innovation, Ontario Innovation Trust, Natural Sciences and Engineering Research Council of Canada, Ontario Centres of Excellence and York University. We would also like to thank Itay Yavin of McMaster University for helpful discussions and Adam Carew of York University for building phase-locked loops.

\appendix*
\renewcommand{\theequation}{A\arabic{equation}} 
\setcounter{equation}{0}  

\section*{Appendix}

In this appendix, we derive expressions for the signals generated by the two- and three-pulse interferometers in the presence of a constant external force, $\mathcal{F}$. In \Ref \citenum{Weel-PRA-2006}, a similar calculation for the two-pulse signal is given, in which only two ground state sub-levels are considered, and effects due to spontaneous emission are ignored. Here, we account for $2F+1$ magnetic sub-levels in the field scattered from the atoms, as well as spontaneous emission during the excitation pulses. Both of these effects are crucial for an accurate description of these interferometers. We also give a general expression for the signal generated by an $N$-pulse AI from which all classes of grating-echo interferometers can be realized.

The potential is assumed to have the form $\hat{U}(z) = -\hat{\mathcal{M}} z$, where $\hat{\mathcal{M}} = -\partial \hat{U}/\partial z$ is an operator that computes with $z$ and $p$, and acts on the basis states $\ket{F\;m_F}$ with eigenvalues $m_F \mathcal{F}$. Here, $\mathcal{F}$ is a constant with units of force. We proceed by computing the ground state wave function after the application of each sw pulse at times $t = T_1$ and $T_2$, with a period of evolution before, between and after each pulse (with durations $T_1$, $T_2 - T_1$ and $t - T_2$, respectively) in the presence of the force. During the application of each sw pulse, the kinetic and potential energy terms in the Hamiltonian are ignored by assuming the pulses are sufficiently short such that the atom does not move significantly (Raman-Nath approximation). In this manner, the sw pulses are treated as Dirac $\delta$-function excitations, even though they are given durations $\tau_j$ for the purposes of the calculation.

The interferometer signal is defined as the back-scattered electric field amplitude at the time of an echo, which is proportional to the amplitude of the $q$-Fourier harmonic of the density distribution at these times. The results for the two-pulse AI signal are then generalized for an $N$-pulse AI, from which we compute the three-pulse AI signal.

The Hamiltonian for the ground state $\ket{F\;m_F}$ in the presence of a sw field and an external potential, $\hat{U}(z)$, can be approximated by \cite{Beattie-PRA-2008, Barrett-PRA-2010}
\be
  \label{eqn:Hg}
  \hat{H}_{m_F} = \frac{p^2}{2M} + \hbar \chi_{m_F} e^{i\theta} \cos(q z) + \hat{U}(z),
\ee
where $\theta$ is a phase associated with spontaneous emission during the sw pulse
\be
  \theta = \tan^{-1} \left( -\frac{\gamma}{\Delta} \right),
\ee
and $\chi_{m_F}$ is a two-photon Rabi frequency given by
\be
  \chi_{m_F} = \frac{\Omega_0^2}{2\Delta} \left( 1 + \frac{\gamma^2}{\Delta^2} \right)^{-1/2} \left( C^{F\;\;\;\;1\;\;\;F+1}_{m_F\;q_L\;m_F + q_L} \right)^2.
\ee
Here, $\Omega_0$ is the on-resonance Rabi frequency for a two-level atom, $\Delta = \omega_L - \omega_0$ is the atom-field detuning with atomic resonance frequency $\omega_0$ and laser frequency $\omega_L$, $\gamma$ is half of the spontaneous emission rate, and $(C^{F\;\;\;\;1\;\;\;F+1}_{m_F\;q_L\;m_F + q_L})$ is a Clebsch-Gordan coefficient for a light field with a polarization state $q_L$. We ignore the excited state in this treatment, since the field is assumed to be relatively weak and far off-resonance $(|\Delta| \gg \Omega_0,\, \gamma)$. We also neglect the Zeeman shift of magnetic sub-levels by assuming $|\Delta| \gg g_F \mu_B B/\hbar$.

The amplitude of the ground state wave function at $t = 0$ can be written as a superposition of spin states:
\be
  a(z,0) = \sum_{m_F} a_{m_F}(z,0) \ket{F\;m_F},
\ee
where the amplitude of each spin state is
\begin{subequations}
\bea
  a_{m_F}(z,0) & = & \frac{\alpha_{m_F}}{\sqrt{2\pi\hbar}} e^{i p_0 z/\hbar}, \\
  a_{m_F}(p,0) & = & \alpha_{m_F} \delta(p - p_0).
\eea
\end{subequations}
Here, $p_0$ is the initial momentum of the atom along the $z$-direction, $|\alpha_{m_F}|^2$ is the population of state $\ket{F\;m_F}$, with $\sum_{m_F} |\alpha_{m_F}|^2 = 1$, and $a_{m_F}(p,0)$ is the amplitude of the spin state in momentum space.

The main challenge in this calculation is evolving the wave function between sw pulses in the presence of the additional potential energy, $\hat{U}(z)$. In the absence of this potential, it is straightforward to integrate the Schr\"{o}dinger equation in momentum space. However, with $\hat{U}(z)$ present, we have the following equation of motion:
\be
  i \hbar \frac{\partial a_{m_F}}{\partial t} = \left( \frac{p^2}{2M} - \hat{\mathcal{M}} z \right) a_{m_F}(p,t).
\ee
One can integrate this equation to find
\be
  \label{eqn:aexp(z+p2)}
  a_{m_F}(p,t) = e^{-i(-\hat{\mathcal{M}} z + p^2/2M) t/\hbar} a_{m_F}(p,0),
\ee
but some care must be taken when evaluating the right hand side. The challenge arises from the fact that $z$ and $p = -i\hbar \partial/\partial z$ are non-commuting operators. As a result, the exponential in \Eq \refeqn{eqn:aexp(z+p2)} is really a matrix exponential of non-commuting matrices $\hat{A}$ and $\hat{B}$. In general $e^{\hat{A} + \hat{B}} \not= e^{\hat{A}}e^{\hat{B}}$, but one can use the Zassenhaus formula \cite{Suzuki-CommunMathPhys-1977} to expand the matrix exponential as
\begin{align}
\begin{split}
  \label{eqn:Zassenhaus}
  e^{\xi(\hat{A} + \hat{B})}
  & = e^{\xi\hat{A}} e^{\xi\hat{B}} e^{-\xi^2 [\hat{A},\hat{B}]/2} \\
  & \times e^{\xi^3([\hat{A},[\hat{A},\hat{B}]] - 2[[\hat{A},\hat{B}],\hat{B}])/6} \cdots,
\end{split}
\end{align}
where $\xi$ is an arbitrary constant. The higher order factors (represented by $\cdots$ in the above equation) vanish if $[[\hat{A},\hat{B}],\hat{B}]$ and $[\hat{A},[\hat{A},\hat{B}]]$ commute with all higher order nested commutators. Choosing $\hat{A} = -\hat{\mathcal{M}} z$ and $\hat{B} = p^2/2M$ \footnote{This choice is not arbitrary. Since the $p$-space wave function is an eigenstate of the operator $p^2/2M$, but not $-\hat{\mathcal{M}} z$, we save ourselves some effort by choosing $\hat{B} = p^2/2M$ since $e^{\hat{B}}$ operates on the wave function before $e^{\hat{A}}$.}, and using the commutation relations $[z,p^2] = i 2\hbar p$, $[z, p] = i\hbar$, we find:
\begin{subequations}
\label{eqn:commutators}
\bea
  \left[-\hat{\mathcal{M}} z, \frac{p^2}{2M} \right] & = & -i\frac{\hbar \hat{\mathcal{M}}}{M} p, \\
  \left[-\hat{\mathcal{M}} z, \left[-\hat{\mathcal{M}} z, \frac{p^2}{2M} \right] \right] & = & - \frac{\hbar^2 \hat{\mathcal{M}}^2}{M}, \\
  \left[ \left[ \frac{p^2}{2M},-\hat{\mathcal{M}} z \right], \frac{p^2}{2M} \right] & = & 0.
\eea
\end{subequations}
\begin{widetext} 
Using \Eq \refeqn{eqn:Zassenhaus} with $\xi = -i t /\hbar$ and the commutators in \Eqs \refeqn{eqn:commutators}, \Eq \refeqn{eqn:aexp(z+p2)} becomes
\be
  \label{eqn:aexp-expanded}
  a_{m_F}(p,t) = e^{i \hat{\mathcal{M}} t z/\hbar} e^{-i p^2 t/2M\hbar} e^{-i\hat{\mathcal{M}} p t^2/2M\hbar} e^{-i\hat{\mathcal{M}}^2 t^3/6M\hbar} a_{m_F}(p,0).
\ee
Since $e^{\xi (\hat{\mathcal{M}})^n} \ket{F\,m_F} = e^{\xi (m_F \mathcal{F})^n} \ket{F\,m_F}$, it follows that the amplitude of the state $\ket{F\;m_F}$ before the onset of the first sw pulse is
\begin{subequations}
\bea
  a_{m_F}(p,t)
  & = & \alpha_{m_F} e^{i (m_F \mathcal{F}) t \, z/\hbar} e^{-i p^2 t/2M\hbar} e^{-i (m_F \mathcal{F}) p \, t^2/2M\hbar} e^{-i (m_F \mathcal{F})^2 t^3/6M\hbar} \delta(p - p_0), \\
  a_{m_F}(z,t)
  & = & \frac{\alpha_{m_F}}{\sqrt{2\pi\hbar}}
  e^{i (p_0 + m_F \mathcal{F} t) z/\hbar} e^{-i \epsilon_0 t/\hbar} e^{-i (m_F \mathcal{F}) p_0 t^2/2M\hbar} e^{-i (m_F \mathcal{F})^2 t^3/6M\hbar},
\eea
\end{subequations}
where $\epsilon_0 = p_0^2/2M$ is the initial kinetic energy of the atom.

The first sw pulse, applied at $t = T_1$, diffracts the atom into a superposition of momentum states. The wave function is computed in position space using the Raman-Nath approximation and integrating the Schr\"{o}dinger equation to obtain
\begin{subequations}
\bea
  a_{m_F}^{(1)}(z,T_1)
  & = & a_{m_F}(z,T_1) \sum_n (-i)^n J_n(\Theta^{(1)}_{m_F}) e^{inqz}, \\
  \label{eqn:a_mF(1)(p,T_1)}
  a_{m_F}^{(1)}(p,T_1)
  & = & \alpha_{m_F} e^{-i\epsilon_0 T_1/\hbar} e^{-i (m_F \mathcal{F}) p_0 T_1^2 / 2 M \hbar} e^{-i(m_F \mathcal{F})^2 T_1^3/6 M \hbar} \\
  & \times & \sum_n (-i)^n J_n(\Theta_{m_F}^{(1)}) \delta(p - p_0 - m_F \mathcal{F} T_1 - n\hbar q). \notag
\eea
\end{subequations}
Here, $\Theta_{m_F}^{(1)} \equiv u_{m_F}^{(1)} e^{i\theta}$ is the (complex) area of pulse 1, $u_{m_F}^{(1)} = \chi_{m_F} \tau_1$, $\tau_1$ is the duration of the pulse, and $a_{m_F}^{(1)}(p,T_1)$ is the wave function in momentum space. The superscript $(1)$ on $a_{m_F}^{(1)}$ denotes the number of sw pulses that have been applied to the atom so far. We use the prescription of \Eq \refeqn{eqn:aexp-expanded} to evolve the amplitude in momentum space [\Eq \refeqn{eqn:a_mF(1)(p,T_1)}] until the onset of the second pulse
\begin{align}
\begin{split}
  a_{m_F}^{(1)}(p,t)
  & = \alpha_{m_F}  e^{i (m_F \mathcal{F}) (t - T_1) z/\hbar} e^{-i [p_0^2 T_1 + p^2 (t - T_1)]/2M\hbar}
  e^{-i (m_F \mathcal{F}) [p_0 T_1^2 + p (t - T_1)^2]/2M\hbar} \\
  & \times e^{-i(m_F \mathcal{F})^2 [T_1^3 + (t - T_1)^3]/6M\hbar}
  \sum_n (-i)^n J_n(\Theta_{m_F}^{(1)}) \delta(p - p_0 - m_F \mathcal{F} T_1 - n\hbar q).
\end{split}
\end{align}
To apply the next sw pulse to the wave function, it is convenient to transform back to position space:
\begin{align}
\begin{split}
  a_{m_F}^{(1)}(z,t)
  & = \frac{\alpha_{m_F}}{\sqrt{2\pi\hbar}} e^{i(p_0 + m_F \mathcal{F} t) z /\hbar} e^{-i[p_0^2 T_1 + (p_0 - m_F \mathcal{F} T_1)^2(t - T_1)]/2M\hbar} \\
  & \times e^{-i (m_F \mathcal{F}) [p_0 T_1^2 + (p_0 + m_F \mathcal{F} T_1) (t - T_1)^2]/2M\hbar} e^{-i(m_F \mathcal{F})^2 [T_1^3 + (t - T_1)^3]/6M\hbar} \\
  & \times \sum_n (-i)^n J_n(\Theta_{m_F}^{(1)}) e^{i n q z} e^{-i n q v_0 (t - T_1)} e^{-i n^2 \omega_q (t - T_1)}
  e^{-i n q (m_F \mathcal{F}) [(t - T_1)^2 + 2 T_1 (t - T_1)]/2M}.
\end{split}
\end{align}
Here, $v_0 = p_0/M$ is the initial velocity of the atom and $\omega_q = \hbar q^2/2M$ is the two-photon recoil frequency. Applying the second pulse at $t = T_2$, the wave function becomes
\begin{align}
\begin{split}
  a_{m_F}^{(2)}(z,T_2)
  & = \frac{\alpha_{m_F}}{\sqrt{2\pi\hbar}} e^{i(p_0 + m_F \mathcal{F} T_2) z /\hbar} e^{-i \epsilon_0^2 T_2/\hbar}
  e^{-i p_0 (m_F \mathcal{F}) T_2^2/2M\hbar} e^{-i (m_F \mathcal{F})^2 T_2^3/6M\hbar} \\
  & \times \sum_{n,m} (-i)^{(n+m)} J_n(\Theta_{m_F}^{(1)}) J_m(\Theta_{m_F}^{(2)}) e^{i(n+m)qz} e^{-i n q v_0 (T_2 - T_1)} e^{-i n^2 \omega_q (T_2 - T_1)} e^{-i n q (m_F \mathcal{F}) (T_2^2 - T_1^2)/2M}.
\end{split}
\end{align}
To evolve the wave function in the presence of the external force until time $t$, once again we transform into $p$-space and use \Eq \refeqn{eqn:aexp-expanded} to obtain
\begin{align}
\begin{split}
  a_{m_F}^{(2)}(p,t)
  & = \alpha_{m_F} e^{i(m_F \mathcal{F}) (t - T_2) z /\hbar} e^{-i [p_0^2 T_2 + p^2 (t - T_2)]/2M\hbar}
  e^{-i (m_F \mathcal{F}) [p_0 T_2^2 + p (t - T_2)^2]/2M\hbar} e^{-i (m_F \mathcal{F})^2 [T_2^3 + (t - T_2)^3]/6M\hbar} \\
  & \times \sum_{n,m} (-i)^{(n+m)} J_n(\Theta_{m_F}^{(1)}) J_m(\Theta_{m_F}^{(2)}) e^{-i n q v_0 (T_2 - T_1)} e^{-i n^2 \omega_q (T_2 - T_1)} e^{-i n q (m_F \mathcal{F}) (T_2^2 - T_1^2)/2M} \\
  & \times \delta[p - p_0 - m_F \mathcal{F} T_2 - (n + m) \hbar q].
\end{split}
\end{align}
Finally, the amplitude in position space after the second pulse can be shown to be
\begin{align}
\begin{split}
  a_{m_F}^{(2)}(z,t)
  & = \frac{\alpha_{m_F}}{\sqrt{2\pi\hbar}} e^{i(p_0 + m_F \mathcal{F} t) z /\hbar} e^{-i \epsilon_0 t/\hbar}
  e^{-i (m_F \mathcal{F}) p_0 t^2/2M\hbar} e^{-i (m_F \mathcal{F})^2 t^3/6M\hbar} \\
  & \times \sum_{n,m} (-i)^{(n+m)} J_n(\Theta_{m_F}^{(1)}) J_m(\Theta_{m_F}^{(2)}) e^{i(n+m)qz} e^{-i q v_0 [n (T_2 - T_1) + (n+m) (t - T_2)]} \\
  & \times e^{-i \omega_q [n^2 (T_2 - T_1) + (n+m)^2 (t - T_2)]} e^{-i q (m_F \mathcal{F})[ n (T_2^2 - T_1^2) + (n+m) (t^2 - T_2^2)]/2M}.
\end{split}
\end{align}
To compute the field scattered from the atomic interference as a function of $t$, we use the $q$-Fourier component of the ground state density, $\rho_{m_Fm_F}^{(2)}(z,t) = |a_{m_F}^{(2)}(z,t)|^2$, which can be shown to be
\begin{align}
\begin{split}
  \label{eqn:rho(2)}
  \rho_{m_F m_F}^{(2)}(z,t)
  & =
  \frac{|\alpha_{m_F}|^2}{2\pi\hbar} \sum_{n,m,n',m'} (-i)^{n+m-n'-m'} J_n(\Theta_{m_F}^{(1)}) J_m(\Theta_{m_F}^{(2)}) J_{n'}(\Theta_{m_F}^{(1)\,*}) J_{m'}(\Theta_{m_F}^{(2)\,*}) e^{i(n+m-n'-m')qz} \\
  & \times
  e^{-i q v_0 [(n-n') (T_2 - T_1) + (n+m-n'-m') (t - T_2)]} e^{-i \omega_q \{(n^2-n'^2) (T_2 - T_1) + [(n+m)^2 - (n'+m')^2](t - T_2)\}} \\
  & \times
  e^{-i q (m_F \mathcal{F})[(n-n') (T_2^2 - T_1^2) + (n+m-n'-m') (t^2 - T_2^2)]/2M}.
\end{split}
\end{align}
Since the density distribution contains frequency components that depend only on the \emph{difference} between interfering momentum states, we recast the sums over $n'$ and $m'$ in terms of $\nu\bar{N} = n-n'$ and $\nu = n'+m'-n-m$ (the integer difference between momentum states after the first and second pulses, respectively):
\begin{align}
\begin{split}
  \label{eqn:rho(2)-reduced}
  \rho_{m_F m_F}^{(2)}(z,t)
  & =
  -\frac{|\alpha_{m_F}|^2}{2\pi\hbar} \sum_{\nu,\bar{N},n,m} i^\nu J_n(\Theta_{m_F}^{(1)}) J_{n-\nu\bar{N}}(\Theta_{m_F}^{(1)\,*}) J_m(\Theta_{m_F}^{(2)}) J_{m+\nu(\bar{N}+1)}(\Theta_{m_F}^{(2)\,*}) e^{-i\nu qz} \\
  & \times
  e^{i \nu q v_0 [(t - T_2) - \bar{N} (T_2 - T_1)]}
  e^{i \nu \omega_q \{[2(n+m)+\nu](t - T_2) - \bar{N}(2n-\nu\bar{N}) (T_2 - T_1)\}} \\
  & \times
  e^{i \nu q (m_F \mathcal{F}) [(t^2-T_2^2) - \bar{N}(T_2^2 - T_1^2)]/2M}.
\end{split}
\end{align}
The scattered field is proportional to the $q$-Fourier harmonic of $\rho_{m_F m_F}^{(2)}(z,t)$ [the coefficient of the $e^{-i \nu q z}$ term in \Eq \refeqn{eqn:rho(2)-reduced}, with $\nu = 1$]. Summing over all magnetic sub-levels in the ground state, one can show that
\be
  E^{(2)}_{\mathcal{F}}(t;\bm{T}) = \sum_{m_F} E^{(2)}_{m_F}(t;\bm{T}) e^{i m_F \phi^{(2)}_{\mathcal{F}}(t;\bm{T})},
\ee
where
\begin{align}
\begin{split}
  \label{eqn:E(2)_mF}
  E^{(2)}_{m_F}(t;\bm{T})
  & \propto |\alpha_{m_F}|^2 \left( C^{F\;\;\;\;1\;\;\;F+1}_{m_F\;q_L\;m_F + q_L} \right)^2 \sum_{\bar{N}} (-1)^{\bar{N}+1} e^{-\left[ \left(t - t_{\rm{echo}}^{(2)}\right)/\tau_{\rm{coh}} \right]^2} e^{i q v_0 \left( t - t^{(2)}_{\rm{echo}} \right)} \\
  & \times J_{\bar{N}} \left( 2 u_{m_F}^{(1)} \sqrt{\sin(\varphi_1 - \theta) \sin(\varphi_1 + \theta)} \right)
  J_{\bar{N}+1} \left( 2 u_{m_F}^{(2)} \sqrt{\sin(\varphi_2 - \theta) \sin(\varphi_2 + \theta)} \right) \\
  & \times \left( \frac{\sin(\varphi_1 + \theta)}{\sin(\varphi_1 - \theta)} \right)^{\bar{N}/2} \left( \frac{\sin(\varphi_2 - \theta)}{\sin(\varphi_2 + \theta)} \right)^{(\bar{N}+1)/2}
\end{split}
\end{align}
is the field scattered from each magnetic sub-level, with recoil phases
\begin{subequations}
\bea
  \varphi_1(t;\bm{T}) & = & \omega_q \left(t - t_{\rm{echo}}^{(2)} \right), \\
  \varphi_2(t;\bm{T}) & = & \omega_q (t - T_2),
\eea
\end{subequations}
and $m_F \phi^{(2)}_{\mathcal{F}}$ is the phase shift of the density grating produced in the ground state $\ket{F\,m_F}$ due to the presence of the external force, $\mathcal{F}$, with $\phi^{(2)}_{\mathcal{F}}$ given by
\be
  \label{eqn:phi(2)_F}
  \phi^{(2)}_{\mathcal{F}}(t;\bm{T}) = \frac{q\mathcal{F}}{2M} \big[ (t^2 - T_2^2) - \bar{N} (T_2^2 - T_1^2) \big].
\ee
In deriving \Eq \refeqn{eqn:E(2)_mF} we have made use of the Bessel function summation theorem \cite{Gradshteyn-Book-2007, Beattie-PRA-2008, Barrett-PRA-2010}
\be
  \sum_n J_n \left( u e^{i\theta} \right) J_{n+\eta} \left( u e^{-i\theta} \right) e^{i(2n+\eta)\phi} =
  i^{\eta} J_{\eta} \left( 2 u \sqrt{\sin(\phi-\theta) \sin(\phi+\theta)} \right) \left( \frac{\sin(\phi - \theta)}{\sin(\phi + \theta)} \right)^{\eta/2},
\ee
and we averaged over the velocity distribution of the sample assuming a Maxwellian distribution centered at $v_0$ with $e^{-1}$ width $\sigma_v = \sqrt{2 k_B \mathcal{T}/M}$. In this way, we account for the possibility of an initial launch of the atomic cloud and for the dephasing of the echo due to the distribution of Doppler phases in the sample. An additional factor of $\left( C^{F\;\;\;\;1\;\;\;F+1}_{m_F\;q_L\;m_F + q_L} \right)^2$ was added to the scattered field to account for the atom-field coupling by the read-out pulse. The scattered field lasts for a time $\tau_{\rm{coh}} = 2 /q \sigma_v$---called the coherence time---about each echo, which occur at times $t_{\rm{echo}}^{(2)} = \bar{N} (T_2 - T_1) + T_2$. The phase $\theta$ in \Eq \refeqn{eqn:E(2)_mF}, associated with spontaneous emission during the excitation pulses, affects only the recoil-dependent component of the signal \cite{Barrett-PRA-2010}.

These results can be generalized for the case of an $N$-pulse interferometer with a set of onset times $\bm{T} = \{ T_1, T_2, \ldots, T_N \}$ for which $T_{j+1} > T_j$. After $N$ sw pulses, each with pulse area $u_{m_F}^{(j)}$, the total scattered field at time $t$ is
\be
  \label{eqn:E(N)}
  E^{(N)}_{\mathcal{F}}(t;\bm{T}) = \sum_{m_F} E^{(N)}_{m_F}(t;\bm{T}) e^{i m_F \phi^{(N)}_{\mathcal{F}}(t;\bm{T})}
\ee
where
\begin{align}
\begin{split}
  \label{eqn:E(N)_mF}
  E^{(N)}_{m_F}(t;\bm{T})
  & \propto -|\alpha_{m_F}|^2 \left( C^{F\;\;\;\;1\;\;\;F+1}_{m_F\;q_L\;m_F + q_L} \right)^2 \sum_{l_1, l_2, \ldots, l_{N-1}} e^{-\left[ \left(t - t_{\rm{echo}}^{(N)}\right)/\tau_{\rm{coh}} \right]^2} e^{i q v_0 \left( t - t^{(N)}_{\rm{echo}} \right)} \\
  & \times \prod_{j=1}^{N} J_{(l_j - l_{j-1})} \left( 2 u_{m_F}^{(j)} \sqrt{\sin(\varphi_j - \theta) \sin(\varphi_j + \theta)} \right) \left( \frac{\sin(\varphi_j - \theta)}{\sin(\varphi_j + \theta)} \right)^{(l_j - l_{j-1})/2}.
\end{split}
\end{align}
Here, $\bm{l} = \{l_1, l_2, \ldots, l_N\}$ denotes the set of momentum states that interfere after the pulse sequence, where $l_j$ is the difference between interfering momentum states (in units of $\hbar q$) after pulse $j$. The echo times and the recoil phases are given by
\begin{subequations}
\bea
  \label{eqn:techo(N)}
  t_{\rm{echo}}^{(N)}(\bm{T}) & = & T_N - \frac{1}{l_N} \sum_{j=1}^{N-1} l_j (T_{j+1} - T_j), \\
  \label{eqn:varphi(N)}
  \varphi_j(t;\bm{T}) & = & \omega_q \sum_{k=j}^N l_k (T_{k+1} - T_k),
\eea
\end{subequations}
and the contribution to the phase of the grating due to the force, $\mathcal{F}$, is
\be
  \label{eqn:phi(N)}
  \phi_{\mathcal{F}}^{(N)}(t;\bm{T}) = \frac{q\mathcal{F}}{2 M} \sum_{j=1}^N l_j (T_{j+1}^2 - T_j^2).
\ee
In \Eqs \refeqn{eqn:E(N)_mF}--\refeqn{eqn:phi(N)} $l_N = 1$, which corresponds to the scattered field from the $q$-Fourier harmonic of the density formed after the sw pulses, and it is understood that $l_0 = 0$ and $T_{N+1} = t$.

We now use the formalism for the $N$-pulse echo signal [\Eq \refeqn{eqn:E(N)_mF}] to obtain an expression for the three-pulse interferometer signal discussed in \Sec \ref{sec:Theory}. We begin by setting $N = 3$ and $\bm{T} = \{ T_1, T_1 + T_{21}, T_1 + T_{21} + T_{32} \}$. For an echo to occur at $t_{\rm{echo}}^{(3)} = T_1 + T_{32} + (\bar{N} + 1) T_{21}$ for any $T_1$, $T_{32}$ and $T_{21}$, \Eq \refeqn{eqn:techo(N)} dictates the set of $l_j$ to be $\bm{l} = \{-\bar{N}, 0, 1\}$. Then, it can be shown that the scattered field is given by
\begin{align}
\begin{split}
  \label{eqn:E(3)_mF}
  E^{(3)}_{m_F}(t; \bm{T})
  & \propto |\alpha_{m_F}|^2 \left( C^{F\;\;\;\;1\;\;\;F+1}_{m_F\;q_L\;m_F + q_L} \right)^2 \sum_{\bar{N}} (-1)^{\bar{N}+1} e^{-\left[ \left(t - t_{\rm{echo}}^{(3)}\right)/\tau_{\rm{coh}} \right]^2} e^{i q v_0 \left( t - t^{(3)}_{\rm{echo}} \right)} \\
  & \times
  J_{\bar{N}} \left( 2 u_{m_F}^{(1)} \sqrt{\sin(\varphi_1 - \theta) \sin(\varphi_1 + \theta)} \right)
  J_{\bar{N}} \left( 2 u_{m_F}^{(2)} \sqrt{\sin(\varphi_2 - \theta) \sin(\varphi_2 + \theta)} \right) \\
  & \times J_1 \left( 2 u_{m_F}^{(3)} \sqrt{\sin(\varphi_3 - \theta) \sin(\varphi_3 + \theta)} \right)
  \left( \frac{\sin(\varphi_1 + \theta)}{\sin(\varphi_1 - \theta)} \right)^{\bar{N}/2}
  \left( \frac{\sin(\varphi_2 - \theta)}{\sin(\varphi_2 + \theta)} \right)^{\bar{N}/2}
  \left( \frac{\sin(\varphi_3 - \theta)}{\sin(\varphi_3 + \theta)} \right)^{1/2},
\end{split}
\end{align}
where the recoil phases in this case are
\begin{subequations}
\bea
  \varphi_1 & = & \omega_q \left( t - t_{\rm{echo}}^{(3)} \right), \\
  \varphi_2 & = & \varphi_3 = \omega_q \left( t - t_{\rm{echo}}^{(3)} + \bar{N} T_{21} \right),
\eea
\end{subequations}
and the grating phase due to $\mathcal{F}$ is
\bea
  \phi_{\mathcal{F}}^{(3)}(t;\bm{T})
  & = & \frac{q \mathcal{F}}{2M} \left[ -\bar{N} (T_2^2 - T_1^2) + (t^2 - T_3^2) \right] \notag \\
  & = & \frac{q \mathcal{F}}{2M} \left\{ \bar{N} (\bar{N} + 1) T_{21}^2 + 2\bar{N} T_{32} T_{21} + 2 \big[ T_1 + T_{32} + (\bar{N}+1) T_{21} \big] \Delta t + \Delta t^2 \right\}.
\eea
\end{widetext} 

\bibliographystyle{apsrev4-1}
\bibliography{MasterBib}

\begin{thebibliography}{44}%
\makeatletter
\providecommand \@ifxundefined [1]{%
 \@ifx{#1\undefined}
}%
\providecommand \@ifnum [1]{%
 \ifnum #1\expandafter \@firstoftwo
 \else \expandafter \@secondoftwo
 \fi
}%
\providecommand \@ifx [1]{%
 \ifx #1\expandafter \@firstoftwo
 \else \expandafter \@secondoftwo
 \fi
}%
\providecommand \natexlab [1]{#1}%
\providecommand \enquote  [1]{``#1''}%
\providecommand \bibnamefont  [1]{#1}%
\providecommand \bibfnamefont [1]{#1}%
\providecommand \citenamefont [1]{#1}%
\providecommand \href@noop [0]{\@secondoftwo}%
\providecommand \href [0]{\begingroup \@sanitize@url \@href}%
\providecommand \@href[1]{\@@startlink{#1}\@@href}%
\providecommand \@@href[1]{\endgroup#1\@@endlink}%
\providecommand \@sanitize@url [0]{\catcode `\\12\catcode `\$12\catcode
  `\&12\catcode `\#12\catcode `\^12\catcode `\_12\catcode `\%12\relax}%
\providecommand \@@startlink[1]{}%
\providecommand \@@endlink[0]{}%
\providecommand \url  [0]{\begingroup\@sanitize@url \@url }%
\providecommand \@url [1]{\endgroup\@href {#1}{\urlprefix }}%
\providecommand \urlprefix  [0]{URL }%
\providecommand \Eprint [0]{\href }%
\@ifxundefined \urlstyle {%
  \providecommand \doi  [0]{\begingroup \@sanitize@url \@doi}%
  \providecommand \@doi [1]{\endgroup \@@startlink {\doibase
  #1}doi:\discretionary {}{}{}#1\@@endlink }%
}{%
  \providecommand \doi  [0]{doi:\discretionary{}{}{}\begingroup
  \urlstyle{rm}\Url }%
}%
\providecommand \doibase [0]{http://dx.doi.org/}%
\providecommand \Doi [0]{\begingroup \@sanitize@url \@Doi }%
\providecommand \@Doi  [1]{\endgroup\@@startlink{\doibase#1}\@@Doi}%
\providecommand \@@Doi [1]{#1\@@endlink}%
\providecommand \selectlanguage [0]{\@gobble}%
\providecommand \bibinfo  [0]{\@secondoftwo}%
\providecommand \bibfield  [0]{\@secondoftwo}%
\providecommand \translation [1]{[#1]}%
\providecommand \BibitemOpen [0]{}%
\providecommand \bibitemStop [0]{}%
\providecommand \bibitemNoStop [0]{.\EOS\space}%
\providecommand \EOS [0]{\spacefactor3000\relax}%
\providecommand \BibitemShut  [1]{\csname bibitem#1\endcsname}%
\bibitem [{\citenamefont {Kasevich}\ and\ \citenamefont
  {Chu}(1991)}]{Kasevich-PRL-1991}%
  \BibitemOpen
  \bibfield  {author} {\bibinfo {author} {\bibfnamefont {M.}~\bibnamefont
  {Kasevich}}\ and\ \bibinfo {author} {\bibfnamefont {S.}~\bibnamefont {Chu}},\
  }\href@noop {} {\bibfield  {journal} {\bibinfo  {journal} {Phys. Rev.
  Lett.},\ }\textbf {\bibinfo {volume} {67}},\ \bibinfo {pages} {181} (\bibinfo
  {year} {1991})}\BibitemShut {NoStop}%
\bibitem [{\citenamefont {Peters}\ \emph {et~al.}(1999)\citenamefont {Peters},
  \citenamefont {Chung},\ and\ \citenamefont {Chu}}]{Peters-Nature-1999}%
  \BibitemOpen
  \bibfield  {author} {\bibinfo {author} {\bibfnamefont {A.}~\bibnamefont
  {Peters}}, \bibinfo {author} {\bibfnamefont {K.~Y.}\ \bibnamefont {Chung}}, \
  and\ \bibinfo {author} {\bibfnamefont {S.}~\bibnamefont {Chu}},\ }\href@noop
  {} {\bibfield  {journal} {\bibinfo  {journal} {Nature},\ }\textbf {\bibinfo
  {volume} {400}},\ \bibinfo {pages} {849} (\bibinfo {year}
  {1999})}\BibitemShut {NoStop}%
\bibitem [{\citenamefont {Peters}\ \emph {et~al.}(2001)\citenamefont {Peters},
  \citenamefont {Chung},\ and\ \citenamefont {Chu}}]{Peters-Metrologia-2001}%
  \BibitemOpen
  \bibfield  {author} {\bibinfo {author} {\bibfnamefont {A.}~\bibnamefont
  {Peters}}, \bibinfo {author} {\bibfnamefont {K.~Y.}\ \bibnamefont {Chung}}, \
  and\ \bibinfo {author} {\bibfnamefont {S.}~\bibnamefont {Chu}},\ }\href@noop
  {} {\bibfield  {journal} {\bibinfo  {journal} {Metrologia},\ }\textbf
  {\bibinfo {volume} {38}},\ \bibinfo {pages} {25} (\bibinfo {year}
  {2001})}\BibitemShut {NoStop}%
\bibitem [{\citenamefont {Hughes}\ \emph {et~al.}(2009)\citenamefont {Hughes},
  \citenamefont {Burke},\ and\ \citenamefont {Sackett}}]{Hughes-PRL-2009}%
  \BibitemOpen
  \bibfield  {author} {\bibinfo {author} {\bibfnamefont {K.~J.}\ \bibnamefont
  {Hughes}}, \bibinfo {author} {\bibfnamefont {J.~H.~T.}\ \bibnamefont
  {Burke}}, \ and\ \bibinfo {author} {\bibfnamefont {C.~A.}\ \bibnamefont
  {Sackett}},\ }\href@noop {} {\bibfield  {journal} {\bibinfo  {journal} {Phys.
  Rev. Lett.},\ }\textbf {\bibinfo {volume} {102}},\ \bibinfo {pages} {150403}
  (\bibinfo {year} {2009})}\BibitemShut {NoStop}%
\bibitem [{\citenamefont {Poli}\ \emph {et~al.}(2011)\citenamefont {Poli},
  \citenamefont {Wang}, \citenamefont {Tarallo}, \citenamefont {Alberti},
  \citenamefont {Prevedelli},\ and\ \citenamefont {Tino}}]{Poli-PRL-2011}%
  \BibitemOpen
  \bibfield  {author} {\bibinfo {author} {\bibfnamefont {N.}~\bibnamefont
  {Poli}}, \bibinfo {author} {\bibfnamefont {F.-Y.}\ \bibnamefont {Wang}},
  \bibinfo {author} {\bibfnamefont {M.~G.}\ \bibnamefont {Tarallo}}, \bibinfo
  {author} {\bibfnamefont {A.}~\bibnamefont {Alberti}}, \bibinfo {author}
  {\bibfnamefont {M.}~\bibnamefont {Prevedelli}}, \ and\ \bibinfo {author}
  {\bibfnamefont {G.~M.}\ \bibnamefont {Tino}},\ }\href@noop {} {\bibfield
  {journal} {\bibinfo  {journal} {Phys. Rev. Lett.},\ }\textbf {\bibinfo
  {volume} {106}},\ \bibinfo {pages} {038501} (\bibinfo {year}
  {2011})}\BibitemShut {NoStop}%
\bibitem [{\citenamefont {Snadden}\ \emph {et~al.}(1998)\citenamefont
  {Snadden}, \citenamefont {McGuirk}, \citenamefont {Bouyer}, \citenamefont
  {Haritos},\ and\ \citenamefont {Kasevich}}]{Snadden-PRL-1998}%
  \BibitemOpen
  \bibfield  {author} {\bibinfo {author} {\bibfnamefont {M.~J.}\ \bibnamefont
  {Snadden}}, \bibinfo {author} {\bibfnamefont {J.~M.}\ \bibnamefont
  {McGuirk}}, \bibinfo {author} {\bibfnamefont {P.}~\bibnamefont {Bouyer}},
  \bibinfo {author} {\bibfnamefont {K.~G.}\ \bibnamefont {Haritos}}, \ and\
  \bibinfo {author} {\bibfnamefont {M.~A.}\ \bibnamefont {Kasevich}},\
  }\href@noop {} {\bibfield  {journal} {\bibinfo  {journal} {Phys. Rev.
  Lett.},\ }\textbf {\bibinfo {volume} {81}},\ \bibinfo {pages} {971} (\bibinfo
  {year} {1998})}\BibitemShut {NoStop}%
\bibitem [{\citenamefont {McGuirk}\ \emph {et~al.}(2002)\citenamefont
  {McGuirk}, \citenamefont {Foster}, \citenamefont {Fixler}, \citenamefont
  {Snadden},\ and\ \citenamefont {Kasevich}}]{McGuirk-PRA-2002}%
  \BibitemOpen
  \bibfield  {author} {\bibinfo {author} {\bibfnamefont {J.~M.}\ \bibnamefont
  {McGuirk}}, \bibinfo {author} {\bibfnamefont {G.~T.}\ \bibnamefont {Foster}},
  \bibinfo {author} {\bibfnamefont {J.~B.}\ \bibnamefont {Fixler}}, \bibinfo
  {author} {\bibfnamefont {M.~J.}\ \bibnamefont {Snadden}}, \ and\ \bibinfo
  {author} {\bibfnamefont {M.~A.}\ \bibnamefont {Kasevich}},\ }\href@noop {}
  {\bibfield  {journal} {\bibinfo  {journal} {Phys. Rev. A},\ }\textbf
  {\bibinfo {volume} {65}},\ \bibinfo {pages} {033608} (\bibinfo {year}
  {2002})}\BibitemShut {NoStop}%
\bibitem [{\citenamefont {Yu}\ \emph {et~al.}(2006)\citenamefont {Yu},
  \citenamefont {Kohel}, \citenamefont {Kellogg},\ and\ \citenamefont
  {Maleki}}]{Yu-ApplPhysB-2006}%
  \BibitemOpen
  \bibfield  {author} {\bibinfo {author} {\bibfnamefont {N.}~\bibnamefont
  {Yu}}, \bibinfo {author} {\bibfnamefont {J.~M.}\ \bibnamefont {Kohel}},
  \bibinfo {author} {\bibfnamefont {J.~R.}\ \bibnamefont {Kellogg}}, \ and\
  \bibinfo {author} {\bibfnamefont {L.}~\bibnamefont {Maleki}},\ }\href@noop {}
  {\bibfield  {journal} {\bibinfo  {journal} {Appl. Phys. B},\ }\textbf
  {\bibinfo {volume} {84}},\ \bibinfo {pages} {647} (\bibinfo {year}
  {2006})}\BibitemShut {NoStop}%
\bibitem [{\citenamefont {Gustavson}\ \emph {et~al.}(1997)\citenamefont
  {Gustavson}, \citenamefont {Bouyer},\ and\ \citenamefont
  {Kasevich}}]{Gustavson-PRL-1997}%
  \BibitemOpen
  \bibfield  {author} {\bibinfo {author} {\bibfnamefont {T.~L.}\ \bibnamefont
  {Gustavson}}, \bibinfo {author} {\bibfnamefont {P.}~\bibnamefont {Bouyer}}, \
  and\ \bibinfo {author} {\bibfnamefont {M.~A.}\ \bibnamefont {Kasevich}},\
  }\href@noop {} {\bibfield  {journal} {\bibinfo  {journal} {Phys. Rev.
  Lett.},\ }\textbf {\bibinfo {volume} {78}},\ \bibinfo {pages} {2046}
  (\bibinfo {year} {1997})}\BibitemShut {NoStop}%
\bibitem [{\citenamefont {Wu}\ \emph {et~al.}(2007)\citenamefont {Wu},
  \citenamefont {Su},\ and\ \citenamefont {Prentiss}}]{Wu-PRL-2007}%
  \BibitemOpen
  \bibfield  {author} {\bibinfo {author} {\bibfnamefont {S.}~\bibnamefont
  {Wu}}, \bibinfo {author} {\bibfnamefont {E.}~\bibnamefont {Su}}, \ and\
  \bibinfo {author} {\bibfnamefont {M.}~\bibnamefont {Prentiss}},\ }\href@noop
  {} {\bibfield  {journal} {\bibinfo  {journal} {Phys. Rev. Lett.},\ }\textbf
  {\bibinfo {volume} {99}},\ \bibinfo {pages} {173201} (\bibinfo {year}
  {2007})}\BibitemShut {NoStop}%
\bibitem [{\citenamefont {Burke}\ and\ \citenamefont
  {Sackett}(2009)}]{Burke-PRA-2009}%
  \BibitemOpen
  \bibfield  {author} {\bibinfo {author} {\bibfnamefont {J.~H.~T.}\
  \bibnamefont {Burke}}\ and\ \bibinfo {author} {\bibfnamefont {C.~A.}\
  \bibnamefont {Sackett}},\ }\href@noop {} {\bibfield  {journal} {\bibinfo
  {journal} {Phys. Rev. A},\ }\textbf {\bibinfo {volume} {80}},\ \bibinfo
  {pages} {061603} (\bibinfo {year} {2009})}\BibitemShut {NoStop}%
\bibitem [{\citenamefont {Cahn}\ \emph {et~al.}(1997)\citenamefont {Cahn},
  \citenamefont {Kumarakrishnan}, \citenamefont {Shim}, \citenamefont
  {Sleator}, \citenamefont {Berman},\ and\ \citenamefont
  {Dubetsky}}]{Cahn-PRL-1997}%
  \BibitemOpen
  \bibfield  {author} {\bibinfo {author} {\bibfnamefont {S.~B.}\ \bibnamefont
  {Cahn}}, \bibinfo {author} {\bibfnamefont {A.}~\bibnamefont
  {Kumarakrishnan}}, \bibinfo {author} {\bibfnamefont {U.}~\bibnamefont
  {Shim}}, \bibinfo {author} {\bibfnamefont {T.}~\bibnamefont {Sleator}},
  \bibinfo {author} {\bibfnamefont {P.~R.}\ \bibnamefont {Berman}}, \ and\
  \bibinfo {author} {\bibfnamefont {B.}~\bibnamefont {Dubetsky}},\ }\href@noop
  {} {\bibfield  {journal} {\bibinfo  {journal} {Phys. Rev. Lett.},\ }\textbf
  {\bibinfo {volume} {79}},\ \bibinfo {pages} {784} (\bibinfo {year}
  {1997})}\BibitemShut {NoStop}%
\bibitem [{\citenamefont {Strekalov}\ \emph {et~al.}(2002)\citenamefont
  {Strekalov}, \citenamefont {Turlapov}, \citenamefont {Kumarakrishnan},\ and\
  \citenamefont {Sleator}}]{Strekalov-PRA-2002}%
  \BibitemOpen
  \bibfield  {author} {\bibinfo {author} {\bibfnamefont {D.~V.}\ \bibnamefont
  {Strekalov}}, \bibinfo {author} {\bibfnamefont {A.}~\bibnamefont {Turlapov}},
  \bibinfo {author} {\bibfnamefont {A.}~\bibnamefont {Kumarakrishnan}}, \ and\
  \bibinfo {author} {\bibfnamefont {T.}~\bibnamefont {Sleator}},\ }\href@noop
  {} {\bibfield  {journal} {\bibinfo  {journal} {Phys. Rev. A},\ }\textbf
  {\bibinfo {volume} {66}},\ \bibinfo {pages} {023601} (\bibinfo {year}
  {2002})}\BibitemShut {NoStop}%
\bibitem [{Note1()}]{Note1}%
  \BibitemOpen
  \bibinfo {note} {This condition is true for far off-resonant excitation
  fields only. For fields closer to resonance, both the AC Stark effect and the
  Zeeman effect can induce a relative shift between the ground and excited
  states, thus affecting the response of the interferometer in a systematic
  way.}\BibitemShut {Stop}%
\bibitem [{\citenamefont {Su}\ \emph {et~al.}(2010)\citenamefont {Su},
  \citenamefont {Wu},\ and\ \citenamefont {Prentiss}}]{Su-PRA-2010}%
  \BibitemOpen
  \bibfield  {author} {\bibinfo {author} {\bibfnamefont {E.~J.}\ \bibnamefont
  {Su}}, \bibinfo {author} {\bibfnamefont {S.}~\bibnamefont {Wu}}, \ and\
  \bibinfo {author} {\bibfnamefont {M.~G.}\ \bibnamefont {Prentiss}},\
  }\href@noop {} {\bibfield  {journal} {\bibinfo  {journal} {Phys. Rev. A},\
  }\textbf {\bibinfo {volume} {81}},\ \bibinfo {pages} {043631} (\bibinfo
  {year} {2010})}\BibitemShut {NoStop}%
\bibitem [{\citenamefont {Beattie}\ \emph {et~al.}(2008)\citenamefont
  {Beattie}, \citenamefont {Barrett}, \citenamefont {Weel}, \citenamefont
  {Chan}, \citenamefont {Mok}, \citenamefont {Cahn},\ and\ \citenamefont
  {Kumarakrishnan}}]{Beattie-PRA-2008}%
  \BibitemOpen
  \bibfield  {author} {\bibinfo {author} {\bibfnamefont {S.}~\bibnamefont
  {Beattie}}, \bibinfo {author} {\bibfnamefont {B.}~\bibnamefont {Barrett}},
  \bibinfo {author} {\bibfnamefont {M.}~\bibnamefont {Weel}}, \bibinfo {author}
  {\bibfnamefont {I.}~\bibnamefont {Chan}}, \bibinfo {author} {\bibfnamefont
  {C.}~\bibnamefont {Mok}}, \bibinfo {author} {\bibfnamefont {S.~B.}\
  \bibnamefont {Cahn}}, \ and\ \bibinfo {author} {\bibfnamefont
  {A.}~\bibnamefont {Kumarakrishnan}},\ }\href@noop {} {\bibfield  {journal}
  {\bibinfo  {journal} {Phys. Rev. A},\ }\textbf {\bibinfo {volume} {77}},\
  \bibinfo {pages} {013610} (\bibinfo {year} {2008})}\BibitemShut {NoStop}%
\bibitem [{\citenamefont {Barrett}\ \emph {et~al.}(2010)\citenamefont
  {Barrett}, \citenamefont {Yavin}, \citenamefont {Beattie},\ and\
  \citenamefont {Kumarakrishnan}}]{Barrett-PRA-2010}%
  \BibitemOpen
  \bibfield  {author} {\bibinfo {author} {\bibfnamefont {B.}~\bibnamefont
  {Barrett}}, \bibinfo {author} {\bibfnamefont {I.}~\bibnamefont {Yavin}},
  \bibinfo {author} {\bibfnamefont {S.}~\bibnamefont {Beattie}}, \ and\
  \bibinfo {author} {\bibfnamefont {A.}~\bibnamefont {Kumarakrishnan}},\
  }\href@noop {} {\bibfield  {journal} {\bibinfo  {journal} {Phys. Rev. A},\
  }\textbf {\bibinfo {volume} {82}},\ \bibinfo {pages} {023625} (\bibinfo
  {year} {2010})}\BibitemShut {NoStop}%
\bibitem [{\citenamefont {Mossberg}\ \emph {et~al.}(1979)\citenamefont
  {Mossberg}, \citenamefont {Kachru}, \citenamefont {Hartmann},\ and\
  \citenamefont {Flusberg}}]{Mossberg-PRA-1979}%
  \BibitemOpen
  \bibfield  {author} {\bibinfo {author} {\bibfnamefont {T.~W.}\ \bibnamefont
  {Mossberg}}, \bibinfo {author} {\bibfnamefont {R.}~\bibnamefont {Kachru}},
  \bibinfo {author} {\bibfnamefont {S.~R.}\ \bibnamefont {Hartmann}}, \ and\
  \bibinfo {author} {\bibfnamefont {A.~M.}\ \bibnamefont {Flusberg}},\
  }\href@noop {} {\bibfield  {journal} {\bibinfo  {journal} {Phys. Rev. A},\
  }\textbf {\bibinfo {volume} {20}},\ \bibinfo {pages} {1976} (\bibinfo {year}
  {1979})}\BibitemShut {NoStop}%
\bibitem [{\citenamefont {Bord\'{e}}\ \emph {et~al.}(1984)\citenamefont
  {Bord\'{e}}, \citenamefont {Salomon}, \citenamefont {Avrillier},
  \citenamefont {{Van Lerberghe}}, \citenamefont {Br\'{e}ant}, \citenamefont
  {Bassi},\ and\ \citenamefont {Scoles}}]{Borde-PRA-1984}%
  \BibitemOpen
  \bibfield  {author} {\bibinfo {author} {\bibfnamefont {C.~J.}\ \bibnamefont
  {Bord\'{e}}}, \bibinfo {author} {\bibfnamefont {C.}~\bibnamefont {Salomon}},
  \bibinfo {author} {\bibfnamefont {S.}~\bibnamefont {Avrillier}}, \bibinfo
  {author} {\bibfnamefont {A.}~\bibnamefont {{Van Lerberghe}}}, \bibinfo
  {author} {\bibfnamefont {C.}~\bibnamefont {Br\'{e}ant}}, \bibinfo {author}
  {\bibfnamefont {D.}~\bibnamefont {Bassi}}, \ and\ \bibinfo {author}
  {\bibfnamefont {G.}~\bibnamefont {Scoles}},\ }\href@noop {} {\bibfield
  {journal} {\bibinfo  {journal} {Phys. Rev. A},\ }\textbf {\bibinfo {volume}
  {30}},\ \bibinfo {pages} {1836} (\bibinfo {year} {1984})}\BibitemShut
  {NoStop}%
\bibitem [{\citenamefont {Allen}\ and\ \citenamefont
  {Eberly}(1987)}]{Allen-BOOK-1987}%
  \BibitemOpen
  \bibfield  {author} {\bibinfo {author} {\bibfnamefont {L.}~\bibnamefont
  {Allen}}\ and\ \bibinfo {author} {\bibfnamefont {J.~H.}\ \bibnamefont
  {Eberly}},\ }\href@noop {} {\emph {\bibinfo {title} {Optical Resonance and
  Two-Level Atoms}}}\ (\bibinfo  {publisher} {Dover},\ \bibinfo {address} {New
  York},\ \bibinfo {year} {1987})\BibitemShut {NoStop}%
\bibitem [{\citenamefont {Dubetsky}\ \emph {et~al.}(1992)\citenamefont
  {Dubetsky}, \citenamefont {Berman},\ and\ \citenamefont
  {Sleator}}]{Dubetsky-PRA(R)-1992}%
  \BibitemOpen
  \bibfield  {author} {\bibinfo {author} {\bibfnamefont {B.}~\bibnamefont
  {Dubetsky}}, \bibinfo {author} {\bibfnamefont {P.~R.}\ \bibnamefont
  {Berman}}, \ and\ \bibinfo {author} {\bibfnamefont {T.}~\bibnamefont
  {Sleator}},\ }\href@noop {} {\bibfield  {journal} {\bibinfo  {journal} {Phys.
  Rev. A},\ }\textbf {\bibinfo {volume} {46}},\ \bibinfo {pages} {R2213}
  (\bibinfo {year} {1992})}\BibitemShut {NoStop}%
\bibitem [{\citenamefont {Barrett}\ \emph
  {et~al.}(2011){\natexlab{a}}\citenamefont {Barrett}, \citenamefont {Chan},
  \citenamefont {Mok}, \citenamefont {Carew}, \citenamefont {Yavin},
  \citenamefont {Kumarakrishnan}, \citenamefont {Cahn},\ and\ \citenamefont
  {Sleator}}]{Barrett-Advances-2011}%
  \BibitemOpen
  \bibfield  {author} {\bibinfo {author} {\bibfnamefont {B.}~\bibnamefont
  {Barrett}}, \bibinfo {author} {\bibfnamefont {I.}~\bibnamefont {Chan}},
  \bibinfo {author} {\bibfnamefont {C.}~\bibnamefont {Mok}}, \bibinfo {author}
  {\bibfnamefont {A.}~\bibnamefont {Carew}}, \bibinfo {author} {\bibfnamefont
  {I.}~\bibnamefont {Yavin}}, \bibinfo {author} {\bibfnamefont
  {A.}~\bibnamefont {Kumarakrishnan}}, \bibinfo {author} {\bibfnamefont
  {S.~B.}\ \bibnamefont {Cahn}}, \ and\ \bibinfo {author} {\bibfnamefont
  {T.}~\bibnamefont {Sleator}},\ }\href@noop {} {\emph {\bibinfo {title} {Time
  Domain Interferometry With Laser Cooled Atoms}}},\ edited by\ \bibinfo
  {editor} {\bibfnamefont {E.}~\bibnamefont {Arimondo}}, \bibinfo {editor}
  {\bibfnamefont {P.~R.}\ \bibnamefont {Berman}}, \ and\ \bibinfo {editor}
  {\bibfnamefont {C.~C.}\ \bibnamefont {Lin}},\ \bibinfo {series} {Advances in
  Atomic, Molecular and Optical Physics}, Vol.~\bibinfo {volume} {60}\
  (\bibinfo  {publisher} {Elsevier},\ \bibinfo {year} {2011})\ Chap.~\bibinfo
  {chapter} {3}\BibitemShut {NoStop}%
\bibitem [{\citenamefont {Weel}\ and\ \citenamefont
  {Kumarakrishnan}(2003)}]{Weel-PRA(R)-2003}%
  \BibitemOpen
  \bibfield  {author} {\bibinfo {author} {\bibfnamefont {M.}~\bibnamefont
  {Weel}}\ and\ \bibinfo {author} {\bibfnamefont {A.}~\bibnamefont
  {Kumarakrishnan}},\ }\href@noop {} {\bibfield  {journal} {\bibinfo  {journal}
  {Phys. Rev. A},\ }\textbf {\bibinfo {volume} {67}},\ \bibinfo {pages}
  {061602(R)} (\bibinfo {year} {2003})}\BibitemShut {NoStop}%
\bibitem [{\citenamefont {Beattie}\ \emph
  {et~al.}(2009){\natexlab{a}}\citenamefont {Beattie}, \citenamefont {Barrett},
  \citenamefont {Chan}, \citenamefont {Mok}, \citenamefont {Yavin},\ and\
  \citenamefont {Kumarakrishnan}}]{Beattie-PRA(R)-2009}%
  \BibitemOpen
  \bibfield  {author} {\bibinfo {author} {\bibfnamefont {S.}~\bibnamefont
  {Beattie}}, \bibinfo {author} {\bibfnamefont {B.}~\bibnamefont {Barrett}},
  \bibinfo {author} {\bibfnamefont {I.}~\bibnamefont {Chan}}, \bibinfo {author}
  {\bibfnamefont {C.}~\bibnamefont {Mok}}, \bibinfo {author} {\bibfnamefont
  {I.}~\bibnamefont {Yavin}}, \ and\ \bibinfo {author} {\bibfnamefont
  {A.}~\bibnamefont {Kumarakrishnan}},\ }\href@noop {} {\bibfield  {journal}
  {\bibinfo  {journal} {Phys. Rev. A},\ }\textbf {\bibinfo {volume} {79}},\
  \bibinfo {pages} {021605(R)} (\bibinfo {year}
  {2009}{\natexlab{a}})}\BibitemShut {NoStop}%
\bibitem [{\citenamefont {Beattie}\ \emph
  {et~al.}(2009){\natexlab{b}}\citenamefont {Beattie}, \citenamefont {Barrett},
  \citenamefont {Chan}, \citenamefont {Mok}, \citenamefont {Yavin},\ and\
  \citenamefont {Kumarakrishnan}}]{Beattie-PRA-2009}%
  \BibitemOpen
  \bibfield  {author} {\bibinfo {author} {\bibfnamefont {S.}~\bibnamefont
  {Beattie}}, \bibinfo {author} {\bibfnamefont {B.}~\bibnamefont {Barrett}},
  \bibinfo {author} {\bibfnamefont {I.}~\bibnamefont {Chan}}, \bibinfo {author}
  {\bibfnamefont {C.}~\bibnamefont {Mok}}, \bibinfo {author} {\bibfnamefont
  {I.}~\bibnamefont {Yavin}}, \ and\ \bibinfo {author} {\bibfnamefont
  {A.}~\bibnamefont {Kumarakrishnan}},\ }\href@noop {} {\bibfield  {journal}
  {\bibinfo  {journal} {Phys. Rev. A},\ }\textbf {\bibinfo {volume} {80}},\
  \bibinfo {pages} {013618} (\bibinfo {year} {2009}{\natexlab{b}})}\BibitemShut
  {NoStop}%
\bibitem [{\citenamefont {Barrett}\ \emph
  {et~al.}(2011){\natexlab{b}}\citenamefont {Barrett}, \citenamefont {Beattie},
  \citenamefont {Carew}, \citenamefont {Chan}, \citenamefont {Mok},
  \citenamefont {Yavin},\ and\ \citenamefont
  {Kumarakrishnan}}]{Barrett-SPIE-2011}%
  \BibitemOpen
  \bibfield  {author} {\bibinfo {author} {\bibfnamefont {B.}~\bibnamefont
  {Barrett}}, \bibinfo {author} {\bibfnamefont {S.}~\bibnamefont {Beattie}},
  \bibinfo {author} {\bibfnamefont {A.}~\bibnamefont {Carew}}, \bibinfo
  {author} {\bibfnamefont {I.}~\bibnamefont {Chan}}, \bibinfo {author}
  {\bibfnamefont {C.}~\bibnamefont {Mok}}, \bibinfo {author} {\bibfnamefont
  {I.}~\bibnamefont {Yavin}}, \ and\ \bibinfo {author} {\bibfnamefont
  {A.}~\bibnamefont {Kumarakrishnan}},\ }in\ \href@noop {} {\emph {\bibinfo
  {booktitle} {ICONO 2010: International Conference on Coherent and Nonlinear
  Optics}}},\ \bibinfo {series} {Proc. of SPIE}, Vol.\ \bibinfo {volume}
  {7993},\ \bibinfo {editor} {edited by\ \bibinfo {editor} {\bibfnamefont
  {C.}~\bibnamefont {Fabre}}, \bibinfo {editor} {\bibfnamefont
  {V.}~\bibnamefont {Zadkov}}, \ and\ \bibinfo {editor} {\bibfnamefont
  {K.}~\bibnamefont {Drabovich}}}\ (\bibinfo {year} {2011})\ pp.\ \bibinfo
  {pages} {79930Y--1}\BibitemShut {NoStop}%
\bibitem [{\citenamefont {Weel}\ \emph {et~al.}(2006)\citenamefont {Weel},
  \citenamefont {Chan}, \citenamefont {Beattie}, \citenamefont
  {Kumarakrishnan}, \citenamefont {Gosset},\ and\ \citenamefont
  {Yavin}}]{Weel-PRA-2006}%
  \BibitemOpen
  \bibfield  {author} {\bibinfo {author} {\bibfnamefont {M.}~\bibnamefont
  {Weel}}, \bibinfo {author} {\bibfnamefont {I.}~\bibnamefont {Chan}}, \bibinfo
  {author} {\bibfnamefont {S.}~\bibnamefont {Beattie}}, \bibinfo {author}
  {\bibfnamefont {A.}~\bibnamefont {Kumarakrishnan}}, \bibinfo {author}
  {\bibfnamefont {D.}~\bibnamefont {Gosset}}, \ and\ \bibinfo {author}
  {\bibfnamefont {I.}~\bibnamefont {Yavin}},\ }\href@noop {} {\bibfield
  {journal} {\bibinfo  {journal} {Phys. Rev. A},\ }\textbf {\bibinfo {volume}
  {73}},\ \bibinfo {pages} {063624} (\bibinfo {year} {2006})}\BibitemShut
  {NoStop}%
\bibitem [{\citenamefont {Andersen}\ and\ \citenamefont
  {Sleator}(2009)}]{Andersen-PRL-2009}%
  \BibitemOpen
  \bibfield  {author} {\bibinfo {author} {\bibfnamefont {M.~F.}\ \bibnamefont
  {Andersen}}\ and\ \bibinfo {author} {\bibfnamefont {T.}~\bibnamefont
  {Sleator}},\ }\href@noop {} {\bibfield  {journal} {\bibinfo  {journal} {Phys.
  Rev. Lett.},\ }\textbf {\bibinfo {volume} {103}},\ \bibinfo {pages} {070402}
  (\bibinfo {year} {2009})}\BibitemShut {NoStop}%
\bibitem [{\citenamefont {Wicht}\ \emph {et~al.}(2002)\citenamefont {Wicht},
  \citenamefont {Hensley}, \citenamefont {Sarajlic},\ and\ \citenamefont
  {Chu}}]{Wicht-PhysScr-2002}%
  \BibitemOpen
  \bibfield  {author} {\bibinfo {author} {\bibfnamefont {A.}~\bibnamefont
  {Wicht}}, \bibinfo {author} {\bibfnamefont {J.~M.}\ \bibnamefont {Hensley}},
  \bibinfo {author} {\bibfnamefont {E.}~\bibnamefont {Sarajlic}}, \ and\
  \bibinfo {author} {\bibfnamefont {S.}~\bibnamefont {Chu}},\ }\href@noop {}
  {\bibfield  {journal} {\bibinfo  {journal} {Phys. Scr.},\ }\textbf {\bibinfo
  {volume} {T102}},\ \bibinfo {pages} {82} (\bibinfo {year}
  {2002})}\BibitemShut {NoStop}%
\bibitem [{\citenamefont {Chan}\ \emph {et~al.}(2011)\citenamefont {Chan},
  \citenamefont {Barrett},\ and\ \citenamefont
  {Kumarakrishnan}}]{Chan-PRA-2011}%
  \BibitemOpen
  \bibfield  {author} {\bibinfo {author} {\bibfnamefont {I.}~\bibnamefont
  {Chan}}, \bibinfo {author} {\bibfnamefont {B.}~\bibnamefont {Barrett}}, \
  and\ \bibinfo {author} {\bibfnamefont {A.}~\bibnamefont {Kumarakrishnan}},\
  }\href@noop {} {\bibfield  {journal} {\bibinfo  {journal} {Phys. Rev. A},\
  }\textbf {\bibinfo {volume} {84}},\ \bibinfo {pages} {032509} (\bibinfo
  {year} {2011})}\BibitemShut {NoStop}%
\bibitem [{\citenamefont {Davis}\ and\ \citenamefont
  {Narducci}(2008)}]{Davis-JModOpt-2008}%
  \BibitemOpen
  \bibfield  {author} {\bibinfo {author} {\bibfnamefont {J.~P.}\ \bibnamefont
  {Davis}}\ and\ \bibinfo {author} {\bibfnamefont {F.~A.}\ \bibnamefont
  {Narducci}},\ }\href@noop {} {\bibfield  {journal} {\bibinfo  {journal} {J.
  Mod. Opt.},\ }\textbf {\bibinfo {volume} {55}},\ \bibinfo {pages} {3173}
  (\bibinfo {year} {2008})}\BibitemShut {NoStop}%
\bibitem [{\citenamefont {Roach}(2004)}]{Roach-JPhysB-2004}%
  \BibitemOpen
  \bibfield  {author} {\bibinfo {author} {\bibfnamefont {T.~M.}\ \bibnamefont
  {Roach}},\ }\href@noop {} {\bibfield  {journal} {\bibinfo  {journal} {J.
  Phys. B},\ }\textbf {\bibinfo {volume} {37}},\ \bibinfo {pages} {3551}
  (\bibinfo {year} {2004})}\BibitemShut {NoStop}%
\bibitem [{\citenamefont {Weiss}\ \emph {et~al.}(1989)\citenamefont {Weiss},
  \citenamefont {Riss}, \citenamefont {Shevy}, \citenamefont {Ungar},\ and\
  \citenamefont {Chu}}]{Weiss-JOSAB-1989}%
  \BibitemOpen
  \bibfield  {author} {\bibinfo {author} {\bibfnamefont {D.~S.}\ \bibnamefont
  {Weiss}}, \bibinfo {author} {\bibfnamefont {E.}~\bibnamefont {Riss}},
  \bibinfo {author} {\bibfnamefont {Y.}~\bibnamefont {Shevy}}, \bibinfo
  {author} {\bibfnamefont {P.~J.}\ \bibnamefont {Ungar}}, \ and\ \bibinfo
  {author} {\bibfnamefont {S.}~\bibnamefont {Chu}},\ }\href@noop {} {\bibfield
  {journal} {\bibinfo  {journal} {J. Opt. Soc. Am. B},\ }\textbf {\bibinfo
  {volume} {6}},\ \bibinfo {pages} {2072} (\bibinfo {year} {1989})}\BibitemShut
  {NoStop}%
\bibitem [{\citenamefont {Vorozcovs}\ \emph {et~al.}(2005)\citenamefont
  {Vorozcovs}, \citenamefont {Weel}, \citenamefont {Beattie}, \citenamefont
  {Cauchi},\ and\ \citenamefont {Kumarakrishnan}}]{Vorozcovs-JOSAB-2005}%
  \BibitemOpen
  \bibfield  {author} {\bibinfo {author} {\bibfnamefont {A.}~\bibnamefont
  {Vorozcovs}}, \bibinfo {author} {\bibfnamefont {M.}~\bibnamefont {Weel}},
  \bibinfo {author} {\bibfnamefont {S.}~\bibnamefont {Beattie}}, \bibinfo
  {author} {\bibfnamefont {S.}~\bibnamefont {Cauchi}}, \ and\ \bibinfo {author}
  {\bibfnamefont {A.}~\bibnamefont {Kumarakrishnan}},\ }\href@noop {}
  {\bibfield  {journal} {\bibinfo  {journal} {J. Opt. Soc. Am. B},\ }\textbf
  {\bibinfo {volume} {22}},\ \bibinfo {pages} {943} (\bibinfo {year}
  {2005})}\BibitemShut {NoStop}%
\bibitem [{\citenamefont {Slama}\ \emph {et~al.}(2006)\citenamefont {Slama},
  \citenamefont {von Cube}, \citenamefont {Kohler}, \citenamefont
  {Zimmermann},\ and\ \citenamefont {Courteille}}]{Slama-PRA-2006}%
  \BibitemOpen
  \bibfield  {author} {\bibinfo {author} {\bibfnamefont {S.}~\bibnamefont
  {Slama}}, \bibinfo {author} {\bibfnamefont {C.}~\bibnamefont {von Cube}},
  \bibinfo {author} {\bibfnamefont {M.}~\bibnamefont {Kohler}}, \bibinfo
  {author} {\bibfnamefont {C.}~\bibnamefont {Zimmermann}}, \ and\ \bibinfo
  {author} {\bibfnamefont {P.~W.}\ \bibnamefont {Courteille}},\ }\href@noop {}
  {\bibfield  {journal} {\bibinfo  {journal} {Phys. Rev. A},\ }\textbf
  {\bibinfo {volume} {73}},\ \bibinfo {pages} {023424} (\bibinfo {year}
  {2006})}\BibitemShut {NoStop}%
\bibitem [{\citenamefont {Schilke}\ \emph {et~al.}(2011)\citenamefont
  {Schilke}, \citenamefont {Zimmermann}, \citenamefont {Courteille},\ and\
  \citenamefont {Guerin}}]{Schilke-PRL-2011}%
  \BibitemOpen
  \bibfield  {author} {\bibinfo {author} {\bibfnamefont {A.}~\bibnamefont
  {Schilke}}, \bibinfo {author} {\bibfnamefont {C.}~\bibnamefont {Zimmermann}},
  \bibinfo {author} {\bibfnamefont {P.~W.}\ \bibnamefont {Courteille}}, \ and\
  \bibinfo {author} {\bibfnamefont {W.}~\bibnamefont {Guerin}},\ }\href@noop {}
  {\bibfield  {journal} {\bibinfo  {journal} {Phys. Rev. Lett.},\ }\textbf
  {\bibinfo {volume} {106}},\ \bibinfo {pages} {223903} (\bibinfo {year}
  {2011})}\BibitemShut {NoStop}%
\bibitem [{\citenamefont {Anthony}\ and\ \citenamefont
  {Sebastian}(1994)}]{Anthony-PRA-1994}%
  \BibitemOpen
  \bibfield  {author} {\bibinfo {author} {\bibfnamefont {J.~M.}\ \bibnamefont
  {Anthony}}\ and\ \bibinfo {author} {\bibfnamefont {K.~J.}\ \bibnamefont
  {Sebastian}},\ }\href@noop {} {\bibfield  {journal} {\bibinfo  {journal}
  {Phys. Rev. A},\ }\textbf {\bibinfo {volume} {49}},\ \bibinfo {pages} {192}
  (\bibinfo {year} {1994})}\BibitemShut {NoStop}%
\bibitem [{Note2()}]{Note2}%
  \BibitemOpen
  \bibinfo {note} {Measurements of the $B$-gradient from the scattered field
  intensity are not sensitive to the sign of $\beta $. However, the sign can be
  determined using a heterodyne technique to measure the scattered electric
  field amplitude.}\BibitemShut {Stop}%
\bibitem [{\citenamefont {Niebauer}\ \emph {et~al.}(1995)\citenamefont
  {Niebauer}, \citenamefont {Sasagawa}, \citenamefont {Faller}, \citenamefont
  {Hilt},\ and\ \citenamefont {Klopping}}]{Niebauer-Metrologia-1995}%
  \BibitemOpen
  \bibfield  {author} {\bibinfo {author} {\bibfnamefont {T.~M.}\ \bibnamefont
  {Niebauer}}, \bibinfo {author} {\bibfnamefont {G.~S.}\ \bibnamefont
  {Sasagawa}}, \bibinfo {author} {\bibfnamefont {J.~E.}\ \bibnamefont
  {Faller}}, \bibinfo {author} {\bibfnamefont {R.}~\bibnamefont {Hilt}}, \ and\
  \bibinfo {author} {\bibfnamefont {F.}~\bibnamefont {Klopping}},\ }\href@noop
  {} {\bibfield  {journal} {\bibinfo  {journal} {Metrologia},\ }\textbf
  {\bibinfo {volume} {32}},\ \bibinfo {pages} {159} (\bibinfo {year}
  {1995})}\BibitemShut {NoStop}%
\bibitem [{\citenamefont {{Le Gou\"{e}t}}\ \emph {et~al.}(2008)\citenamefont
  {{Le Gou\"{e}t}}, \citenamefont {Mehlst\"{a}ubler}, \citenamefont {Kim},
  \citenamefont {Merlet}, \citenamefont {Clairon}, \citenamefont {Landragin},\
  and\ \citenamefont {Santos}}]{LeGouet-ApplPhysB-2006}%
  \BibitemOpen
  \bibfield  {author} {\bibinfo {author} {\bibfnamefont {J.}~\bibnamefont {{Le
  Gou\"{e}t}}}, \bibinfo {author} {\bibfnamefont {T.~E.}\ \bibnamefont
  {Mehlst\"{a}ubler}}, \bibinfo {author} {\bibfnamefont {J.}~\bibnamefont
  {Kim}}, \bibinfo {author} {\bibfnamefont {S.}~\bibnamefont {Merlet}},
  \bibinfo {author} {\bibfnamefont {A.}~\bibnamefont {Clairon}}, \bibinfo
  {author} {\bibfnamefont {A.}~\bibnamefont {Landragin}}, \ and\ \bibinfo
  {author} {\bibfnamefont {F.~P.~D.}\ \bibnamefont {Santos}},\ }\href@noop {}
  {\bibfield  {journal} {\bibinfo  {journal} {Appl. Phys. B},\ }\textbf
  {\bibinfo {volume} {92}},\ \bibinfo {pages} {133} (\bibinfo {year}
  {2008})}\BibitemShut {NoStop}%
\bibitem [{\citenamefont {Young}\ \emph {et~al.}(2007)\citenamefont {Young},
  \citenamefont {Bonomi}, \citenamefont {Patterson}, \citenamefont {Roller},
  \citenamefont {Tran}, \citenamefont {Vitouchkine}, \citenamefont
  {Gustavson},\ and\ \citenamefont {Kasevich}}]{Young-OSA-2007}%
  \BibitemOpen
  \bibfield  {author} {\bibinfo {author} {\bibfnamefont {B.}~\bibnamefont
  {Young}}, \bibinfo {author} {\bibfnamefont {D.~S.}\ \bibnamefont {Bonomi}},
  \bibinfo {author} {\bibfnamefont {T.}~\bibnamefont {Patterson}}, \bibinfo
  {author} {\bibfnamefont {F.}~\bibnamefont {Roller}}, \bibinfo {author}
  {\bibfnamefont {T.}~\bibnamefont {Tran}}, \bibinfo {author} {\bibfnamefont
  {A.}~\bibnamefont {Vitouchkine}}, \bibinfo {author} {\bibfnamefont
  {T.}~\bibnamefont {Gustavson}}, \ and\ \bibinfo {author} {\bibfnamefont
  {M.}~\bibnamefont {Kasevich}},\ }in\ \href@noop {} {\emph {\bibinfo
  {booktitle} {Proceedings of the International Conference on Laser Science}}}\
  (\bibinfo  {publisher} {Optical Society of America},\ \bibinfo {year}
  {2007})\ p.\ \bibinfo {pages} {LTuH1}\BibitemShut {NoStop}%
\bibitem [{\citenamefont {Suzuki}(1977)}]{Suzuki-CommunMathPhys-1977}%
  \BibitemOpen
  \bibfield  {author} {\bibinfo {author} {\bibfnamefont {M.}~\bibnamefont
  {Suzuki}},\ }\href@noop {} {\bibfield  {journal} {\bibinfo  {journal}
  {Commun. Math. Phys.},\ }\textbf {\bibinfo {volume} {57}},\ \bibinfo {pages}
  {193} (\bibinfo {year} {1977})}\BibitemShut {NoStop}%
\bibitem [{Note3()}]{Note3}%
  \BibitemOpen
  \bibinfo {note} {This choice is not arbitrary. Since the $p$-space wave
  function is an eigenstate of the operator $p^2/2M$, but not $-\protect
  \mathaccentV {hat}05E{\protect \mathcal {M}} z$, we save ourselves some
  effort by choosing $\protect \mathaccentV {hat}05E{B} = p^2/2M$ since
  $e^{\protect \mathaccentV {hat}05E{B}}$ operates on the wave function before
  $e^{\protect \mathaccentV {hat}05E{A}}$.}\BibitemShut {Stop}%
\bibitem [{\citenamefont {Gradshteyn}\ and\ \citenamefont
  {Ryzhik}(2007)}]{Gradshteyn-Book-2007}%
  \BibitemOpen
  \bibfield  {author} {\bibinfo {author} {\bibfnamefont {I.~S.}\ \bibnamefont
  {Gradshteyn}}\ and\ \bibinfo {author} {\bibfnamefont {I.~M.}\ \bibnamefont
  {Ryzhik}},\ }\href@noop {} {\emph {\bibinfo {title} {Tables of Integrals,
  Series, and Products}}},\ \bibinfo {edition} {7th}\ ed.\ (\bibinfo
  {publisher} {Elsevier},\ \bibinfo {year} {2007})\ p.\ \bibinfo {pages}
  {940}\BibitemShut {NoStop}%
\end{thebibliography}%
\end{document}